\documentclass[english,aps,prper,reprint,showpacs,titlepage,longbibliography]{revtex4-2}   
\usepackage{soul}
\usepackage{xcolor}
\soulregister{\textit}{1}

\definecolor{lightgray}{gray}{0.9}

\sethlcolor{lightgray}
\usepackage[utf8]{inputenc}
\usepackage[compatibility=false]{caption} % due to error note 
\usepackage{booktabs} % For better-looking tables
\usepackage{tabularx} % For flexible column widths
\usepackage{ragged2e} % For text alignment
\usepackage{caption} % For caption customization
\usepackage[T1]{fontenc}	
\usepackage{geometry}
\usepackage{times}
\usepackage{hyperref}  
\hypersetup{colorlinks=true,urlcolor=blue,citecolor=blue,linkcolor=blue} 
\usepackage{array}
\usepackage{enumerate}
\usepackage{enumitem}
\usepackage{amsmath}
\usepackage{caption}
\usepackage{amssymb}
\usepackage{tikz}
\usepackage{graphicx}
\usepackage{multirow}
\usepackage{tcolorbox}
\usepackage{ragged2e}
\usepackage{float}
\usepackage[utf8]{inputenc}

\usepackage[normalem]{ulem}
\usepackage{setspace} % Include the setspace package for line spacing control
\usepackage{soul}  % Add this line for text highlighting

\definecolor{lightyellow}{rgb}{1.0, 1.0, 0.6}  % Light yellow color
\definecolor{lightgreen}{rgb}{0.6, 1.0, 0.6}  % Light green color
\usepackage{placeins} % Add this package to control float placement
\usepackage{adjustbox} % For adjusting text width
\begin{document}
\begin{titlepage}

\title{Investigating the Design--Science Connection in a multi-week Engineering Design (ED)-based introductory physics laboratory task}

 \author{Ravishankar Chatta Subramaniam}
 \affiliation{Department of Physics and Astronomy, Purdue University, 525 Northwestern Ave, West Lafayette, IN 47907, U.S.A.} 

 \author{Nikhil Borse}
 \affiliation{Dept. of Physics and Astronomy, Purdue University, West Lafayette, IN 47907, U.S.A.} 
 \author{Amir Bralin}
 \affiliation{Dept. of Physics and Astronomy, Purdue University, West Lafayette, IN 47907, U.S.A.} 
 \author{Jason W. Morphew}
 \affiliation{School of Engineering Education, Purdue University, 610 Purdue Mall, West Lafayette, IN 47907, U.S.A.}
   
 \author{Carina M. Rebello}
 \affiliation{Dept. of Physics, Toronto Metropolitan University, 350 Victoria Street, Toronto, ON M5B 2K3, Canada}
  
 \author{N. Sanjay Rebello}
 \affiliation{Dept. of Physics and Astronomy / Dept. of Curriculum \& Instruction, Purdue University, West Lafayette, IN 47907, U.S.A.} 


\begin{abstract}
Reform documents advocate for innovative pedagogical strategies to enhance student learning. A key innovation is the integration of science and engineering practices through Engineering Design (ED)-based physics laboratory tasks, where students tackle engineering design problems by applying physics principles. While this approach has its benefits, research shows that students do not always effectively apply scientific concepts, but instead rely on trial-and-error approaches, and end up `gadgeteering' their way to a solution. This leads to what is commonly referred to as the ``design--science gap'' - that students do not always consciously apply science concepts while solving a design problem. However, as obvious as the notion of a `gap' may appear, there seems to exist no consensus on the definitions of `design' and `science', further complicating the understanding of this `gap'. This qualitative study addresses the notion of the design-science gap by examining student-groups' discussions and written lab reports from a multi-week ED-based undergraduate introductory physics laboratory task. Building on our earlier studies, we developed and employed a nuanced, multi-layered coding scheme inspired by the Gioia Framework to characterize `design thinking' and `science thinking'. We discuss how student-groups engage in various aspects of design and how they apply physics concepts and principles to solve the problem. In the process, we demonstrate the interconnectedness of students' design thinking and science thinking. We advocate for the usage of the term ``design--science connection'' as opposed to ``design--science gap'' to deepen both design and science thinking. Our findings offer valuable insights for educators in design-based science education. 

    \clearpage
  \end{abstract}

\maketitle
\end{titlepage}
\maketitle
\section{Introduction}

In its 2012 report, the President's Council of Advisors on Science and Technology (PCAST) recommended \textit{``replacing standard laboratory courses with discovery-based research courses''}~\cite[p.16]{pcast_2012}. Engaging college students in STEM (science, technology, engineering, and mathematics)~\cite{kelley2016conceptual, roehrig2021beyond} disciplines can be achieved by providing \textit{``contemporary, authentic research''}~\cite[p.87]{pcast_2012} experiences. The report also recommends the expansion of engineering design-based courses \textit{``in the early years of post-secondary education''}~\cite[p.26]{pcast_2012}. 

In the backdrop of the above-mentioned recommendations, successful preparation of the 21st century STEM workforce largely depends on pedagogical innovations at the post-secondary level~\cite{nrc_dber_2012, cooper2015challenge}. One such innovation is the integration of science and engineering practices with a focus on real-world problems, particularly in the context of an undergraduate introductory physics laboratory~\cite{chao2017bridging, NGSS2013}. Integrating science and engineering can help students become proficient in discipline-specific approaches and, with proper guidance, develop expert-level understanding~\cite{nrc_dber_2012}.

However, despite the evident benefits, numerous studies~\cite{chao2017bridging, vattam2008foundations, kolod_punt_2003, berland2013student} have shown that students often fail to perceive the relevance of science, particularly physics, in engineering design (ED)~\cite{capobianco2018characterizing}. Instead of delving deeply into physics concepts, students frequently tend to employ trial and error approaches to \textit{``gadgeteer''}~\cite[p.63]{berland2013student} (to create practical contraptions and solutions without deeply integrating underlying scientific principles) solutions, leading to what is called the  \textit{``design--science gap''}~\cite[p.1050]{chao2017bridging}, \cite{vattam2008foundations, kolod_punt_2003}. 

Motivated by the aforementioned studies, we began our investigations by searching for the design--science gap in a multi-week, engineering design-based laboratory task where students tackled a prescribed ill-structured problem~\cite{shekoyan2007introducing}. Expanding on earlier work~\cite{subramaniam2023narst, ravi_perc_2023}, our goal was to gain insights into how students approach solving such problems. We soon discovered that there is an absence of consensus on what exactly `design' and `science' may imply, thus muddying the notion of a `gap'. This marked the transition of our thinking from `design--science gap' to `design--science connection', thereby advocating for a more holistic and integrated perspective.

We employed methods of qualitative analysis~\cite{otero2009getting}, leveraging student artifacts such as audio recordings and written reports, submitted in connection with the prescribed ill-structured problem, to gain insights into students' thinking. For our data analysis, we were guided by qualitative research methodologies such as thematic analysis~\cite{braun2006using} and thick description~\cite{brink1987reliability, stahl2020expanding, ponterotto2006brief}, allowing us to explore and contextualize the experiences of the participants. To maintain a structured approach to our analysis, we adopted the Gioia framework~\cite[p.15, 21]{gioia2013seeking}, for conducting and presenting inductive research. This approach allowed us to remain flexible and responsive to the data, enabling a thorough and nuanced understanding of the phenomena under study.

Our objectives include: (i) investigating how students approach a prescribed design problem and how they apply physics principles in their solutions, (ii) developing operational notions to characterize design thinking and science thinking, and (iii) examining the interplay between between students' design thinking and science thinking. Additionally, we also examine the role of scaffolds we had provided during the multi-week ED task with the aim of guiding students to iterate on their solutions. 

Our study offers valuable insights for educators and researchers engaged in integrating engineering design and  science practices. The findings suggest that further research is needed to identify appropriate design-based problems, implement effective instructional interventions, develop robust pedagogical strategies and scaffolds, create relevant assessments, and train graduate teaching assistants to support design-based learning. Additionally, by employing a variety of qualitative research techniques, we contribute to the literature on qualitative methods in physics education research. We hope this study reinforces the importance of bridging design and science in STEM education, and in the process, also provide for an illuminating philosophical reflection.

\section{Literature Review} 

Close on the heels of the 2012 PCAST report, the National Research Council's 2013 report, \textit{Adapting to a Changing World: Challenges and Opportunities in Undergraduate Physics Education}, emphasizes the need for innovative strategies to address the evolving demands of undergraduate physics education~\cite{national2013adapting}. The report urged educators to develop instructional methods that promote active student engagement and create assessments that encompass expert physics learning~\cite[p.100]{national2013adapting}, such as effective oral and written communication, group collaboration, and problem-solving in complex situations, among other skills~\cite[chapter 4]{national2013adapting}.

\subsection{Integrating Engineering and Science Practices}

Though our context is an undergraduate first-year university physics course, we found it valuable to be guided by \textit{A Framework for K-12 Science Education: Practices, Crosscutting Concepts, and Core Ideas}~\cite{NRC2012}. According to this document `science' typically refers to the conventional natural sciences, including physics, chemistry, biology, and, more recently, earth, space, and environmental sciences, while `engineering' is understood in a broad sense as any systematic design practice aimed at solving specific human challenges. The report conceptualizes \textit{``practices''} as integrating both knowledge and skills~\cite[p.41]{NRC2012}, offering an in-depth exploration of engineering and science practices. The report asserts that incorporating engineering practices, the application of science, and the interplay between science, engineering, and technology is essential for all students' science learning~\cite[p.12]{NRC2012}.

In agreement with the above-mentioned report, Cooper {\em et al.}~\cite{cooper2015challenge} advocate for the thoughtful design of learning activities and assessments that may enable instructors to obtain deeper insights into students' understanding. Building on this idea, recent efforts have increasingly emphasized that engineering practices are firmly grounded in science. The goal is to ensure engineering is not merely a  \textit{``build-test-design''} practice, but is a thoughtful and rational scientific endeavor~\cite[p.1, 3]{malmqvist2004lessons}. To reflect this spirit, pedagogical innovations should: integrate science concepts and engineering design (ED), highlight real-world applications to enhance student motivation~\cite{redish2008looking, pintrich2004conceptual}, foster meta-cognitive thinking~\cite{campione2013forms}, and promote evidence-based reasoning~\cite{davidowitz2003enabling, hofsteinchem2004}. These innovations can potentially facilitate learners’ identities as scientists and engineers~\cite{gee2014discourse}, and improve academic success~\cite{schnotz2009some}. Engineering design-based activities can provide an authentic context to promote science learning. Guided by this perspective, and as a departure from the much-debated (and criticized)  \textit{``cookbook''} labs~\cite{karelina2007acting, may2023historical}, we infused engineering design into the physics laboratory to foster the development of science and engineering practices.

\subsection{Engineering Design (ED)}

In the context of our study, we were guided by Capobianco {\em et al.}'s definition of engineering design as \textit{``a recursive activity that results in artifacts $-$ physical or virtual $-$ as well as processes''}~\cite[p.348]{capobianco2018characterizing}. We acknowledge that this definition is broad. In our context, we operationalize this definition to include the following processes of solving an engineering design problem: problem scoping and information gathering, idea generation, project realization, communication and documentation of performance results, and optimization~\cite[p.46, 204]{NRC2012}. 

An `engineering design problem' is client-driven and goal-oriented, and set in an authentic context to engage students. It includes constraints such as cost, time, materials, familiar resources, and tools. The solution is a tangible product or process. Another important characteristic of ED problems is that they have multiple solution paths, requiring teamwork to creatively balance constraints and resources~\cite[p.60, 61]{capobianco2013shedding}. Similar terms such as  \textit{``ill-structured''}~\cite[p.444]{dringenberg2018experiences} and  \textit{``ill-defined''}~\cite[p.226]{cross1982designerly} problems also find mention in the literature, and, for our purposes, we shall not worry about any subtle differences that may exist. 

\subsection{Expansive Framing and Transfer of Learning}

Among our pedagogical goals is to provide appropriate learning supports (or, scaffolds)~\cite{puntambekar2005toward} to maximize science learning, particularly in physics, within the context of the prescribed ED-problem. Strong science learning can lead to better design outcomes, while the pursuit of an effective design can motivate learners to grasp new concepts and expand their physics knowledge. This process involves the transfer of learning not only between design and physics but also across other disciplines. In these aspects, we were guided by the concepts of  \textit{``Expansive Framing''} and  \textit{``Transfer of Learning''}~\cite{engle2012does, sirnoorkar2016students,sirnoorkar2020towards,radloff2018using,mestre2006transfer}.

Expansive Framing involves creating educational opportunities that encourage students to explore and reflect on the application of scientific and mathematical principles to various contexts~\cite[p.31]{radloff2018using}. 

Transfer of Learning occurs when learners apply knowledge or skills acquired in one context to a different, yet related, context~\cite[p.461]{engle2006framing}. In addressing the prescribed ED-problem, students utilize various scientific concepts and principles learned from multiple sources. Additionally, the challenge of solving the problem may motivate them to explore ideas and concepts beyond the syllabus.

\subsection{Design--Science Gap: Situating our Work}

Despite the popularity of ED-based tasks in STEM education~\cite[p.441]{chase2019learning}, studies also reveal that students often do not see the relevance of science in ED~\cite{kolod_punt_2003}. While engineering majors may learn ED practices in their first-year engineering courses, undergraduate physics courses that integrate both ED practices and the learning of science and mathematics, especially within laboratory settings, are noticeably scarce.

Even when ED tasks are introduced in science classrooms, students often fail to consciously apply and explicitly demonstrate their understanding of related science and math concepts during problem-solving activities~\cite{berland2013student, chase2019learning}. Berland {\em et al.}~\cite{berland2013student} caution that students might resort to \textit{``gadgeteering''}—solving design problems through trial and error—without engaging deeply with the underlying scientific and mathematical principles. This tendency, documented across various studies~\cite{chao2017bridging, chase2019learning}, has contributed to the emergence of the term  \textit{``design--science gap''}~\cite{vattam2008foundations}.

However, a closer analysis of these studies reveals a lack of clear definitions for what constitutes design and science thinking, which complicates the notion of a gap. We found no analytic frameworks that clearly elaborate on the definitions of science, design, and the supposed gap between them, particularly in undergraduate physics. 

\subsection{Design Thinking and Science Thinking}\label{sec:II.E}

The terms `design', `design thinking', and `design-based thinking' have generated various interpretations in the literature~\cite{razzouk2012design, li2019design, dalsgaard2014pragmatism, english2023ways}. At its core, design thinking encompasses the way designers perceive, reason, and approach problems. It is both an analytic and a creative process that engages individuals in opportunities to experiment, create prototypes, gather feedback, and refine solutions~\cite[p.330, 334]{razzouk2012design}. In his scholarly essay, Charles Owen~\cite{owen2007design} identifies key traits of design thinking, including its emphasis on invention, its focus on human and environmental concerns, and its aim to create multiple benefits from proposed solutions. Design thinking also involves a holistic approach to problem-solving, relies on diverse forms of communication—written, visual, and mathematical—and is inherently collaborative.

The above definitions align closely with the concept of `engineering design' as outlined in the \textit{Framework for K-12 Science Education}, which describes engineering design as a systematic process grounded in scientific knowledge and models of the physical world~\cite{english2023ways, NRC2012}. Solutions are developed by weighing various factors such as functionality, technological feasibility, cost, safety, aesthetics, and legal compliance. Typically, there is no singular optimal solution, but rather a range of potential solutions, with the best choice determined by the evaluation criteria applied~\cite[p.52]{NRC2012}. While we acknowledge the various terms such as `design', `design thinking', `design-based thinking', and `engineering design (--based thinking)', for the purposes of our paper, we will consider these terms synonymous. Throughout the paper, we will use the term `design thinking'. 

Despite the plethora of definitions on design thinking, Kimbell makes a surprising observation that: 
\begin{quote}
\textit{``Even on a cursory inspection, just what design thinking is supposed to be is not well understood, either by the public or those who claim to practice it''}~\cite[p.288]{kimbell2011rethinking}.
\end{quote}

This uncertainty could arguably be extended to the notion of `science thinking'. While we found some literature defining `science' and `scientific thinking'~\cite{lehrer2006scientific, slavit2021student}, we found just one study~\cite{denick2013stem} which explicitly makes use of the term `science thinking'. We resonate with Denick {\em et al.}'s assertion that \textit{``science thinking revolves around the dynamic refinement of scientific understanding of the natural world''}~\cite[p.3]{denick2013stem}. Lehrer and Schauble~\cite{lehrer2007scientific} describe `scientific thinking' to encompass three dimensions: (i) logical reasoning, (ii) the ability to revise inherent theories, and (iii) involvement in practical activities. The boundaries between these dimensions being evidently blurry~\cite[p.656]{janouvskova2023scientific}. While we recognize the importance of the terms `science thinking' and `scientific thinking', for the purposes of our paper, we will consider these terms synonymous. Throughout the rest of the paper, we will use `science thinking' to refer to thinking based on science-related (specifically physics-related, in our context) ideas, concepts, and principles.

Though our view is that there may not exist a single, universally acceptable one-size-fits-all definition for the above-mentioned terms, for the purposes of this study, we found it useful to draw upon the definition provided by Vattam and Kolodner~\cite{vattam2008foundations}, which distinguishes `design' as the concrete world of direct experiences and product creation, and `science' as the abstract world of physical laws and causal explanations, with the caveat that these terms are open to interpretation and may vary depending on context, disciplinary focus, educational objectives, and the specific goals of a study or project. The operational definitions and application of these terms in this empirical study will be illustrated through the coding schemes presented in Tables~\ref{tab:WoT} and~\ref{tab:BoS}, and Appendices A and B. 

\subsection{Onward to Design--Science Connection}
The absence of universally accepted definitions for terms such as design thinking and science thinking is not necessarily a limitation but rather an opportunity. These terms, which serve as conceptual tools, invite diverse interpretations, extensions, and applications, reflecting the complexity of integrating design and science in educational contexts. Our contention is that if definitions themselves are not universal, the notion of a gap between design and science is inherently fluid and context-dependent.

Moreover, conceptualizing this issue merely as a gap may unintentionally perpetuate a deficit mindset, wherein the focus is on what is lacking rather than on how to enhance the integration between design and science. As researchers and educators, we advocate for a paradigm shift from identifying and addressing a gap to actively fostering a `design--science connection'. This shift involves directing our efforts towards the development of instructional strategies that incorporate well-structured scaffolding and effective pedagogical practices~\cite{lee2013science}. By bridging the conceptual and practical (and apparent) divide between Design (the problem-solving tasks students engage with) and Science (the fundamental physical laws or principles underlying these tasks), we can create a more cohesive, enriching, and integrative learning experience. This approach not only addresses the limitations of a gap-focused perspective but also enhances the potential for meaningful and applied learning in STEM education.

\subsection{Data Sources and Rationale}
Research has shown that by observing what students are doing, such as engaging in peer interactions, it is possible to ascertain how they are thinking~\cite{luna2018teachers}. Peer interactions enable learners to leverage diverse expertise, encounter varied viewpoints, question, explain, exchange ideas, and articulate their reasoning~\cite{etkina2014thinking, firetto2023embracing}. Small-group discussions help students~\textit{``explore their ideas and move from understandings that may often be naive towards more valid scientific ideas and explanations''}~\cite[p.70]{bennett2010talking}, apart from contributing to STEM achievement, motivation, engagement, and problem-solving~\cite{wieman2014similarities}. Our earlier studies~\cite{subramaniam2023narst, ravi_perc_2023} reveal that capturing students' `in-the-moment' thinking provides rich data on their approach to the ED problem. Group discussions provide insights into students'  \textit{``interthinking''}~\cite[p.143]{wilkinson2010developing} in naturalistic settings. At the same time, the inherent value of writing in science \cite{alev2010perceived} cannot be understated. Writing helps students develop their scientific ideas and engage in analytical writing by incorporating mathematical knowledge that may be difficult to express orally, among other benefits. In this study, we use both transcripts of students' group discussions and written reports, which also provides data triangulation for our analysis.

\section{Research Questions}
 
This study investigates how student-groups engage in design and science thinking in the given engineering design (ED) task. Though our initial focus was to identify the design--science gap, we transitioned to thinking in terms of design--science connection as the research evolved. We sought to characterize student-groups' design and science thinking and investigate the emerging themes. By analyzing how student-groups’ thinking evolved during a multi-week task, we highlight the dynamic interplay between design and science thinking. Additionally, we examined how the laboratory activities, provided as `scaffolds' through the multi-week task, influenced student-groups' thinking, with the goal of informing more effective educational strategies.

The specific research questions (RQs) are:

\vspace{0.3cm}

\indent
\begin{minipage}{0.90\linewidth}
    \textbf{RQ1:} \textit{In the context of the prescribed engineering design (ED)-based laboratory task, in what ways can we characterize student-groups' `design thinking' and `science thinking', and in what ways do the connections between design and science unfold?}
    \label{RQ1}
\end{minipage}

\vspace{1em}  % Adjust the space between the two blocks as needed
\begin{minipage}{0.90\linewidth}
    \textbf{RQ2:} \textit{In what ways do the scaffolds influence student-groups' design and science thinking, and students' progression towards a solution?}
    \label{RQ2}
\end{minipage}

\section{Methodology}

\subsection{Context of Study}

\begin{figure}[!htbp]
\caption{Timeline of the ED-based task following the ED Process Model~\cite{atman2007engineering, mosborg2005conceptions, subramaniam2023narst}. Each laboratory session lasted 110 minutes. Refer Tables \ref{ED_schedule} and \ref{prompts} for details.}
\fbox{\includegraphics[width=0.96\linewidth]{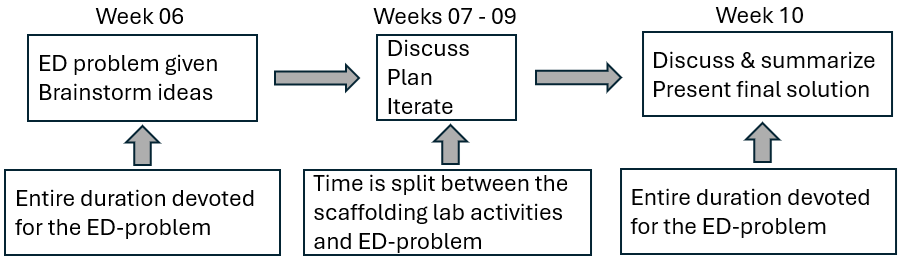}}
\label{EDP_this_study}
\end{figure}

This study took place in a large-enrollment, first-semester, calculus-based, undergraduate physics course at a large U.S. Midwestern land-grant university. A significant reform in this course over the last five years has been integrating ED into the laboratory component of the course. In the 2021-22 academic year, the course had an enrollment of about 1000 students in the Fall semester and 1200 students in the Spring semester. Students in the course include about 25\% women, 10\% under-represented minorities, and about 10\% international students. The current study is situated in the Fall of 2021.

\begin{figure}[tb]
\centering
\begin{tcolorbox}
\justify{``Pristine natural habitats of endangered species such as the gorillas in the Congo River basin are becoming increasingly rare. Today, these habitats and the endangered species that inhabit them need to be not only protected but even sustained by humans. As a member of a team of engineers volunteering for a non-profit organization, you are asked to design a system that can launch a payload of food to an island in the Congo River and land it safely for the gorillas. Each payload is about 50 kg, and it must be delivered to a habitat area located on an island in the Congo River that is about 150 m away from the riverbank. To avoid contributing to global warming, the client wants you to use a means that would minimize the carbon footprint of the delivery. Furthermore, the client also wants to ensure that the habitat remains pristine, so that neither humans nor a robotic machine must disturb the flora and fauna of the habitat while delivering the food.''}   
\end{tcolorbox}
\caption{Engineering Design problem statement provided to the students. Images taken from Wikipedia ~\cite{wikimedia_iamge_gorilla, wikipedia_image_congo} were also provided.}
\label{ED_problem_statement}
\end{figure}

A vast majority (over 80\%) of the students in this course aspired to be future engineers, while the remaining were science majors. 
Engineering majors are concurrently enrolled in a first-year engineering course focused on ED, which creates a unified engineering design experience for these students spanning multiple courses. Non-engineering majors are provided tutorials on ED prior to starting the ED problem. 

The course adopts a principle-based approach~\cite{chabay2015matter}, with the content divided into three units, each focused on fundamental physics principles: momentum, energy, and angular momentum. Common threads include a focus on systems thinking, modeling, and making assumptions and approximations. The Momentum Principle~\cite[ch.2]{chabay2015matter} relates the change in momentum (\( \Delta \vec{p} \)) of a system to the net force (\( \vec{F}_{\text{net}} \)) acting on it, multiplied by the interaction duration (\( \Delta t \)). The time interval must be small enough for the net force to remain nearly constant, i.e., \( \Delta \vec{p} = \vec{F}_{\text{net}} \cdot \Delta t \). The Energy Principle~\cite[ch.6]{chabay2015matter} states that the change in energy of the system (\( \Delta E_{\text{system}} \)) plus the change in energy of the surroundings (\( \Delta E_{\text{surroundings}} \)) equals zero, i.e., \( \Delta E_{\text{system}} + \Delta E_{\text{surroundings}} = 0 \). The Angular Momentum Principle~\cite[ch.11]{chabay2015matter} relates a change in angular momentum (\( \Delta \vec{L} \)) of a system to the net torque (\( \vec{\tau}_{\text{net}} \)) acting on it, with the interaction duration (\( \Delta t \)), i.e., \( \Delta \vec{L} = \vec{\tau}_{\text{net}} \cdot \Delta t \).

The weekly schedule includes two 50-minute lectures, one 110-minute laboratory, and one 50-minute recitation focused on problem-solving. The laboratory segment spanned 13 sessions, with this study focusing on weeks 06 to 10. The laboratory schedule may be seen in Tables~\ref{sem_schedule} and~\ref{ED_schedule}. 

\begin{table}[!htbp]
\captionsetup{justification=raggedright,singlelinecheck=false} % Caption left justified
\caption{Schedule of labs for the semester. This study focuses on ED problem II, occurring in weeks 06-10, hereafter referred to as `ED-problem'.}
\begin{ruledtabular}
\begin{tabular}{lll} % Adjust the width as needed
\textbf{Weeks} & \textbf{Lab activities based on} & \textbf{Design Challenge} \\ 
\hline
01 -- 05 & Momentum Principle &  ED problem I\\
06 -- 10 & Energy Principle & ED problem  II \\
11 -- 13 & Angular Momentum Principle &  ED problem III\\
\end{tabular}
\label{sem_schedule}
\end{ruledtabular}
\end{table}
 
\subsection{Participants}

The participants of this study were undergraduate students in two laboratory sections for which the lead author served as the Graduate Teaching Assistant (GTA). From a total of 28 teams, each consisting of 2 or 3 students, data from 14 teams was selected based on our earlier studies~\cite{subramaniam2023narst, ravi_perc_2023}. The lead author's position as the TA for the course allowed us to select 14 teams that were representative of the variability of student experiences, thus addressing issues of saturation, under-representation, and over-representation \cite{lopez2013sampling}.

\renewcommand{\arraystretch}{1}
\begin{table*}[!htbp]
\begin{center}
\captionsetup{justification=raggedright} % Caption left justified
\caption{Schedule of lab activities, pertaining to the ED-task, provided as scaffolds within the laboratory in weeks 06 to 10. Audio-recordings of group discussions and written reports submitted by the students form the data for this study.}
\begin{ruledtabular}
\label{ED_schedule} 
\begin{tabular}{p{0.05\linewidth} p{0.40\linewidth} p{0.50\linewidth}}
\textbf{Week} & \textbf{Lab activities (Scaffolds)} & \textbf{Deliverables} \\
\hline
06 & Students presented with the ED problem. Student-groups brainstorm ideas for possible solution approaches. & Written lab report + an audio recording of the group discussion in response to prompts - Problem scoping and solution generation. \\
07 & Hands-on task: Exploring springs: Force and Energy - Motivating a launch
mechanism (Initial stage). VPython simulation-1. & Written lab report for the hands-on task. Audio recording of group discussions in response to ED problem related prompts -- Discussing the initial stage. \\
08 & Hands-on task: Exploring Drag Force: Motivating use of a parachute (Middle stage). Dropping a coffee filter or firing a projectile under the influence of air drag. VPython simulation-2. PhET simulation. & Written lab report for the hands-on task. Audio recording of group discussions in response to ED problem related prompts -- Discussing the middle stage \\
09 & Hands-on task: Exploring Coefficient of Restitution (COR). Motivating soft-
landing the payload (Final stage). VPython simulation-3. & Written lab report for the hands-on task. Audio recording of group discussions in response to ED problem related prompts -- Discussing the final stage. \\
10 & Final review and submission of ED problem solutions. VPython simulation-4. & Summarize the several iterations of the group's solution and submit a written lab report containing group's proposed solution to the ED problem. \\
\end{tabular}
\end{ruledtabular}
\end{center}
\vspace{-0.5cm}
\end{table*}

\subsection{ED problem}\label{sec:IV.C}

Seven essential characteristics of engineering design, as outlined by Capobianco {\em et al.}~\cite{capobianco2013shedding}, guided the development of our ED problem: client-driven and goal-oriented, authentic context, constraints, use of familiar materials, resources, and tools, solution is an artifact or process, multiplicity of solutions, and teamwork.

Students worked in teams to first identify the overall context of the problem, then generated possible ideas or solutions using what they knew about the problem as well as using relevant physics knowledge. Even as the teams executed the scaffolded laboratory activities, in parallel, they developed a plan, applied their scientific knowledge, exchanged ideas with other teams, received feedback from the GTA, made iterations, refined their solution approaches, and presented their final solution in the week 10 lab report. The ED-process model~\cite{atman2007engineering, mosborg2005conceptions, subramaniam2023narst} which guided this study is presented in Figure \ref{EDP_this_study}.

Although the ED problem (see Figure~\ref{ED_problem_statement})~\cite{wikimedia_iamge_gorilla, wikipedia_image_congo} was situated in the laboratory component of the course, the scaffolding was not confined to the lab alone. Expansive Framing~\cite{engle2012does} was used to integrate and scaffold learning experiences into the course, including prompting students during lectures to reflect on the relevance of physics concepts to the ED problem and ensuring that weekly recitation problems were related to the same challenge.

Most of the lab sessions, to facilitate the integration of science in ED,  were comprised of two interwoven parts designed to aid student groups' progress towards their solutions. The first part involved a traditional hands-on inquiry-based experiment using PASCO equipment~\cite{pasco_me5300} and sensors, with data collection and analysis performed using PASCO Capstone™ software~\cite{pasco_capstone}. The second part included computational activities using PhET~\cite{PhET} and VPython~\cite{vpython}, which is an extension of the Python programming language particularly useful for visualizing physical simulations. 

Weeks 06 and 10 were entirely made available for students to engage with the ED problem, while in weeks 07 - 09 the class time was roughly equally divided between lab activities and the ED task. In week 06, they engaged in problem scoping and solution generation (brainstorming). In weeks 07-09, they completed inquiry-based activities related to the ED problem, using hands-on equipment, VPython code, and PhET simulations to help build conceptual and computational models for: (i) launching a payload along an inclined ramp using a spring-loaded device (week 07), (ii) dropping a coffee filter or firing a projectile under the influence of air drag (week 08), and (iii) bouncing a falling object off a hard surface with a certain coefficient of restitution (week 09). Each week, groups were asked to revisit the ED problem and audio record their group discussions in response to a set of prompts presented to them. Students presented their solutions in a combined group report in week 10. The details may be seen in Tables~\ref{ED_schedule} and~\ref{prompts}. 

As is standard practice in most laboratory sessions, the lead author provided a short introduction to the students about the laboratory activities and the deliverables for the week. TAs are expected to circulate around the class, answering any questions students may have. The lead author was assisted by an undergraduate teaching assistant. 

\begin{figure*}[htbp]
\caption{Coding Process - Stage 1. Transcripts were coded by textual segmentation. Refer Section~\ref{sec:IV.E(i)}, Tables~\ref{tab:WoT} and~\ref{tab:BoS}.}
\fbox{\includegraphics[width=0.96\linewidth]{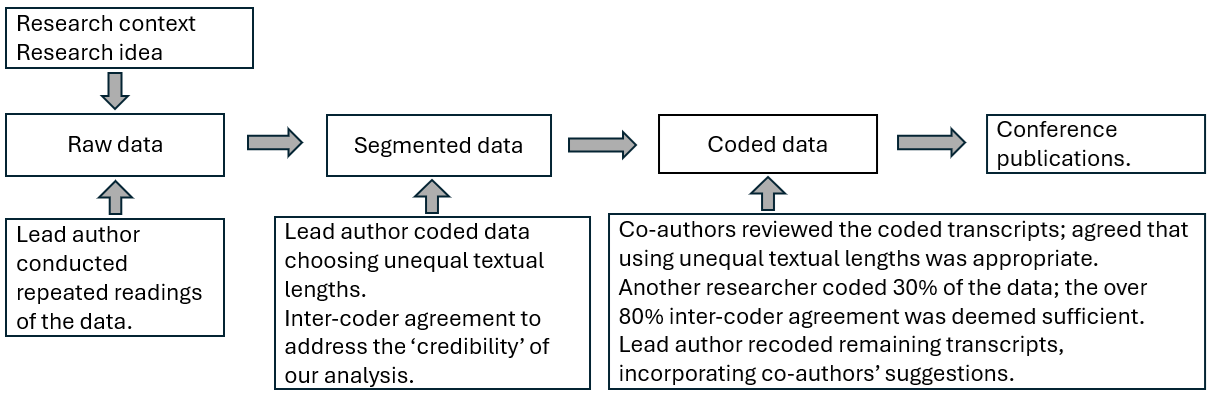}}
\label{coding_stage_1}
\end{figure*}

\subsection{Data Collection}

\renewcommand{\arraystretch}{1}
\begin{table*}[!htbp]
\begin{center}
\captionsetup{justification=raggedright} % Caption left justified
\caption{ED task-related prompts presented to students from weeks 06 to 10. During weeks 06 to 09, responses were captured through audio recordings of group-discussions, while week 10’s response was submitted as a written report.}
\begin{ruledtabular}
\label{prompts} 
\begin{tabular}{p{0.04\linewidth} p{0.92\linewidth}}
\textbf{Week} & \textbf{Prompts} \\
\hline
06 & Brainstorm multiple potential solutions to this problem. Focus on key ideas, principles/concepts involved, and limitations. Complete solutions will be developed in subsequent weeks. \\
07 & Reflect on how today’s lab activity contributes to your understanding of the ED problem. In what ways did you approach the initial stage (the launch) of food delivery to the gorillas, and how might you approach it differently now? Did your ideas change, and what aspects of the lab made you think differently about the design challenge? \\
08 & Reflect on how today’s lab activity contributes to your understanding of the ED problem. In what ways did you approach the middle stage (after the launch) of food delivery to the gorillas, and how might you approach it differently now? Did your ideas change, and what aspects of the lab made you think differently about the design challenge? \\
09 & Reflect on how today’s lab activity contributes to your understanding of the ED problem. In what ways did you approach the final stage (the landing) of food delivery to the gorillas, and how might you approach it differently now? Did your ideas change, and what aspects of the lab made you think differently about the design challenge? \\
10 & Restate the problem. Include the design context, identifying stakeholders, mention the criteria and constraints, and outline the significance of the problem. 
Describe the iterations in the problem-solving approach, list relevant physics principles and concepts, outline your assumptions and approximations., and state the limitations in your design. Present a sketch for your design. 
\\
\end{tabular}
\end{ruledtabular}
\end{center}
\vspace{-0.5cm}
\end{table*}

In each lab, from week 06 to week 09, each team of students was asked to audio-record their discussions using the mobile phone of one of the team members.  Students were provided a specific set of prompts (see Table~\ref{prompts}) to guide their team discussion. We expected that most student-groups would adopt a launch mechanism (we realized later that we may have unintentionally preempted an approach) for delivering the payload. The simulation, hands-on activities, and prompts were all designed with this assumption in mind. However, students were explicitly informed that they could pursue any solution approach. 

One team, among the 14 selected, did not submit a written report in the final week. This left us with 56 audio recordings (14 per week) and 13 written reports, providing a substantial volume of data for our qualitative study. Teams were renumbered from Team-01 to Team-14 and will be referred to by these designations for the remainder of this paper.

\renewcommand{\arraystretch}{1}
\begin{table*}
\begin{center}
\captionsetup{justification=raggedright} % Caption left justified
\caption{Codes for `Ways of Thinking', code descriptions, and example quotes. \newline{DST - Design Thinking; SCT - Science Concepts; MAT - Mathematical Thinking; MER - Metacognitive Reflection.}}
\begin{ruledtabular}
\label{tab:WoT} 
\begin{tabular}{p{0.05\linewidth} p{0.25\linewidth} p{0.40\linewidth} p{0.25\linewidth}}
\textbf{Code} & \textbf{Code Descriptor} & \textbf{Illustrative Quote} & \textbf{Code Justification} \\
\hline
DST & Define the problem, identify constraints, brainstorm multiple solutions, iterate, select the best solution, consider design issues / aspects, prototype the solution, test, communicate.  & \hl{\textit{``an electric drone which could carry enough weight. It could fly a payload over and drop off and leave without really interacting with anything. The problem could be the charge of the drone. Yeah we don’t know how long the distance is and it might be difficult to charge it in the middle of the Congo. It might be hard to find a place to take off and to land. Lifting 50 kg will be hard for a drone to lift. It’d have to be one heck of an electric drone.''}} \newline{(Team-05; Week 06)} &  Identifies a design constraint - Can the drone carry 50 kg? \newline{Addresses design issues - How to charge the drone?} \newline{Addresses a design criteria - Not to disturb the habitat.} \\
\hline
SCT & Organization, cause and effect, systems, scale – absolute and relative, models, change and rate of change. In our context, it may be expected that the students refer to concepts of projectile motion, drag force among other things. & \hl{\textit{``I would say that this spring system is probably not the best idea because at least with what we’ve found with our data the displacement of the spring with the amount of mass we were using was not enough to get a whole lot of force. And with the velocity they’re trying to achieve for the actual problem itself this does not seem a very likely solution unless we had a very long launching system.''}} (Team-07; Week 07) & Referring to physics concepts - Considering if it may be practical for a spring to have large enough compression and spring constant to generate enough speed for launch. Energy conversion implied. 
\newline{There is some element of DST in terms of what type of a spring may be required for the design.}\\
\hline
MAT & It may refer to a scientific statement in terms of a relation among several variables and constants. Proportional reasoning is acceptable. Units analysis. Use of explicit expression / equation.  & \hl{\textit{``I am just estimating here … and shooting it that way so you get more altitude … with more altitude and the shape of the object considered more air resistance and further slowing down.''}} \newline{(Team-03; Week 08)} & Proportional reasoning - How drag force increases with distance of fall. 
\newline{There is probably some SCT in the form of an indirect reference to the formula for drag force.} \\
\hline
MER  & Reflect on their ideas / recall previous notions and comment / reflect and modify / critically review.  & \hl{\textit{``so after this lab what we’ve realized is that either this new discovery of the coefficient of restitution …will either hurt launching an object or help it because we could either destroy the environment by not bouncing or we can retain some energy and bounce it multiple times instead of letting out all the energy on the environment.''}} \newline{(Team-06; Week 09)} & Reflecting on their learning post the lab activity with bouncing balls.
\newline{Also coded for - }
\newline{DST - Paying attention to the design criteria that habitat's integrity should not be compromised.}
\newline{SCT - Thinking about the energy loss due to impact.}\\
\end{tabular}
\end{ruledtabular}
\end{center}
\vspace{-0.5cm}
\end{table*}

\subsection{Data Analysis}

Student-groups recorded discussions in the laboratory or corridors, and the resulting background noise made it extremely challenging to uniquely associate conversation segments with individual speakers. Moreover, the final report in week 10 was a group effort. These factors, along with the fact that students worked in groups in all labs, influenced our decision to choose the  \textit{``grain size''} or  \textit{``unit of analysis''}~\cite[p.10]{otero2009getting} to be that of a group. Each recording, ranging from 3 to 5 minutes, was manually transcribed and coded by the lead author. This traditional `pen and paper'  \textit{``codus operandi''}~\cite[p.22]{saldana2009introduction} allowed for direct engagement with the data, offering valuable insights through close interaction with the material. The text length of the transcripts ranged from about 400 to 800 words each. The lead author meticulously took handwritten notes while reviewing the raw data multiple times, documenting thoughts and decisions in Word documents and Excel spreadsheets. He drew inspiration from the words of B. O’Dwyer: 
\begin{quote}
\textit{``Whatever process of analysis you use, there is no substitute for knowing your data intimately''}~\cite[p.404]{odwyer_qlr_messy_intimate}.
\end{quote}

To address  \textit{``reliability''} and for  \textit{``ensuring rigor''} in our qualitative inquiry, we were guided by Morse {\em et al.}~\cite[p.14]{morse2002verification}. The lead author engaged in periodic  \textit{``peer debriefing''}~\cite[p.14]{morse2002verification} sessions with the co-authors, not  only to review and refine the coding schema, but also to develop an iterative  \textit{``constructive''}~\cite[p.15]{morse2002verification} procedure for the study. The data analysis evolved through two stages.

\subsubsection{Stage 1 Coding}\label{sec:IV.E(i)}

In Stage 1 (see Figure~\ref{coding_stage_1}), our initial efforts were focused on investigating the design--science gap~\cite{subramaniam2023narst, ravi_perc_2023}. In this stage we looked into students' `ways of thinking'~\cite{ravi_perc_2024, english2023ways}, and the basis of students' statements. The codes with detailed descriptors and examples are presented in Tables~\ref{tab:WoT} and~\ref{tab:BoS}. In both cases, textual segments of varying lengths were chosen as coding units. This flexible approach was adopted after careful consideration of factors such as the complexity of student discussions and the unique conversation styles within each group~\cite{nihas2020, otero2009getting}. The selection of coding units was guided by factors such as communication styles, thematic shifts, speaker transitions (this being the most challenging to decipher from the audio), contextual completeness, and, above all, the nature of our research questions. According to Elliot~\cite[p.2856]{elliott2018thinking}, \textit{``there is no simple answer''} to the choice of \textit{``the chunk of data''} to be coded, and \textit{``it depends on the study and your aims within it''}. Furthermore, fixed text lengths might overlook the richness and diversity of these discussions, as some groups engaged in dynamic exchanges while others spoke more linearly.  It is important to point out that multiple codes (code co-occurrence)~\cite{guest2012introduction},~\cite[p.2853]{elliott2018thinking} can be applied to a single segment. To adjust for the variable lengths of the transcripts and the use of unequal textual lengths, we calculated fractional coding frequencies for each team and averaged them across all groups. 

An important consideration in our analysis was whether to include error bars in the bar charts. In an earlier study~\cite{ravi_perc_2023}, we included error bars because our arguments relied on code frequencies, necessitating the establishment of statistical significance for differences in those frequencies. However, for the current study, the qualitative nature of our work posed challenges in meaningfully defining error, particularly given that the researcher functions \textit{``as an instrument''}~\cite{wa2020researcher, creswell2017research, guba1981criteria, patton2002qualitative} in qualitative research. Standard practices for error bars, such as using standard deviation, standard error, or confidence intervals, are not directly applicable. Although custom error bars could be considered, they would still require a clearly defined notion of coding error, which remains ambiguous, perhaps even questionable, in qualitative research.

Despite extensive searches, we found no clear guidance in the literature on the use of error bars for bar charts in qualitative research. Given our focus on the interpretive aspects of qualitative data rather than the statistical significance of code percentages, we decided not to include error bars in Figures~\ref{ways_of_thinking} and~\ref{basis_statements}. Out of curiosity, we queried AI platforms such as Perplexity AI, ChatGPT, and MS Copilot regarding the appropriateness of error bars for such data. Their responses were notably consistent, stating that \textit{``Error bars on bar charts of code frequencies from qualitative data are generally inappropriate unless quantifiable measures of variability are included''}~\cite{perplexity_ai, openai_chatgpt, ms_copilot}. While insights from AI platforms may not be definitive, we believe this perspective may be of interest to the research community. These considerations further supported our decision to exclude error bars in this study.

The code percentages~\cite[p.2853]{elliott2018thinking} from Stage 1 coding are majorly used only in two of the Findings \&\ Discussions sections, namely: Sections~\ref{sec:V.B} and~\ref{sec:V.C}.

\begin{figure*}
\caption{Coding Process - Stage 2. Data Reduction inspired by the Gioia Chart~\cite[p.21]{gioia2013seeking}. Written reports included for `Triangulation'. Refer to Appendices A and B.}
\fbox{\includegraphics[width=0.96\linewidth]{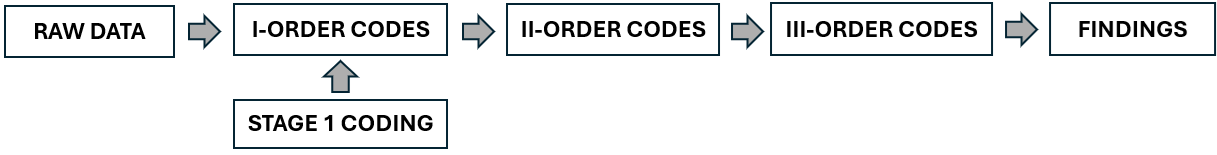}}
\label{data_reduction_stage_2}
\end{figure*}
 
\subsubsection{Stage 2 Coding}\label{sec:IV.E(ii)}
Initially guided by the first stage of coding, we aimed to enhance the level of detail. To address concerns about the unequal lengths of text segments in stage one coding, we inductively developed a more comprehensive and multi-layered coding scheme, grounded in the data. For data triangulation~\cite[p.37]{brink1987reliability}, we used multiple data sources—audio recordings and written reports. The coding process for this stage is depicted in Figures~\ref{data_reduction_stage_2} and~\ref{coding_stage_2}. Separate coding schemata developed for the transcripts and written reports  may be seen in Appendices A and B (or Tables~\ref{tab:stage_2_code_transcripts} and~\ref{tab:stage_2_code_reports}). 

For data reduction~\cite[p.13]{guest2012introduction}, we were guided by the data structure model proposed by Gioia {\em et al.}~\cite[p.21]{gioia2013seeking}, which is illustrated in Figure~\ref{data_reduction_stage_2}. This detailed, multi-layered approach helped to enrich our qualitative analysis, providing deeper insights into our research questions, particularly concerning the `design--science connection'.

The coding schema underwent several revisions, and only the final version is presented in this study. Two co-authors independently assisted with the coding after being briefed on the project’s goals. An inter-coder reliability (ICR) exceeding 80 percent was deemed sufficient in both stages of the coding process, a decision taken after much debate and research. As to what may be an appropriate percentage of ICR agreement, we observe that \textit{``there is little consensus on how to approach reliability in qualitative research''}~\cite[p.72:18]{mcdonald2019reliability}. McDonald {\em et al.} even caution that, for some studies, an ICR test may be \textit{``potentially harmful''}~\cite[p.72:19]{mcdonald2019reliability}, a view echoed by O'Connor and Joffe who argue that an ICR test may not be appropriate for every qualitative study and should not be considered a \textit{``magic bullet''}~\cite[p.11]{oconnor_icr} to automatically ensure reliability.  We were committed to documenting our analysis and while Stage 1 coding was useful, its insufficiency led us to engage in more nuanced Stage 2 coding. However, we were wary that relying too heavily on code percentages could undermine the qualitative nature of our study. These concerns led us to prioritize `thick description'~\cite{brink1987reliability, younas2023proposing, stahl2020expanding, ponterotto2006brief, leeds-hurwitz2019} to enhance the trustworthiness of our findings. The minor differences in coding were not deemed significant enough to affect the results, particularly since `thick description' allows \textit{``the reader to enter the research context''}\cite[p.26]{stahl2020expanding},\cite[p.38]{brink1987reliability}. We also questioned whether an ICR was necessary when using thick description for our analysis. Although we did not find satisfactory answers, we chose to retain coding for two key reasons: (i) we began with coding and had detailed results, and (ii) coding helped us verify consistency between themes and thick descriptions~\cite[p.2]{oconnor_icr}. The detailed coding process provided a valuable audit trail that guided our discussions and decisions when finalizing the themes and descriptions. Based on our experience, we believe that coding is a valuable first step for beginners in qualitative research who wish to employ thick description. It provides an organized framework for structuring data and ensures that rich, detailed narratives emerge from the analysis.

\subsubsection{Guiding Frameworks}

Given our context, research goals, research questions, the nature of our data, and the type of analysis we conducted, we found useful guidance in a variety of theoretical or conceptual frameworks in different stages of our analysis. 

Following the principle \textit{``letting the data speak to the researcher''}~\cite[p.8]{urquhart2022grounded}, we strove to let the data guide our analysis. Our approach involved flexible notions of design and science thinking, and the iterative development of a coding scheme from substantial data sources (group discussions and written reports). Given the inductive nature of our analysis and the evolving focus of our study, we received - only for the purposes of data analysis - some guidance from the grounded theory method approach outlined by Urquhart~\cite{urquhart2022grounded}. 

Our coding process followed the Gioia Framework~\cite{gioia2013seeking}, while the presentation of our findings was guided by the MIRACLE framework for Thick Description. In our thematic analysis~\cite{braun2006using},~\cite[p.8] {peel2020beginner}, we employ thick description (with the authors' interpretation) and direct student quotes (to capture the informants' voices). The MIRACLE framework's seven key criteria~\cite[p.7, 8, and 9]{younas2023proposing} informed the articulation of our results: Meaningful (clear articulation of participants' thinking), Interpretative (balanced consideration of participant perspectives and researchers' interpretations), Rational (alignment of methods, data, and findings within the research context), Authentic (genuine representation of participants' thinking), Contextualized (interpretations drawn from the data, with researchers' biases made transparent), Linked (cohesive connections between methods, research questions, and findings), and Emic (detailed, context-centered account).

\renewcommand{\arraystretch}{1}
\begin{table*}
\begin{center}
\captionsetup{justification=raggedright} % Caption left justified
\caption{Codes for `Behind the Words: The Basis of Student Statements', code descriptions, and example quotes. \newline{HT - Hands-on Task; VP - VPython and PhET Simulations; PK - Physics Knowledge; GK - General Knowledge.}}
\begin{ruledtabular}
\label{tab:BoS} 
\begin{tabular}{p{0.04\linewidth} p{0.25\linewidth} p{0.44\linewidth} p{0.25\linewidth}}
\textbf{Code} & \textbf{Code Descriptor} & \textbf{Illustrative Quote} & \textbf{Code Justification}\\
\hline
HT & Students make references to the hands-on laboratory experiments using PASCO hardware and software. Could be the current lab / the previous, including those in the weeks 01–05.  & \hl{\textit{``This was basically Hooke's law. So we’re going to be using that. Apply to the catapult with the spring force and the stretch. We’ve to calculate the force to the stretch ratio and hopefully do the calculations.''}} \newline{(Team-05; Week 07)} & Directly referring to the hands-on laboratory activity. 
\newline{This was also coded for PK, probably gained from coursework.}\\
VP & Students make references to the PhET simulations and / or VPython simulations. Since it would be hard to decipher which simulation activity the students may be referring to, it was decided to club them together.  & \hl{\textit{``Also how inconsistent our actual data is from simulations and stuff…because our actual data from capstone is very different than the Python simulation …yeah that's true we had a technical error yeah …so calculating drag force .. like by looking at a video link manually is much more difficult than …digitally ….obviously. ''}} \newline{(Team-01; Week 08)}  & Making references to the simulation activities in the lab. \newline{Also coded for HT - Referring to the hands-on lab activities with data collection via PASCO software.}\\
PK & Statements made based on their physics knowledge gained via coursework in school / university, and from textbooks.  & \hl{\textit{``Do we have a theoretical idea if we use a parachute about how much the speed will reduce by or do we have to do the calculations for that. We probably have to do a timed parachute release.''}} \newline{(Team-03; Week 06)} & Referring to calculation methods possibly based on coursework. \newline{The reference to `timed parachute release' probably reflects GK.}\\
GK & Statements based on common knowledge and everyday experience.   & \hl{\textit{``Do not know how to say it in English…But it’s a place where you can jump. Oh, a trampoline yes yes instead of a parachute so it feels safe and eco-friendly or maybe like an airbag, something like you know how like you know when people jump out of a building they use huge air sacks … when they practice snowboarding.. you do not need to worry about bouncing off. It kind of like absorbs it.''}} \newline{(Team-02; Week 06)} & This is during the brainstorming session of week 06. Statements based on experience or common knowledge.  \\
\end{tabular}
\end{ruledtabular}
\end{center}
\vspace{-0.5cm}
\end{table*}

In examining the interplay between student groups' design and science thinking, Mestre's Transfer of Learning framework~\cite{mestre2006transfer} offered valuable guidance. Furthermore, the concept of  \textit{``distributed scaffolding''}, as articulated by Kolodner and Puntambekar~\cite{kolod_punt_2003}, proved instrumental helping us analyze the impact of scaffolds on students' thinking.

Most importantly, since our data is grounded in a physics laboratory, our analysis and interpretation were strongly guided by our disciplinary understanding of Newtonian Mechanics. While students occasionally drew on ideas from other areas of physics, their science thinking predominantly centered around concepts of motion, force, and energy, framed within the conceptual framework of Newtonian Mechanics. 

The co-authors reviewed the themes and narrative descriptions, and the lead author incorporated their feedback to ensure consistency with the coding. Extensive discussions were held on the most effective ways to present the qualitative findings. Figure~\ref{proces_flow} depicts the process flow diagram from Data to Findings.

\section{Findings \&\ Discussion}

\begin{figure*}[!htbp]
\caption{Coding Process - Stage 2. Multi-layer coding scheme separately developed for transcripts and written reports, guided by the Gioia Framework~\cite{gioia2013seeking}. Incorporated `Thick Description'~\cite{stahl2020expanding}. Refer to Appendices A and B.}
\fbox{\includegraphics[width=0.96\linewidth]{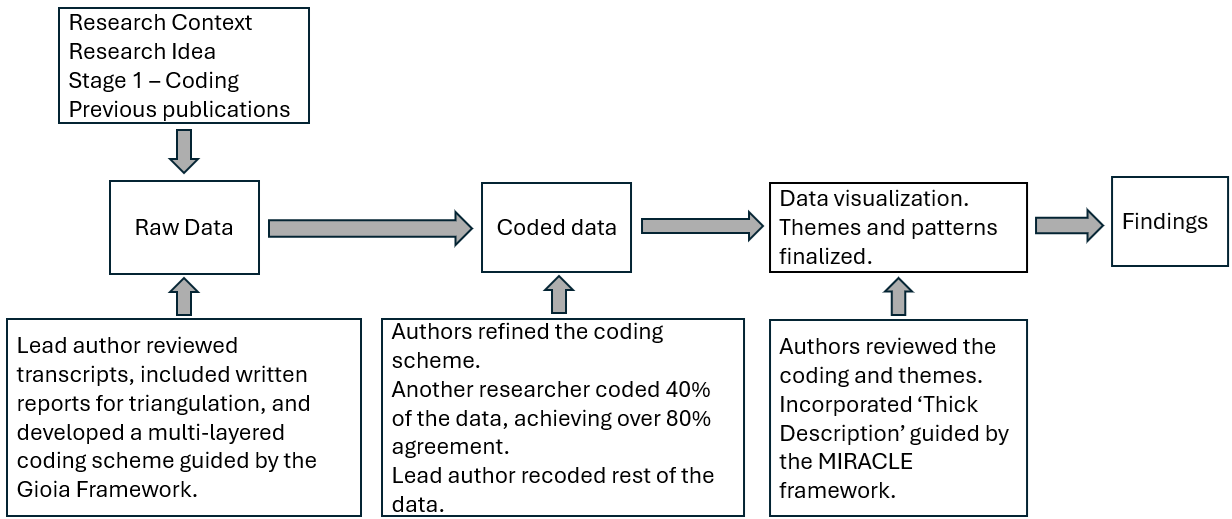}}
\label{coding_stage_2}
\end{figure*}

In this section, we present our findings on how the selected 14 student groups navigated the prescribed engineering design (ED) task. We focus on the emergent themes in students' design and science thinking, and how the design-science connections unfolded in their work, addressing~\hyperref[RQ1]{RQ1}. To address~\hyperref[RQ2]{RQ2}, we examine how the scaffolds—particularly the lab activities—may have influenced students' thinking. These findings are drawn from both stages 1 (see Figure ~\ref{coding_stage_1}) and 2 (see Figure~\ref{coding_stage_2}) of our coding process, with the latter stage playing a more prominent role. It is important to note that our discussion does not rely solely on coding; we majorly employ thick description to convey the researchers' perspectives, and incorporate participants' voices through carefully and thoughtfully chosen direct quotations. As Ponterotto rightly notes:

\begin{quote}
\textit{``A thickly described Discussions section of a qualitative report successfully merges the participants' lived experiences with the researcher's interpretations of these experiences...''}~\cite[p.547]{ponterotto2006brief}.
\end{quote}

In stage 1, we coded only the transcripts, for which the coding schemata may be seen in Tables~\ref{tab:WoT} and~\ref{tab:BoS}. Figures~\ref{ways_of_thinking} and~\ref{basis_statements} visually represent the variation of code percentages (for all groups taken together) across weeks 06 to 09.  In stage 2, as previously mentioned, we coded both transcripts and written reports. The coding schemata for this stage are detailed in Appendices A and B, with Figures~\ref{DST_SCT_disc} and~\ref{DST_SCT_reports} offering visual representations of the coding results.

In Section~\ref{sec:V.A}, we outline the variety of solution approaches considered by the student-groups. Section~\ref{sec:V.B} explores the variation in students' design, science, mathematical, and reflective thinking through the multi-week task. Section~\ref{sec:V.C} examines the factors that may have influenced students' thinking during the task and provides an overarching view of how the provided scaffolds shaped their thought processes. These sections offer partial insights into both of our research questions.

In Sections~\ref{sec:V.D} through~\ref{sec:V.I}, we delve deeper towards addressing our research questions. Section~\ref{sec:V.D} has a particular focus on students' design thinking, while Section~\ref{sec:V.E} investigates their physics thinking. Section \ref{sec:V.F} explores the blurred boundaries between the often-perceived dichotomy of design and science. The importance of Mathematical Thinking in science and engineering can never be discounted, and this aspect is discussed in detail in Section \ref{sec:V.G}. We conclude by offering our perspectives on the notion of the `design-science gap' in Section~\ref{sec:V.I}, ultimately emphasizing the need for educational objectives to enhance the connection between design and science. Although these sections primarily address~\hyperref[RQ1]{RQ1}, the reader will notice how the discussion, quite unavoidably, also intersects with~\hyperref[RQ2]{RQ2}.

Our response to~\hyperref[RQ2]{RQ2} is interwoven throughout most sections, with Sections~\ref{sec:V.A},~\ref{sec:V.C},~\ref{sec:V.D(iii)},~\ref{sec:V.E(iii)},~\ref{sec:V.F(i)}, and~\ref{sec:V.F(iv)} addressing it more explicitly.

Since this was a multi-week activity, we were particularly interested in how students reflected on their design and science thinking throughout the process. Section~\ref{sec:V.H} addresses this, simultaneously drawing connections to both research questions.

As previously noted, students' thinking emerges in complex ways, and readers will observe that all sections maintain some connection to both research questions. Our analysis is guided by a thematic exploration, without being confined to a single team or a specific set of teams. This study aims to uncover patterns that can inform future interventions, while also providing insights to guide pedagogical decisions, with an awareness of the limits of generalization~\cite{maxwell2014generalization}.

\subsection{Variety of Solution Approaches}\label{sec:V.A}
While our analysis remains largely agnostic to the solution approaches taken by student groups, it would be illustrative to highlight some of the initial solution paths considered by students. 

The most popular approach involved launching the load using a catapult or trebuchet, closely followed by the use of drones and ziplines. Other innovative ideas included raft systems, automatic underground tubing systems, amphibious automated guided vehicles, autonomous sky-cranes, boats, tug-boat ferry systems, pressure guns, and airplane drops, among others. These diverse approaches reflect the students' extensive general knowledge, creativity, willingness and eagerness to explore out-of-the-box ideas, and interdisciplinary knowledge. It demonstrates their ability to think critically about the design problem, weigh in various science principles, engage deeply with the project, and explore a wide range of technologies and methods. Among the interesting conversation snippets during the brainstorming session (week 06)  was from Team-03: \hl{\textit{``We could do a composter-cannon. What do you guys think about that? ...Actually what is it? Explain what it is.... Like you’d have a cannon that uses compressed air to launch the payload and you’d have a parachute attached to the payload so that it wouldn’t damage any of the trees or wildlife. The parachute would also be biodegradable...''.}} The reader would note that this conversation segment also captures a thought-provoking interaction among the group members. The manner in which students seamlessly interweave science - not just physics - concepts into their design is palpable.  Interestingly, the idea of a composter-cannon was entirely novel to the lead author before this study, highlighting that there is much we can learn from our students. 

Opinions on adopting a launch mechanism varied widely, ranging from mild criticism to outright rejection. Team-07 offered a gentle and balanced view, noting: \hl{\textit{``I would say that this spring system is probably not the best idea because at least with what we’ve found with our data the displacement of the spring with the amount of mass we were using was not enough to get a whole lot of force''.}} The reader may also observe that the team's thinking is guided by the scaffolding hands-on activity (see Table~\ref{ED_schedule}) in week 07. Conversely, Team-09 in week 07 expressed strong reservations about the launch mechanisms, feeling constrained by the scaffolds. They remarked that insisting on a catapult \hl{\textit{``would require a very big and very powerful spring so even though this lab was informative I don’t think it’s necessarily a good solution to the problem and not necessarily a practical solution.. I mean .. I think they’re going to push us towards springs ..I think that’s what they’re going to give us next. But there are better launching methods''.}}  This `feedback' underscores the importance of remaining open to alternative student solution approaches, in addition to highlighting an important lesson for us: we may have prematurely favored a sub-optimal approach. What is particularly notable in these conversation segments is student-groups' motivation to debate and refine their ideas. It is easy to note that the students are thinking in terms of the magnitude of the spring constant while referring to a ``very powerful'' spring, and about dimensions of their contraption while referring to a ``very big'' spring.  While some teams were uncertain about the best approach, Team-06 clearly identified what to avoid and proceeded with their rocket-based contraption (see Figure~\ref{image_A_12}), which we will address in more detail in the later sections.

The initial brainstorming session facilitated the generation of a wide range of ideas. As educators, we got a sense of the range of solution approaches students explored. It may be noted that the classroom instructions merely asked students ``brainstorm'' their ideas without any specific guidance on how to brainstorm. We wonder how our data and findings might have differed if we had  offered our students a more structured guidance on brainstorming techniques. For our future iterations we intend to make use of the four key rules for effective brainstorming proposed by Kaufman and McCuish~\cite[p.7, 8]{kaufman2002getting}: avoid judgment; prioritize quantity over quality; embrace unconventional ideas; and build upon each other's contributions. 

In their written reports, most teams settled on the catapult approach or a variation of it.  This could be due the influence of the accompanying scaffolds - hands-on and virtual lab activities (see Table~\ref{ED_schedule}) - that may have prompted students to think about the catapult solution. A few of the teams chose alternative solution methods such as drones and ziplines.

We acknowledge that some, or perhaps all, of these solution approaches may not fully meet the task requirements. However, our goal was to remain objective in evaluating the approaches and not be judgemental of students' work. Accordingly, the spirit of our coding schema is to remain respectful to students' choices.

\subsection{Students' Ways of Thinking}\label{sec:V.B}

\begin{figure*}[t] % or [!ht] for better placement control
\centering
\caption{Student-groups' Ways of Thinking. Refer to Table~\ref{tab:WoT} for details.
\newline{DST - Design Thinking; SCT - Science Thinking; MAT - Mathematical Thinking; MER - Metacognitive Reflection}}
\fbox{\includegraphics[width=0.96\textwidth]{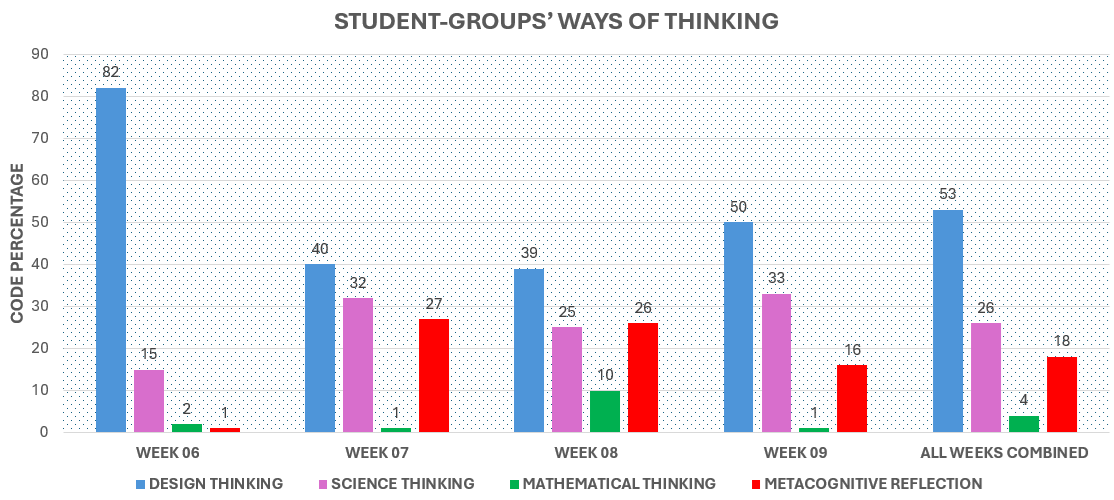}}
\label{ways_of_thinking}
\end{figure*}

In our earlier studies \cite{subramaniam2023narst, ravi_perc_2023, ravi_perc_2024}, as mentioned in Section~\ref{sec:IV.E(i)}, we had explored student-groups' `Ways of Thinking'~\cite{ravi_perc_2024, english2023ways, dalal2021developing, slavit2019stem}. We started the current study by exploring four `Ways of Thinking' students display, namely: Design Thinking (DST), Science Thinking (SCT), Mathematical Thinking (MAT), and Metacognitive Reflection (MER). Figure~\ref{ways_of_thinking}, based on the coding scheme in Table~\ref{tab:WoT}, presents broad trends in the thinking of student groups through the task. 

\begin{figure*}
\caption{From Data to Findings - Process Flow. For Data Reduction, we adopted the Gioia Framework~\cite[p.21]{gioia2013seeking}. For Thick Description, we were informed by the MIRACLE Framework~\cite[p.10]{younas2023proposing}, while for Thematic Analysis we were  guided by the methodology outlined by Braun and Clarke ~\cite{braun2006using}.}
\fbox{\includegraphics[width=0.96\linewidth]{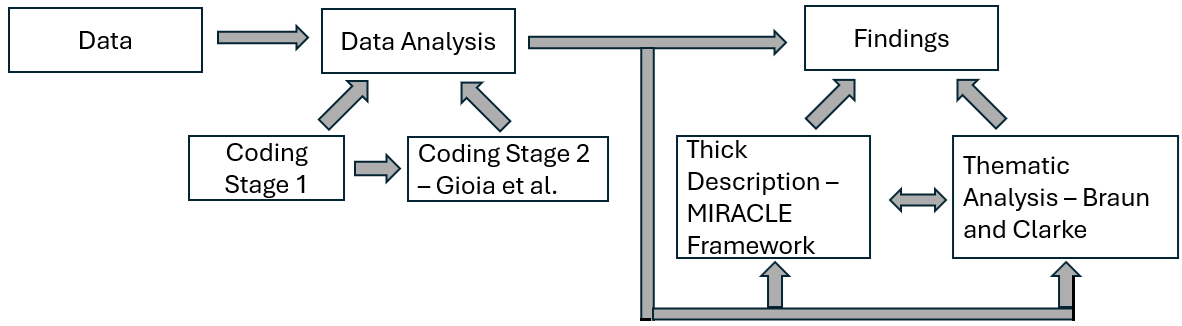}}
\label{proces_flow}
\end{figure*}

A striking feature, which we will elaborate on in the following sections, is the near absence of mathematics in conversations, a trend consistent across all lab sessions. Student groups exhibit fairly consistent metacognitive reflection (MER), except during the initial brainstorming session (week 06), which is understandable as it is the first in the sequence. Design-based thinking dominates students' thinking throughout. In the brainstorming session, students eagerly shared ideas, likely causing the predominance of design thinking. Although the relative proportion of design thinking decreases after the first session, it remains fairly consistent thereafter. Science-based thinking, particularly in physics, is lower during the brainstorming session, which is understandable as students may initially focus less on physics details.

The question we posed is whether these trends, particularly the comparison between design thinking (DST) and science thinking (SCT), are indicative of a design--science gap. However, arguing solely on the basis of relative code frequencies or percentages has a few obvious drawbacks. First, this coding process includes repetitive statements or ideas from different members of a group. Second, if we rely completely on code frequencies or percentages, the issue of statistical significance needs to be addressed. Given our intent to retain a qualitative flavor in our analysis, we did not deem it to be an appropriate approach. Additionally, as Hammer discusses in his essay on the `productiveness' of student conversation~\cite[p.423]{hammer1995student}, a deeper question is whether students were merely making broad statements indicative of design and science thinking, or were they engaging in thoughtful inquiry. Is it not possible that students spend time repeating ideas, engaging in unproductive ways, and making inaccurate statements? It may be premature to discuss any `gap' with the information gathered at this stage, which is why we transitioned to a more nuanced and multi-layered coding scheme, elaborated in the Appendices A and B, the results of which may be seen in Figures~\ref{DST_SCT_disc} and~\ref{DST_SCT_reports}. 

\subsection{Behind the Words: The Basis of Student Statements}\label{sec:V.C}

\begin{figure*}[t]
\centering
\caption{Behind the Words: Basis of Student-groups' Statements. Refer Table~\ref{tab:BoS} for details. 
\newline{ HT - Hands-on Tasks; VP - VPython and PhET Simulations; PK - Physics Knowledge; GK - General Knowledge}}
\fbox{\includegraphics[width=0.96\textwidth]{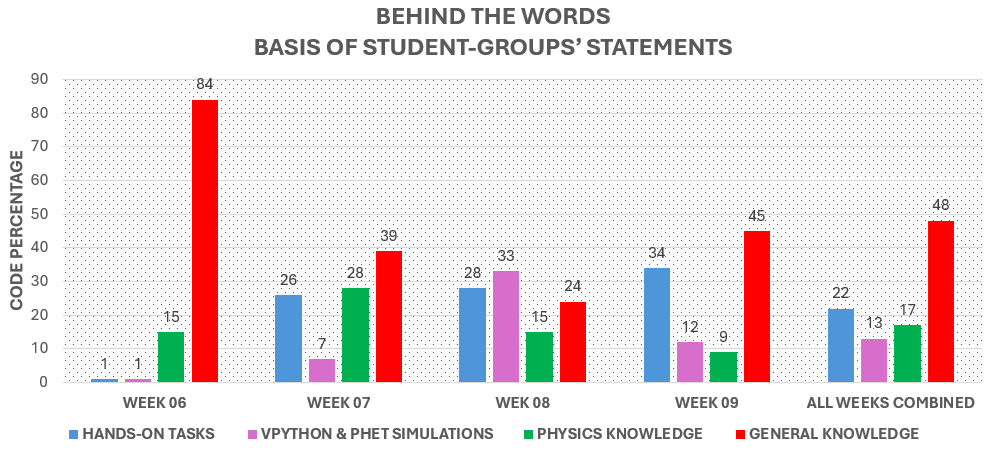}}
\label{basis_statements}
\end{figure*}

As evident in Table~\ref{ED_schedule}, we provided scaffolds to guide students in the ED task. Our approach included hands-on tasks, VPython, and PhET simulations in the laboratory, along with lectures and recitations outside, as illustrated in Figure~\ref{EDP_this_study}. We aimed to explore how these supports influenced student-groups' thinking. Figure~\ref{basis_statements}, based on the coding scheme in Table~\ref{tab:BoS}, offers insights into what may have possibly influenced students' thinking during their conversations. While these findings may not be definitive, they help instructors understand the context behind students' words, and  offer a glimpse into how students `transfer' ideas from different sources. 

Students had the entire duration of week 06 to brainstorm their ideas. With no hands-on task or simulations to work on, it is probably not surprising that their ideas were based on their physics knowledge (from lectures, recitation, textbook, etc.) and general knowledge, with the latter dominating. One question we debated was whether there exists a clear division between what we coded as GK (general knowledge) and PK (physics knowledge). Our experience with the data taught us that the line is indeed blurry. However, the division has practical value at least for the purposes of this study. At the same time, there may not be much value in insisting on a rigid classification. The blurry line between the two is evident in Team-03's statement in week 06: \hl{\textit{``The other idea was using water as a mode of transport and that would be essentially a raft system where we would build the least water invasive raft ..what I mean by water invasive is that ..of course if we stick a motor under a boat it is going to propel the boat forward''}}, and we believe it would be acceptable if one classified this as being based on general knowledge (GK).

An important feature of our coding scheme is the existence of code co-occurrences. The following statement is an example in which codes VP (VPython and PhET simulations) and HT (hands-on task) (see Table~\ref{tab:BoS}) co-occurred.  Team-07 blended its experience with hands-on apparatus and the simulations in week 07 by stating: \hl{\textit{``I would say that this spring system is probably not the best idea because at least with what we’ve found with our data the displacement of the spring with the amount of mass we were using was not enough to get a whole lot of force, and with the velocity they’re trying to achieve for the actual problem itself this does not seem a very likely solution unless we had a very long launching system''.}} This team remained unconvinced that launching the payload is a good approach, and one may note how they are being quite critical about a catapult launch. They are also drawing from their experiences with both the hands-on tasks and simulations. The team displayed design thinking by understanding that a 50 kg payload would need large springs with a high spring constant. They also demonstrated science thinking by indirectly applying the energy principle to argue that the velocity generated would need to be large enough to be launched over a distance of 150 m. Student groups think in complex ways, and this is a fine example blending several codes. The interplay between design and science is quite evident here. As an aside, the reader may also note that breaking off the text in between while coding would interrupt the flow, and would not do justice to the student conveying the idea, illustrating why we chose unequal textual lengths in our first coding stage. 

The general trend was that hands-on tasks had a stronger impact on students' thinking, and the reader may be surprised at the reversal in week 08, as evident in Figure~\ref{basis_statements}. Though one may tend to attribute it to random chance, a look at Table~\ref{ED_schedule} would show that there were two simulation tasks given to the students in that one session in addition to the hands-on task. This could be a possible explanation for the outcome. These simulations were explicitly on terminal velocity, and this is captured by Team-06 as: \hl{\textit{``we have and also one thing we would probably want to have a lower velocity because it requires a lot more force to slow down and likelihood of us reaching terminal velocity with such a heavy payload in such a short distance is unlikely''.}} This team demonstrates a profound conceptual understanding that a heavy falling object, unless it falls through a sufficient distance, may not attain terminal velocity despite the drag force. As much as students discuss the physics of a falling object, one can also observe how they weigh in (pun, intended) the design considerations of \textit{``such a heavy load''}, once again highlighting that students' thinking is too complex to be sorted into silos. 

The observation that students appear to rely more on hands-on tasks than simulations seems to reinforce both  \textit{``anecdotal evidence''}~\cite{greenhalgh2005can, nsfconsulting2024, kauffman2024} and research which suggest that students learn most effectively through direct engagement \cite{moye2014learning, flick1993meanings}. However, there remains an ongoing debate about the relative effectiveness of hands-on versus virtual experiences~\cite{darrah2014virtual}, \cite{ekmekci2015case}. One caveat we would like to add is that it is natural to recall most vividly from the most recent experience, which could introduce  \textit{``recency bias''}~\cite[p.1]{fudenberg2014learning} and potentially influence our findings.

\subsection{Design Thinking - Key Takeaways}\label{sec:V.D}

\begin{figure*}
\centering
\caption{Design Thinking vs. Science Thinking - Variation across weeks 06 - 09. Refer to Appendices A for details. Design Details refers to `Design Considerations' (Section~\ref{sec:V.D(i)}), `Design Metrics' (Section~\ref{sec:V.D(ii)}), and 'Contraption Mechanism' (Section~\ref{sec:V.D(iii)}). Refer Section~\ref{sec:V.E(i)} for `Physics Principles' and Section~\ref{sec:V.E(ii)} for `Physics Vocabulary'. No / Low detail is coded as Level 1; Moderate / High detail is coded Level 2. The descriptive narration will make evident the levels.}
\fbox{\includegraphics[width=0.96\textwidth]{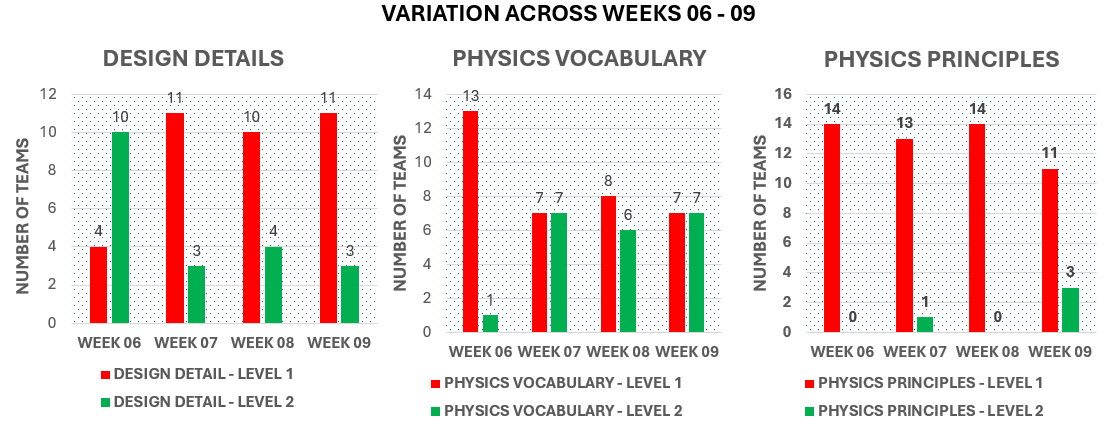}}
\label{DST_SCT_disc}
\end{figure*}

Guided by our coding of transcripts and written reports (see Appendices A and B), we summarize key findings on students' design thinking in the following subsections. Figures~\ref{DST_SCT_disc} and~\ref{DST_SCT_reports} illustrate these findings. In subsection~\ref{sec:V.D(i)}, we examine the extent to which students consider factors such as economic cost, criteria and constraints, habitat integrity, safety, and stakeholders. In subsection~\ref{sec:V.D(ii)}, we analyze the numerical details provided by students regarding the accuracy of payload delivery, payload dimensions, physical aspects of the contraption, carbon footprint assessment, and measures to ensure habitat integrity. In subsection~\ref{sec:V.D(iii)}, we explore how students detail their contraption mechanisms. And finally, in subsection~\ref{sec:V.D(iv)}, we reflect on strategies to enhance students' design thinking.

\subsubsection{Design Considerations - Generic mentions only}\label{sec:V.D(i)}
In our analysis, we refer to `design considerations' to include: design dimensions, economic cost, criteria and constraints, habitat integrity, safety aspects, and stakeholders.  These found mention in the discussions or reports of all the groups, but none of the teams elaborated beyond a cursory mention. This is probably natural in a physics course, as these ideas are seldom discussed in a physics classroom, except perhaps the physical dimensions of the design. None of the teams went on to elaborate on the physical dimensions of their contraption in detail, though some teams did mention that a large and highly stiff spring may be needed to launch a 50 kg load, and that the load may be just too large to be powered by a drone. \hl{\textit{``50 kg is a lot for a drone. Therefore we need a heavy duty electric drone''}}, said Team-06. The contraption, \hl{\textit{``would have to be pretty big in order to get it to the island and making sure it lands safely at the island interrupting any of the environment  in order for this idea to work''}}, said Team-04. Beyond this level, across all teams, there was no further detailing on the dimensions of the contraption, whatever it may be. 

\subsubsection{Design Metrics: Opportunities for Deeper Detailing Missed?}\label{sec:V.D(ii)}
In our analysis, we refer to `design metrics' to encompass specific numeric details of: how to achieve and ensure accuracy of payload delivery, payload dimensions, physical aspects of the contraption, assessing carbon-footprint, and measures to ensure habitat integrity. 
As teams iterated on their solution, we observed that students missed an opportunity to address the above-mentioned aspects of design metrics. There was a lack of detail on how to achieve delivery accuracy, determine payload dimensions, assess carbon footprint, and ensure habitat integrity, though there were direct or indirect but rudimentary references to these issues. As an example, Team-09, in week 08, had this to state about the dimensions of the parachute: \hl{\textit{``this [drag] force is dependent on ... it’s dependent on the two things that we could change about it are the cross sectional area and the coefficient of drag so we wanted to maximize how far it can travel''}}. Drawing upon their hands-on experiences in the lab, this team considered the area of cross section of the parachute as a factor that can affect the drag force, but did not get into numerical details pertaining to the actual size of the parachute. At the same time, expecting such details may be pitching it too high for the students given the limited time and resources at their disposal. 

The trend of providing minimal detail on metrics persisted into the final written reports of week 10. The absence of detailing may be attributed to the lack of specific guidance or prompts to direct students to consider these aspects in their discussions. It could also be partly due to the fact that these aspects are more common in engineering classrooms and not typically the focus in introductory physics classrooms. This finding underscores a disciplinary gap: while detailing design metrics is integral to engineering, it is not typically prioritized in physics education. Addressing this gap may require intentional scaffolding and the inclusion of more specific prompts in future iterations to better integrate these interdisciplinary skills.

\subsubsection{Contraption Mechanisms - Minimal Elaboration}\label{sec:V.D(iii)}
Another observation from the data is that student groups did not think beyond providing basic details for the mechanisms of their contraption. For example, teams that considered using a catapult mentioned that large springs might be needed, the setup might be large, air resistance would need to be overcome, parachutes might need to be deployed, and the landing must be gentle while maintaining habitat integrity. Teams that considered using drones were aware that drones might not be able to carry a heavy load of 50 kg and that their use might not meet the design criteria. Whether it is practical or reasonable to expect more details in an introductory classroom is open to debate. Despite this being a multi-week activity, the teams did not iterate on these aspects. It appears that while the scaffolds may have been effective in advancing students' thinking in physics (which is positive), they did little to promote design-based thinking. 
Lastly, apart from broadly weighing the pros and cons of their design during the brainstorming session, there was only a decreasing reference to design limitations as the weeks progressed, with almost no mention by week 09. In hindsight, we realize that it would have been beneficial if students had been explicitly asked to engage in a `feasibility study' of their design at each stage.

\subsubsection{Design Thinking - Raising the Bar}\label{sec:V.D(iv)}
The findings reveal that the teams were not directly addressing the limitations of their designs, though there were occasional, indirect references suggesting they considered such limitations. For instance, Team-02 initially explored using catapults and parachutes and expressed concerns about the risks of launching a load and the impact of weather on the flight path. They later switched to using a zipline, believing it to be a better solution due to energy considerations in comparison to a launching mechanism. However, they were still concerned about the load causing unintended accidents upon landing. The team continued to evaluate these options into the subsequent lab sessions, debating whether the zipline was preferable because it would avoid wind deflection and be less affected by drag force. This team's deliberations highlights the complexity of real-life problems and the challenges teams faced, especially given the time constraints in a classroom setting. Additionally, while the week 10 written report gave students an opportunity to summarize and develop their ideas further, most teams' details remained skeletal, lacking significant iteration over their discussions, as may be evident from Figure~\ref{DST_SCT_reports}.

In summary, learners would benefit from increased structured guidance and iterative feedback on design thinking from educators. In our future iterations, we intend to incorporate Siverling {\em et al.}'s suggestion that it is important to prompt \textit{``students to justify their design decisions or reflect on scientific reasonings for their design decisions''}~\cite[p.311]{siverling2021initiates}. In our introductory course, learners come with varying backgrounds and may be unaware of what constitutes a good design and how to apply physics concepts and principles. Expectations based on our coding schema in Appendices A and B can be communicated through a rubric to help students progress toward their solutions, while taking care not to overwhelm them.

\subsection{Use of Physics - Key Takeaways}\label{sec:V.E}
Guided by our coding of transcripts and written reports (see Appendices A and B), we summarize key findings on students' use of physics in the following subsections. Figures~\ref{DST_SCT_disc} and~\ref{DST_SCT_reports} illustrate these findings. In subsection~\ref{sec:V.E(i)}, we consider how students applied the fundamental physics principles such as momentum, energy, and angular momentum. In subsection~\ref{sec:V.E(ii)}, we discuss the range and depth to which students used and applied physics terms and concepts. In subsection~\ref{sec:V.E(iii)}, we outline the extent to which students used physics to analyze the payload's take off, flight path, and landing. And finally, in subsection~\ref{sec:V.E(iv)}, we offer some practical instructional strategies that may help enhance physics thinking in students.  

\subsubsection{Physics Principles - Mention vs. Application}\label{sec:V.E(i)}
\begin{figure*}
\centering
\caption{Design Thinking vs. Science Thinking - Written Reports. Refer Appendices A and B for details. 
\newline{L1, L2, L3 \&\ L4 stand for Level 1, Level 2, Level 3 and Level 4 respectively. IT - Iterations; SK - Sketch; AA - Assumptions and Approximations; DL - Design Limitations; FC - (Physics) Facts and Concepts; PP - Physics Principles; MAT - Use of Mathematics}}
\fbox{\includegraphics[width=0.96\textwidth]{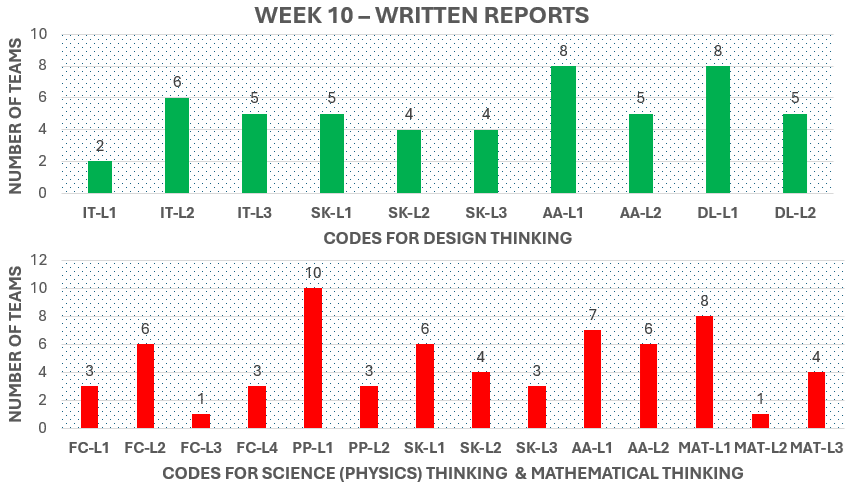}}
\label{DST_SCT_reports}
\end{figure*}

In our analysis we specifically considered the three fundamental principles: momentum, energy, and angular momentum \cite{chabay2015matter}. Student groups mostly focused on momentum and energy principles in their discussions, and angular momentum was rarely mentioned. This may be due to the fact that angular momentum was only formally introduced after week 10, and probably that the ED problem did not yield well to the application of this principle. In group discussions, almost all groups mentioned these principles, but only a minority of teams went on to elaborate and explain their application in the design task. In the final report, there was some improvement, but only three teams elaborated on the principles (see Figure~\ref{DST_SCT_reports}). Only two teams (Team-03 and Team-09) displayed a degree of consistency in both discussions and reports.

In week 07, Team-03, while exploring the application of the energy principle to further their design, stated: \hl{\textit{``If we know the spring constant, a property of the spring, possibly we can use certain energy calculations like PE of the spring, the KE of the payload to ensure it gets to the destination it has to get to''}}. Evidently, the team was relying on the hands-on experiments and simulations in the lab and hinting at determining the launch speed using the energy principle to obtain the desired range for the payload. A few other teams, Team-02, for example, in week 09 did invoke the energy principle but in lesser detail: \hl{\textit{``I think the big part of it is that energy is still like falls..energy conservation..energy gets transferred from the ball or whatever our load is..to basically being spread into the earth or through sound''}}. The contrast between the two examples is noteworthy. In the former, the focus is on the conservation of mechanical energy due to conservative forces, while in the latter, it is on energy dissipation due to collision.

In the written reports, only three teams elaborated on how they applied the principles in their design. \hl{\textit{``Because the velocity is instantaneously reversed when contact is made during a collision, the momentum principle can tell us the inverse change in momentum that will occur''}}, stated Team-06 while invoking the momentum principle to discuss the impulse change due to impact:. Though the team may not have been entirely accurate in assuming that the collision is instantaneous, the students were likely making the assumptions usually made in physics courses. It is notable that this team, rather than merely relying on the concept of the coefficient of restitution, quite unusually, considered momentum changes during the collision. The team also demonstrated a high degree of motivation by incorporating several details into its sketch, as evident in Figure~\ref{image_A_12}.  

\subsubsection{Physics Vocabulary \&\ Concepts - Mention vs. Elaboration}\label{sec:V.E(ii)}
 In our analysis we refer to `vocabulary' as the use of physics terms, concepts, and ideas, as distinct from the physics principles discussed in the earlier section. While there may be some overlap, our focus here is on the breadth of terminology and conceptual understanding.
 
 As seen in Figure~\ref{DST_SCT_disc}, teams demonstrated a broad range of vocabulary during the brainstorming session (week 06), albeit with less detail. This is understandable as students were enthusiastically sharing ideas and focusing on broad solution approaches. In the subsequent weeks, the distribution of vocabulary usage became more balanced between the mere mention of concepts and their detailed presentation. Although student groups employed a wide range of physics concepts, there were missed opportunities for further elaboration. This could be attributed to the absence of specific guidelines in the prompts provided, limited time, or other factors.

 Teams discussed a variety of physics concepts, including variable tension forces in zipline and pulley systems, buoyancy in boats and rafts, and drag forces on parachutes. They also explored stiffness constants of springs, the use of compressed air in composter-cannons, and the application of heat sensors on drones. Other topics included the terminal speed of falling objects, frictional forces on payloads in underground tubing systems, the installation of solar panels on drones, and the height, range, and trajectory of projectiles. Additionally, teams considered the density and elasticity of materials, the coefficient of restitution, and the forces acting on payloads and parachutes, among other concepts. 

 There were notable instances where student teams elaborated on specific concepts. For example, Team-09 in week 08 stated: \hl{\textit{"[to solve] this projectile motion we cannot just use normal kinematic equations like we might have used before, as those might not take into account the air resistance and how that force changes with the velocity''}}. This team demonstrated an understanding that standard kinematic equations apply only to uniform acceleration and are not suitable for motion involving velocity-dependent drag forces. In this sense, the team is employing the momentum principle in an indirect manner. A few teams were thoughtful to point out that the coefficient of restitution depends on both the material of the falling object and the surface of impact. To quote Team-04 in week 09: \hl{\textit{``One thing the coefficient of restitution made us think about was exactly what part of the earth where we’d be dropping the payload at. As we saw by our experiment with the tennis ball if we drop it on the hard concrete it’s gonna bounce with a lot higher velocity''}}.

In the final week's reports, students used newer physics terms, but only three of the 14 teams extended their analysis beyond mere terminology use. Team-03 noted their decision to neglect the payload's size and rotation around its center of mass, and mentioned applying Bernoulli's theorem but without further explanation. Team-06 acknowledged not considering air speed and pressure differences that could impact the payload's path.

Although it is commendable that some student teams reflected on their lab experiences and integrated scientific concepts to advance their designs, we observed that many teams limited their engagement to merely mentioning physics ideas without further elaboration—an issue that instructors may need to address~\cite[p.312]{siverling2021initiates}.

\subsubsection{Payload Delivery - Tracing Students' Thinking Trajectory}\label{sec:V.E(iii)}
We examined students' use of physics concepts and principles during the initiation, transit, and touchdown phases of the payload's journey.

The teams which considered employing drones questioned whether drones could carry a load as heavy as 50 kg. While, Team-11 was quite dismissive of this approach as \hl{\textit{``it doesn’t actually use any of the physics principles we’ve learned in this course''}}, in contrast, another team went on to analyze the forces on the drone to ensure adequate lift. Teams opting for a catapult launch explored the use of springs, and realized that a large (design aspect), stiff spring (physics aspect) was necessary for sufficient kinetic energy. Notably, they discussed energy conservation and conversion without equations, with one team suggesting multiple springs in parallel for added stiffness and others mentioning launch angles without detail. Concerns about potential harm to animals on the island (which we coded for environmental science) were also noted. Teams favoring a zipline discussed variable tension forces in the rope and payload but did not elaborate why they thought the tension forces would vary and how. Their design involved elevating one end of the zipline to convert potential energy into kinetic energy, though they were unsure how to decelerate the payload at the other end. How students' design and science thinking intricately overlap is quite evident through their discussions. 

The discussions on the payload's trajectory revealed several insights. Some student groups recognized that drag forces would influence the path and noted that standard kinematic equations for uniform acceleration were not applicable. They considered factors such as drag coefficient, payload mass, shape, and size, but did not delve into specific metrics. One team implied that the momentum principle might need to be applied in both horizontal and vertical directions. To quote Team-01: \hl{\textit{``You're mostly focusing on the y [direction], but now we have to also figure out how to get it if we're going to drop it, or if we're going to launch it in the x-direction over woods''}}. Team-06 was even able to rope in the language of mathematics by stating: \hl{\textit{``air resistance is a function of cross section times velocity squared''}}. Additionally, one team accounted for wind as a potential factor that could unpredictably alter the trajectory. Another team, which favored the drone approach, expressed concerns about economic losses if wind affected the drones' paths and caused damage (design aspects). It was evident that the provided scaffolds (lab and simulations on drag force) influenced their physics thinking, though teams did not advance to detailed calculations for trajectory accuracy. What is notable is that students' statements are quite often a blend of both physics and design. 

Although a few groups had initially considered using parachutes for a soft landing by week 08, in the following week, all groups engaged deeply with the physics of impact upon landing. Most students focused on minimizing the coefficient of restitution (COR) to prevent the payload from bouncing. Some groups rightly recognized that COR would be influenced by both the packaging material and the nature of the landing surface. A few teams aimed to use biodegradable packaging (which we coded for material science) while ensuring a COR close to zero. Others considered the forces on the parachute (physics thinking) to slow the payload's speed to near zero to ensure soft landing (design thinking). While students addressed the energy changes, COR, and the momentum principle, there was no significant focus on ensuring that the payload landed at the exact intended spot.

In the other methods considered, basic physics terms were mentioned without much detail. Overall, students applied momentum and energy principles to some extent in all the stages. It may be impractical to expect students spending time on calculations during their conversations, but it is noteworthy how they blended design and science. It is also evident from the preceding discussions that the scaffolds guided students in their thinking, though this observation cannot be attributed solely to the scaffolds.

\subsubsection{Physics Thinking - Raising the Bar}\label{sec:V.E(iv)}
The findings reveal that while a smaller fraction of student teams (see Figures~\ref{DST_SCT_disc} and~\ref{DST_SCT_reports}) provided detailed applications of physics principles and concepts, the majority did not offer such elaboration. This does not necessarily indicate a lack of understanding on the part of the students, but may instead may be a result of the non-specific nature of the prompts presented to the students in both discussions and reports. Freshman students, who are new to the university environment, may especially require additional guidance. To address this, a detailed rubric based on our coding schema in Appendices A and B—or a better alternative if available—could be provided to students.

The role of Graduate Teaching Assistants (GTAs) is also crucial. In many large universities, GTAs are responsible for delivering undergraduate instruction, and it is entirely possible that they, at least some of them, may not fully grasp the complexities of design-based science instruction. Typically, most new GTAs may not have prior teaching experience. Professional development~\cite{gretton2017transforming} workshops, regular meetings, and additional training as needed could be offered to GTAs to  \textit{``encourage more productive tutoring styles''} and increase  \textit{``student engagement''}~\cite[p.7]{stang2014interactions}.

Instructional materials should clarify that `physics principles' specifically refer to the fundamental principles of momentum, energy, and angular momentum. Providing examples to illustrate these principles can further support student understanding. Additionally, students should be encouraged to go beyond merely listing physics facts and vocabulary, and instead, elaborate on these concepts. They should also be urged to actively seek feedback from GTAs, reflect on their own statements, and offer constructive feedback to their team members.

It is important to recognize that some students may not realize that real-life problems differ significantly from the typical textbook `toy' problems, which often have exact and clean solution paths. This realization may surprise or even overwhelm some students. Therefore, it is essential to emphasize the importance of making appropriate assumptions and approximations~\cite{mashood2020approximations} without compromising the integrity of the problem.

\subsection{Interplay of Design and Science}\label{sec:V.F}

The coding schema in Appendices A and B along with the Figures~\ref{DST_SCT_disc} and~\ref{DST_SCT_reports} may create the misleading impression that design and science are indeed dichotomous. However, it is crucial to consider the myriad ways in which student groups express their ideas. While coding serves the valuable purpose of organizing and understanding the data, it can sometimes fragment the intricate nature of students' thinking and communication. As may have been evident, our discussion so far, has rested not just on our coding, but also our narrative thick descriptions coupled with a careful and representative choice of the informants' voices. This blended approach is to ensure that the complexities in students' thinking and communication are made evident, and also address the fragmentation that the coding schema may unintentionally convey. 

Given that this study is situated in an introductory physics course for future engineers, it is important to understand how students integrate design thinking and scientific principles. In this section, by analyzing their discussions and reports, we aim to understand how students make connections between design and science. We believe this analysis will help us guide students in acquiring effective problem-solving skills through the design of creative, engaging, and meaningful engineering projects. 

Though we had provided some glimpses into the design--science connections in the preceding sections, in this section we deal it in specific detail. In subsection~\ref{sec:V.F(i)}, we explore how student-groups integrate science and design. Subsection~\ref{sec:V.F(ii)} brings to light some of the inaccuracies in students' understanding of physics concepts which can potentially impact the design outcomes. In subsection~\ref{sec:V.F(iii)}, we delve into how students navigate making assumptions and approximations, which are important for design and science. Finally, in  subsection~\ref{sec:V.F(iv)} we discuss  the role that the provided scaffolds may have played in shaping students' design and science thinking.

\subsubsection{Integration of Design and Science}\label{sec:V.F(i)}
We questioned whether a clear distinction exists between design and science. We believe there may be no clear answer and that the `line', if any, is blurry. It is reasonable to assume that students think in complex ways, integrating design and science rather than considering them as isolated silos. On a philosophical note, we even wondered if students consciously think about these categories when engaging in discussions.

We analyzed our Stage 1 coding (covering weeks 06 - 09) for transcript segments tagged with both design thinking and science (including math). While no clear pattern emerged, all teams demonstrated some overlap between design and science in at least one of the four weeks. Four teams showed overlapping codes in only one of the four weeks, nine teams had overlaps in week 08, and five teams exhibited overlaps in more than two weeks. However, these findings reveal little about the nature of the overlap, a limitation inherent to Stage 1 coding. Stage 2 coding (see Appendices A and B) revealed more details as we had an additional data source in the form of students' written reports, and that the coding was more layered. Despite this, fragmentation is an inevitable part of the coding process. To address this, the lead author revisited the transcripts and reports using Stage 2 coding to construct a cohesive narrative (see Figure~\ref{proces_flow}). The inclusion of students' sketches as sub-data in their reports offered further insights into how they integrated design and science.

In week 06, students were merely expected to brainstorm as many solution approaches as possible. However, some teams immediately delved into the details of their approaches. Team-11, for instance, began by considering physics principles, stating, \hl{\textit{``[We] can use the momentum principle, energy principle, gravity, and air resistance as the parachute slowly goes down''.}} They also manged to slip in a piece of mathematics, albeit somewhat incompletely, by adding, \hl{\textit{``Force times displacement... [for the] work-energy principle''.}} Continuing in the same breath, the team (the same speaker, to be exact) voiced concerns that the entire launch might fail \hl{\textit{``if the launching is not at the right angle or something like that''}} (a design aspect related to the mechanism) and added a note of caution that if the parachute failed to deploy \hl{\textit{``the 50 kg mass of food is gonna fall through the trees which is not the best''}} (safety, another design aspect).

One particularly compelling example of a team seamlessly integrating math, physics, and design comes from Team-12 during week 08, where they expressed:~\hl{\textit{``Yeah terminal velocity against …. something.. aha but we can use .. we can use a graph similar to that to figure out what the coefficient of drag like .. what coefficient of drag lead to what and if we were to like to use a parachute or something we have to like consider it’s like is the material from the parachute ..is gonna stay on the island and..kind of  litter it.. or can we use like some sort of material that’s like biodegradable like you mentioned earlier''}}. This student discussed terminal speed and drag coefficient (physics thinking), referenced graphs (mathematical thinking), considered material choices (material science), reflected on lab experiences (scaffolds) and drew upon others' thinking (metacognitive thinking), to list a few. 

The written reports in week 10 offered teams an opportunity to organize their ideas and present them in greater detail. However, only a smaller fraction (see Figure~\ref{DST_SCT_reports}) demonstrated something substantial when compared to their discussions. Just three teams effectively utilized their sketches, with Team-06 providing a standout example (see Figure~\ref{image_A_12}). A closer look at this figure reveals how the team intricately visualized their design. At the takeoff stage of their rocket-based contraption, they clearly illustrate how the thrust force must exceed the weight to accelerate and achieve the necessary kinetic energy. The team confidently employed mathematical language (see Section~\ref{sec:V.G} for more details) to explain the underlying physics, seamlessly incorporating chemical energy from the burning fuel into their energy equation. They also thoughtfully addressed how the trajectory and speed of the contraption might need adjustment to achieve the desired outcomes. While not all aspects of their design are fully clarified, and the proposed solution may not meet all the prescribed criteria, their effort represents a serious and a natural integration of design and science. Notably, this team was among the few to abandon the anticipated catapult launch mechanism, opting instead to chart their own course. Nuere {\em et al.}~\cite[p.986]{nuere2022sketch}, in their essay on the use of sketches and drawing, insightfully note \textit{``In the end, sketch becomes a tool of thought to convey ideas on a two-dimensional support''}, and Team-06, through this piece of art, perfectly cemented the integration between design and science. 

While reflecting on the variety of ways in which student-groups demonstrated the intricate blend of design and science, the lead author was reminded of Slattery and Langerock's concluding thoughts in their delightful and lyrical essay  \textit{`Blurring Art and Science'}:   

\begin{quote}
\textit{``We reject the assumption that art and science are dichotomous. We choose to work and live within, on the borders of, the aesthetic Deleuzean moment that constitutes art with science. Within this space we find possibilities for overcoming the traditional bifurcations of art and science...''}~\cite[p.355]{slattery2002blurring}.
\end{quote}
Surely, what applies to art must apply to design!

\subsubsection{Design deficiencies vs. Physics conceptual inaccuracies}\label{sec:V.F(ii)}
Given the nature of the problem, design is expected to be an evolving process, making it impractical to expect a complete or rigorously final solution, more so due to time constraints. Students were generally mindful of economic aspects, user needs, criteria, and constraints, and they demonstrated reasonable reflection on their designs. However, we believe that prompting them to actively engage in a feasibility study could have led to improved outcomes.

Scientific concepts are indeed subject to evolution, but in the context of this problem, it is reasonable to assert that the underlying principles are well established. Students applied a broad range of concepts effectively, with a predominant focus on momentum and energy principles. However, we observed some minor conceptual inaccuracies in their discussions. As educators, our goal is to ensure that students have a clear understanding of scientific concepts and apply them accurately.

In week 06, Team-13 stated: \hl{\textit{``we need to structure the boat with quality materials and also take into consideration the density of the material so that it’ll float above water and also that density of the materials must be less than 1000 $kg/m^3$ in order for it to float which means density of the float must less than that of water''}}. The reader may note the usual misconception that for an object to float, its density must be necessarily less than that of the liquid. In another instance, in week 09, Team-01 stated: \hl{\textit{``To ensure a low coefficient of restitution we have to have a low velocity''}}, which reflects yet another misconception that coefficient of restitution `depends on' the velocities of the colliding objects. Over-reliance on the formula or  inadequate understanding of the equation~\cite{sherin2001students} may have led to this inaccurate understanding. In week 08, which was on the drag force, some students tended to think that the presence of drag force meant a reduction in speed. A better statement would be that the net force would cause a decrease in acceleration, as opposed to a decrease in speed. To quote Team-12: \hl{\textit{``so you’re essentially saying that [we] use air resistance to slow it down''}}. To give the benefit of doubt, we also wondered whether the team might have been referring to a specific design intended to slow down the object, in which case the statement would be correct.

Clarifying concepts to students, TAs providing prompt feedback, encouraging critical and reflective thinking, and ensuring fact(or concept)-checking can guide students towards developing a better understanding of the underlying physics principles and thereby resulting in a more accurate application of physics (or science, in general) principles in their designs. The observation of Liu and Fang that misconceptions can  \textit{``negatively affect their [students'] academic performance not only in these courses but also in many subsequent courses such as machine structure and design as well as advanced dynamics''}~\cite[p.19]{liu2016student} is worth noting.

\subsubsection{On Assumptions and Approximations}\label{sec:V.F(iii)}

In the context of physics, \textit{``making valid approximations is a fairly systematic exercise; there is a method in what beginning students might perceive to be adhoc''}~\cite[p.929]{mashood2020approximations}. Yet, the complexity of this seemingly straightforward process becomes evident when juxtaposed with the well-known \textit{``Consider a spherical cow...''} joke~\cite[preface.xiii]{harte1988consider}, where a theoretical physicist famously reduces the shape of a cow to a sphere. While amusing, this joke subtly underscores that making assumptions and approximations is not only foundational but a serious undertaking in both science and design.

At the outset of this study, we expected students to easily navigate such approximations in the engineering design (ED) task, viewing them as routine and straightforward. However, their responses challenged this assumption, revealing a level of difficulty that prompted us to reconsider what we had taken for granted. This surprising complexity highlights the nuanced nature of approximations in design-science contexts, reflecting how even seemingly simple tasks can sometimes pose challenges for learners.

In their written reports, students were asked to specifically discuss their assumptions and approximations. However, many responses were either vague or generic, with only a few being thoughtful (see Appendix B). Notably, there was barely any meaningful focus on approximations.

We found several teams state assumptions such as \hl{\textit{``payload landing safely on its own''}}, \hl{\textit{``neglecting accuracy''}}, and \hl{\textit{``assuming all packages had identical mass and measurements''}}, and we classified them as being `vague' as we were unable to decipher what students meant. A small number of teams made `generic' statements, such as assuming \hl{\textit{``no friction at launch''}}, \hl{\textit{``perfect weather conditions''}}, and a \hl{\textit{``constant drag coefficient''}}. As evident, some were physics-related while some were design-related, but a majority of the teams did not progress beyond making cursory mentions. 

A few teams, like Team-06, offered detailed considerations, acknowledging the impact of friction and the impracticality of ignoring the ``mass of a heavy spring'' (which was actually way beyond their course material - `transfer' from quite unpredictable sources). They also chose to simplify drag forces by assuming static air, avoiding complications from wind. Team-03 revised their initial assumption that drag forces act only in the y-direction and considered additionally the impact of rotational motion on the load. 

Overall, some teams focused primarily on design-related assumptions, others on science-related assumptions, while a few addressed both. However, we did not find any team delve into the reasons for their assumptions. Our findings align with findings that instructors need to emphasize the significance of assumptions in real-world physics projects~\cite[p.20]{verostek2022making} and guide students in engaging with these assumptions more productively~\cite[p.4]{sirnoorkar2023analyzing}.

What was even more striking was that nearly all the teams completely ignored any mention of approximations. This suggests either fatigue at the end of the semester or a lack of clarity in our prompts. It is possible that students were unaware of the importance of, and the differences between, assumptions and approximations in the context of design and science. Our findings underscore the need to clearly define these concepts in future assignments.

\subsubsection{Iterations - Role of scaffolds on Design and Science}\label{sec:V.F(iv)}

This section examines how some of the scaffolds we provided within the laboratory may have influenced students' iterations. We explore the impact of the scaffolds on students' design thinking, the application of scientific principles, and project outcomes. 

Figure~\ref{basis_statements} illustrates the broad impact of scaffolds on students' thinking as they iterated on their designs. Following the brainstorming session (week 06), in the following weeks, hands-on tasks and simulations were consistently represented, with hands-on tasks demonstrating a notably higher impact, which is unsurprising given their practical nature.

Our data suggests that while the scaffolds may have directly advanced the design efforts of teams exploring the catapult system, they were less helpful to the other teams. At the same time, quite surprisingly, even the teams which explored alternative strategies found it useful refer to the scaffolds when they considered various challenges in the delivery process, irrespective of the design. For instance, some teams debated the feasibility of a spring system launching a 50 kg load, while others speculated on the size and rotor requirements for a drone capable of lifting 50 kg. As an example, Team-02, realizing the need for a spring with a large force constant, considered employing springs in parallel: \hl{\textit{``What I learnt from the laboratory is that our spring system must contain parallel springs to make it more efficient and we basically use PE of the spring to be converted to KE''}}. Though the lab activity did not involve use of springs in parallel, these students were probably referring to what they may have learned in lectures or recitations. Or perhaps, the design problem motivated them to look for ways to increase the spring constant. Whether this is reflective of a transfer from science to design or design to science is something we discussed, but were unable to come to any definitive conclusion.  As mentioned earlier, some teams did feel these scaffolds were not aligned with their design idea. Team-06 did not hold back while stating: \hl{\textit{``We felt that this was a rash way to payload to the island as it would have so much velocity coming in which would endanger the species living there''}}. Echoing a similar view, Team-07 felt it best to discard the launch mechanism, and continue with their drone approach as \hl{\textit{``we can control the drone by ourselves''}}. 

Concerns about air resistance and wind were common across all teams. Team-12, referring to the lab activities, stated: \hl{\textit{``Even if we use like a drone ... anything like that flies... we have to consider air resistance like how much energy we have to put in to overcome''}}. Notably, the team considering a zipline (coded for general knowledge) argued that their design would be least affected by drag forces and winds. Momentum and Energy principles found mention in the discussions of almost all the teams at some stage or other, be it the catapult launch of a payload, a drone's take-off, or delivery through a zipline. In many designs, students tried to incorporate a parachute to ensure a soft landing for the payload.  

It is to be noted that the concept of coefficient restitution (COR) is only given a cursory mention in the lectures. In contrast, given that an entire lab (week 09) was devoted to COR, both the hands-on task and simulation had profound impact on students' thinking. Referring to the simulations, Team-03 recalled: \hl{\textit{``We explored some shapes that’d most efficiently dissipate energy''}}. A fine example of how students combine both design and science may be seen in Team-11's statement: \hl{\textit{``One thing we can do is to find a material that has a very low COR''}} so that the impact is softened. Even Team-14 which adopted the zipline approach had this to state: \hl{\textit{``I think this can be applied to any one of our solutions like the zipline''}}, continuing on how to incorporate the concept of COR (scaffold) to ensure soft landing of the payload (design). 

The effectiveness of the scaffolds in guiding student-groups' progression towards the solution appears to be mixed. The scaffolds seem to have fallen short in guiding students on providing specific details about their contraptions, critically evaluating design limitations, articulating assumptions, clarifying intended approximations, or elaborating on the application of physics principles and detailed mathematics. This is evident from the analysis of their written reports, as illustrated in Figure~\ref{DST_SCT_reports}.

Our findings provide evidence that though the iterative processes and scaffolds helped enhance students' ability to integrate design and science, there were several areas where the instructional supports have room for improvement. One notable drawback is that the scaffolds appeared to favor payload delivery using a catapult-launch mechanism, which, in retrospect, we acknowledge may have inadvertently steered students' thinking in that direction. The findings also provide a gentle lesson for us in planning more thoughtful instructional materials.  

\subsection{Role of Math in Design and Science}\label{sec:V.G}
A striking feature is the near absence or minimal presence of mathematics (see Appendices A and B, and Figure~\ref{ways_of_thinking}) in most student-groups' conversations. Instances of presence included cursory references to terms such as statistics and vectors, making basics algebraic manipulations, comparative and proportional reasoning. As an example, Team-04, while reflecting on the hands-on and simulation activities in week 08 demonstrated proportional reasoning by stating: \hl{\textit{``We also played with the diameter of the object and we saw that the bigger the diameter the less distance it went. And also that the we also saw that higher the drag coefficient the shorter the distance''}}. It is interesting how the team uses math while thinking about the dimensions of the parachute (design-based thinking) and how drag coefficient may impact the range (physics-based thinking). Students think in quite complex ways, often blending formal and informal mathematical language with elements of design and physics~\cite{kuo2013students}. As an example, Team-08, in week 09, stated: \hl{\textit{``we have and also one thing we would probably want to have a lower velocity because it requires a lot more force to slow down and likelihood of us reaching terminal velocity with such a heavy payload in such a short distance is unlikely''}}. Team-03, while exploring the use of drones considered installing heat sensors on the drones to locate gorilla groups on the island. They stated: \hl{\textit{``The destination to which the drone goes should be dependent on statistics. Essentially, find points or spots where the gorillas usually populated within the span of the island and then determine the delivery location based off of that''}}, hinting on the scope for data science, which we coded for mathematical thinking. Whether this excerpt from the transcripts should be interpreted as design-based or science-based thinking is an intriguing question that we enjoyed debating, and we ultimately leave it for readers to decide. This is one among the multiple instances where we felt that the perceived dichotomy between design and science blurred. Additionally, it revealed that how and from where students `transfer' learning is indeed unpredictable and complex.

\begin{figure*}[t]
\caption{Sample sketch from a student-group. The teams rolls design, physics, and math - all into one! Nuere {\em et al.}'s insightful statement~\textit{``In the end, sketch becomes a tool of thought to convey ideas on a two-dimensional support''} is evidently at play here~\cite[p.986]{nuere2022sketch}.}
\fbox{\includegraphics[width=0.96\linewidth]{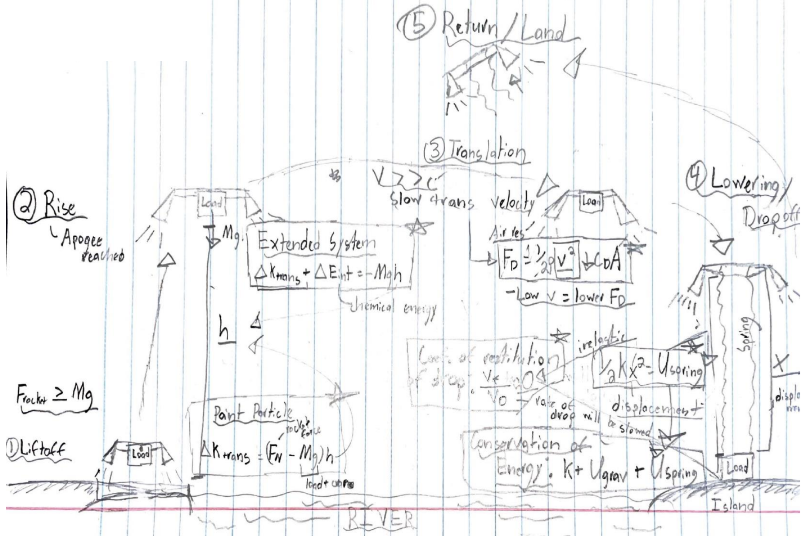}}
\label{image_A_12}
\end{figure*}

The situation was only slightly better in the written reports, with only three teams using mathematics. Notably, Team-06 employed the language of mathematics to convey their thinking by using the Energy Principle, stating: \hl{\textit{``We can break down our system into a point particle system and an extended system. The point particle system for $K_{\text{trans}}$ in vertical flight would be $(F_n - Mg)h$. For the extended system, this would be $K_{\text{trans}} + E_{\text{int}} = -Mgh$ where $E_{\text{int}}$ is the chemical energy stored in our contraption''}}. (The equations describe the energy dynamics of the system. $K_{\text{trans}}$ represents translational kinetic energy, while $(F_n - Mg)h$ represents the work done by the launch force $F_n$ against the weight of the object $Mg$ over a height $h$. For the extended system, $K_{\text{trans}} + E_{\text{int}} = -Mgh$, where $E_{\text{int}}$ is the chemical energy stored in the contraption, and $-Mgh$ represents the gravitational potential energy change during vertical motion). Incidentally, this was the only team to use math both in text and in images, as evident in their `tool of thought' (see Figure~\ref{image_A_12}). 

In summary, the lack of mathematical discussion within student groups points to an area of improvement in the integration of design and science. Educators may need to offer explicit guidance and feedback to help students incorporate mathematical reasoning into their discussions. In our calculus-based course, we could encourage students to incorporate calculus to optimize design parameters, use algebra to write and solve equations related to both design and science, utilize geometry for spatial reasoning, and invoke trigonometry for analyzing vectors. Emphasizing mathematical modeling for making predictions, applying statistical techniques to evaluate design performance, using programming languages like Python to automate calculations, and employing the tools in data science are other effective approaches to engage in a science-based approach to ED-problems~\cite{pines2002integrating, narode2011math, bush2018design, lambert2021udl}. 

\subsection{Metacognitive Reflection: Thinking About Thinking}\label{sec:V.H}
Although we recognize that metacognition can be subject to a range of interpretations, for the purposes of this study, we have adopted a simplistic understanding of the term as evident in Table~\ref{tab:WoT}. Our goal was to observe if and how students reflected on their design decisions and reviewed their science thinking to advance their progress toward a solution.

During weeks 07 to 09, we explicitly prompted students to reflect on their design processes. The data suggests these prompts were somewhat effective in encouraging students to think critically about both design and scientific aspects of their problem-solving approaches. Given the complex nature of the engineering design problem, it was not surprising that several teams remained undecided on their solution paths even by week 08. 

As students engaged in each lab activity, their thinking evolved with the introduction of new concepts. For instance, in week 08, Team-12 stated: \hl{\textit{``After doing this lab and looking at this kind of spring based system where the spring is used to launch off …we might want to rethink our catapult idea and maybe change it to something else using springs instead of a catapult or drone''}}. Initially, they had considered using a catapult but were uncertain about the launching mechanism. The lab activities led them to consider spring-based launching system as a potential solution. However, as they continued with the conversation, they did not detail how they may employ springs, nor did they build on their mention of a drone. This conversation segment of the team's conversation primarily focused on the design aspect, without further elaboration on the scientific principles involved. In the subsequent week, Team-12's thinking was influenced by lab activities focused on drag force: \hl{\textit{``long shot.. interesting it [parachute] increases the coefficient of drag …it’s kind of the same thing but air resistance will have to be something that we can consider .. if..  even if we use like a drone anything like that flies we have to consider air resistance like how much energy we have to put in to overcome''}}. This time, the conversation shifted to incorporate the concept of air resistance into their design, though once again, they referenced a drone without further development in their discussion. 

Students discussed in random ways and sometimes the ideas appeared scattered. They tended to talk about what was important to them, which we view as being most natural. While we appreciate and value their interests, we wondered if more specific guidance could have helped them make more concrete progress toward a solution. However, we also understand there may be a risk there, as students may feel a bit constrained by what we believe is guidance.

The variety of ways students built on their lab experiences was fascinating. Team-10, in week 07, stated: \hl{\textit{``Personally, the first thing that comes to my mind … as it was mentioned earlier… How energy is going to affect our design? Like the use of total energy and being able to calculate it…especially when we’re launching food over a distance…So, being able to calculate that energy so that it doesn’t splash into the river and also it doesn’t explode when it eventually does land''}}. This team's focus was on the energy principle and how they could use it to determine the launch velocity needed to achieve the required range. As for the design aspects, this team focused on habitat integrity and safety. Interestingly, this team spent most of its time discussing physics concepts rather than design aspects for the remainder of the conversation. The trend continued into week 08. Most teams had not initially considered the importance of air resistance in their design. However, the lab activities made them review their thinking and incorporate air resistance. Team-10 reflected on the evolution of their thought as: \hl{\textit{``[we] consider the drag force which is something we didn’t really consider before this lab what affects the drag force and we learnt a lot about it from this lab like the diameter, the mass... all of that affects the like the projectile that our load will be travelling so have to take all that into consideration''}}. The rest of the conversation remained on exploring ways to minimize drag force to ensure soft landing, striking a balance between design and science. 

By and large, although most teams demonstrated reasonable reflection on both design and science aspects, we found ourselves questioning whether the data might have revealed deeper insights had we provided more specific prompts for student reflection. In retrospect, this reflection aligns with an insightful recommendation we only recently encountered:

\begin{quote} \textit{``[Students] should not be instructed to merely use `metacognition' but instead should be directed toward specific, higher-level analyses in which they explain how they are solving the problem''}~\cite[p.79]{hacker2003not}. \end{quote}

For our next iteration, we would strongly consider this suggestion. Providing more structured prompts that guide students to higher-level analysis of their problem-solving processes could enhance their engagement with both design and science thinking, leading to more meaningful reflections and stronger connections between the two.

\subsection{What about the Design--Science Gap then?}\label{sec:V.I}

The literature is replete with diverse definitions, interpretations, and perspectives on terms such as `design', `design thinking', `design-based thinking', `engineering design', `science', and `scientific thinking' (see Section~\ref{sec:II.E}). Indeed, we question whether a singular, fixed definition is appropriate, as such an approach might not fully capture the richness and diversity of these concepts across various educational contexts. Drawing on Gee's perspective, our view is that these terms serve as useful \textit{``tools of inquiry''} or \textit{``thinking devices''}~\cite[p.37]{gee2014discourse} that guide our understanding. In this light, flexible interpretations of these terms can enhance our ability to engage with them in meaningful ways.  Be this as it may, we have yet to consider the various interpretations and notions of these concepts that students are entitled to develop! This diversity complicates the task of arriving at a universal definition for the `design--science gap'. 

Although this study draws inspiration by the works of Chao {\em et al.}~\cite{chao2017bridging}, Kolodner {\em et al.}~\cite{ kolod_punt_2003}, Vattam and Kolodner~\cite{vattam2008foundations}, and Chase {\em et al.}~\cite{chase2019learning}, we sought a more analytic approach that better aligns with the specific requirements of our undergraduate physics context. Our efforts to operationalize design and science thinking, in the context of this study, were steps towards understanding the notion of design--science gap. 

In one of his scholarly essays, Nigel Cross~\cite[p.226]{cross1982designerly} identifies five key aspects of designerly thinking: Designers tackle  \textit{``ill-defined''} problems, adopt a  \textit{``solution-focused''} approach to problem-solving, think  \textit{``constructively''}, and use  \textit{``codes''} that translate abstract requirements into concrete objects. If we contend that these aspects are equally applicable to contemporary science practitioners, it is unlikely that we would encounter substantial opposition. 

Given such ambiguities, discussing a `gap' between design and science may not be particularly meaningful. Furthermore, interpretations of these terms are often practitioner-dependent, which is rightfully so. For instance, a business professional might prioritize economic considerations, an engineer might focus on practical implementation with relatively less concern for underlying scientific principles, an environmentalist may be more inclined towards habitat integrity, safety etc., while a physics educator might emphasize the understanding and application of physics concepts. This is despite the fact that all practitioners would agree that money matters. This variability extends to mathematicians and practitioners in other scientific disciplines. Furthermore, the variability is twofold: determining `which aspects' of design or science to focus on and deciding  `to what depth' — both of which add to the complexities in the notion of `design--science gap'.

While we found the preceding `philosophical' discussions intellectually stimulating, we were also confronted with the practical need to establish an operational framework for our study. We examined the data within the local context of an introductory physics course, approaching the task as typical physics educators focused on guiding students to apply their physics knowledge to solve real-world problems. Naturally, our emphasis was more on physics than on the various facets of design. However, we would still desist from taking a position on the existence or the absence of a `gap' even within our context. At the most we would state there was scope for students to have delved more into aspects of both design and science, with the caveat that the distinction between the two is not as straightforward as it may appear. 

Moreover, we did not set specific benchmarks for our students to achieve, making it inappropriate to define any gap in achievement. True to the spirit of allowing the data to speak for itself, we adopted an inductive coding process. This approach further motivated our decision to move away from searching for a `gap', and instead think in terms of strengthening the `connection'. 

\section{Conclusions}\label{sec:VI}
\begin{figure}[H]
\centering
\caption{Design Thinking and Science Thinking. Strengthen the connection (see Section~\ref{sec:VI.A}); Raise the Bar (see Section~\ref{sec:VI.B}.)}
\fbox{\includegraphics[width=0.96\columnwidth]{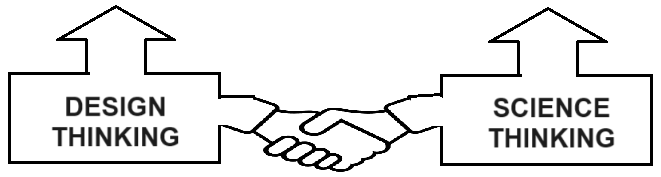}}
\label{dsc}
\end{figure}

To summarize, this study investigated how student-groups engaged in design and science thinking while solving a prescribed ED problem. Given the blurry divide between design and science, we argued that thinking in terms of a `gap' would not be necessarily meaningful, at least in our context. This perspective was instrumental in making us think in terms of `connections', instead. By qualitatively analyzing transcribed group discussions and written reports, we aimed to address our research questions. Figure~\ref{dsc} captures the essence of our responses to the research questions, which are elaborated in the following sections. In Sections~\ref{sec:VI.A} and \ref{sec:VI.B}, we present our conclusions and reflections on \hyperref[RQ1]{RQ1} and \hyperref[RQ2]{RQ2}, respectively. 
\subsection{Pitching for Design--Science Connection}\label{sec:VI.A}

Our first research question \hyperref[RQ1]{RQ1} was about characterizing design and science thinking, and ways in which the connections between them unfold in student-groups' work. Starting from a Ways of Thinking framework (Section~\ref{sec:V.B}), we developed a more detailed coding framework (Appendices A and B) - grounded in our data - which served as a guide to characterize design and science thinking in the context of our study. In the process, we addressed the complexities in the notion of the `design--science gap' (Section~\ref{sec:V.I}).  

In addressing the given problem within the framework of the engineering design (ED) process model, student groups engaged in design thinking by generating multiple solutions (Section~\ref{sec:V.A}), iterating with the provided scaffolds (Section~\ref{sec:V.C}), and offering only basic details about delivery mechanisms and design features (Sections~\ref{sec:V.D(i)},~\ref{sec:V.D(ii)} and~\ref{sec:V.D(iii)}). Notably, we observed that there was room for elaborating on their designs and considering closely the associated limitations. The inclusion of specific prompts to encourage a feasibility study is recommended to yield more comprehensive project outputs from the students (Section~\ref{sec:V.D(iv)}).

Our analysis of students' artifacts about science thinking, revealed that students referenced a variety of physics concepts, vocabulary terms, and ideas but did not elaborate on their applications (Section~\ref{sec:V.E(ii)}). While they primarily invoked the principles of momentum and energy, the angular momentum principle found minimal mention (Section \ref{sec:V.E(i)}). There was certainly space for students to provide a more detailed and physics-based description of the payload's trajectory (Section~\ref{sec:V.E(iii)}). This omission may be attributed to the nature of the problem, their approach to finding solutions, and the fact that angular momentum is introduced in the curriculum after the design challenge, despite some students having prior knowledge of the concept. We recommended instructional supports to enhance students' physics thinking (Section~\ref{sec:V.E(iv)}).  

Our analysis also revealed that students' use of mathematical concepts was minimal during discussions but improved slightly in written reports (Section~\ref{sec:V.G}). Nonetheless, some student groups demonstrated interesting integration of physics and mathematics concepts to inform their designs (Section~\ref{sec:V.F}). We observed a myriad ways in which the connections between design and science played out, blurring the perceived divide between the two. We emphasized the need to look closely at possible misconceptions students may have, as a strong understanding of science principles would have a positive impact on students' design approaches. 

From this study, we assert that defining or characterizing the `design--science gap' is not straightforward and requires a more nuanced response (Section~\ref{sec:V.I}). Factors such as context, field, educational objectives, time, and resources all play a role in determining what specific aspects of, and to what extent or depth, design and science may need to be considered to address the notion of a `gap'. It would perhaps be appropriate to state that the definition of `design--science gap' would depend on the local context. Given that our study is situated within an introductory physics course, in our instructional materials, we focused more on how students apply physics and math concepts to advance their designs. As may be typical of most physics courses, we emphasized less on aspects such as criteria and constraints, habitat integrity, stakeholders, and mechanism details. 

Rather than making a definitive statement on the existence or non-existence of a gap between design and science, we adopt a more neutral view, particularly so since we had no preset levels of what is a good design or what is the expected level of physics thinking. Nor did we inform our students whether our focus was more on design or science. At the same time, given that it is a physics course, it is only natural that students, by default, may have tended to focus less on design and more on science, drawing upon their own interpretations of these terms. It is also entirely possible that students were not even thinking about design and science, but were `merely' (not in any negative sense, but as a matter of a possibility) trying to solve the problem on hand. 

It would also be safe to add that any `design--science gap' one may observe is not in students' thinking, but rather in the data which are dependent on the prompts we had provided. There is potential to design instructional strategies to enhance student outcomes in both design and science. However, when it comes to the specific aspects and levels on which instructors should focus, there may not be a singular answer. The aspects and levels, as evident in our coding framework in Appendices A and B, are specific to our context and data, but we believe our coding framework can guide other researchers to characterize design--science `connection' when applied to their local contexts.

One final comment we would like to make is that one should not hold a dim view if students engage in `trial and error' methods and `gadgeteer' their way towards a solution. As the saying goes, `there may be a method in their madness'~\cite{kanarevnobel, wills1996trial, bell2007teachers, wilson1994there, rajagopalan2008madness}. 

\subsection{Strengthen the Connection; Raise the Bar}\label{sec:VI.B}
As previously mentioned, within the laboratory, we had provided scaffolding activities including hands-on experiments and simulations to guide students thinking. For the operational purposes of this study, we were guided by Puntambekar and Kolodner's indirect reference to scaffolds as instructional \textit{``supports''}~\cite{puntambekar2005toward}. Towards answering our second research question \hyperref[RQ2]{RQ2}, we explored what impact these scaffolds may have had on students' thinking as they progressed towards a solution.  While answering this question, we deem it important to acknowledge that students gather knowledge from a variety of sources, and it would be impossible to assert that students' output was influenced only by what we considered as scaffolds. 

Seeking a deeper understanding, we asked: what exactly is a scaffold? Though the lab activities may be the most immediate scaffolds for students, it would be naive to assume scaffolds to have such a limited meaning. After all, students did draw upon their learning in lectures and recitation (Section~\ref{sec:V.C}). And what to say about general knowledge (Figure~\ref{basis_statements})? Would it be right to assume scaffolds to refer only to the supports provided by the instructional team? To add to the complexity, students' thinking is not bound by such restricted post-facto notions we may have. Though there is evidence (Sections~\ref{sec:V.C},~\ref{sec:V.D},~\ref{sec:V.E}, and~\ref{sec:V.F}) that the lab activities had an impact, it would be disingenuous on our part to claim these were the only contributing factors. It is unlikely that all extraneous, and often unknown, factors can be effectively filtered or controlled in educational research. This raises a somewhat related question: can, or perhaps more importantly, should there ever be a true `control' in educational research? Another question we asked was if the entire ED process model was itself a scaffold. Clearly, the notion of scaffolds is far more complex than the simplistic notions we had. Even as we were riddled with these perplexing questions, we came across the humbling observation by Kember:  

\begin{quote}
\textit{``...experimental designs are not often suited to evaluating the effectiveness of teaching innovations in higher education. Instead of using experimental designs with the (probably illusory) aspiration of showing causation, teacher-researchers might more reasonably aim to establish claims beyond reasonable doubt''}~\cite[p.99]{kember2003control}. 
\end{quote}

As to the influence of scaffolds, we found it illuminating to reflect on the notion of `transfer of learning'~\cite{mestre2006transfer, engle2006framing, chase2019learning}. However, it would be hard to gauge if and how much was the transfer from design to science, science to design, and beyond. Students were drawing ideas from their general knowledge too across all the labs (Section~\ref{sec:V.C}) indicating a transfer from sources which may be near impossible to decipher (Section~\ref{sec:V.F(iv)}). Given that there is no (can there ever be?) consensus on what `transfer' may actually mean, our view is that the notion of `transfer', despite its existence (Sections~\ref{sec:V.D(i)},~\ref{sec:V.E(i)} and~\ref{sec:V.G}) in our context,  merits a more detailed analysis, far beyond what this study can accomplish.

Thinking about scaffolds, as mentioned in Section~\ref{sec:V.A}, we realized post-hoc that we may have unintentionally steered some teams toward a catapult-launch approach by providing lab activities centered on springs, drag force, and the coefficient of restitution (COR). While some teams pursued alternative methods, we suspect that a few others may have reluctantly chosen the catapult approach due to the provided scaffolds. Despite Sections~\ref{sec:V.F(iv)} and~\ref{sec:V.H} revealing the positive impact of the scaffolds in guiding students iterate on their designs approaches, our study highlights a critical lesson: ED-problems should be chosen carefully, with scaffolds thoughtfully aligned to encourage diverse approaches and engagement from all participants~\cite{kpamma2017participatory}.

Another aspect we reflected was role of the prompting questions (Table~\ref{prompts}) we provided. While we did not explicitly consider prompts as scaffolds, it is worth considering the contention of Hathcock {\em et al.} that:  \textit{``the use of inquiry-based questions may scaffold students toward greater creative possibilities''}~\cite[p.745]{hathcock2015scaffolding}. Our data stemmed from discussions and written reports, in which we embedded several prompts. Naturally, we may expect the nature of these prompts to  influence students’ responses and, consequently, our findings. Furthermore, extending the meaning of scaffolds to include prompts would be beyond the scope of this study. Careful choice of prompts is an area that deserves closer attention in our future iterations. Too little guidance in the form of prompts could result in chaotic, unstructured data, while too much might reduce responses to mere bucket-list completion, limiting their value. Thus, as educators, we may forever be left with the Shakespearean dilemma `to guide or not to guide'. 

Despite the inherent challenges, both teaching and educational research remain vital pursuits. These challenges inspire us to persist in developing effective scaffolds—embracing diverse interpretations—while also deepening students' design and science thinking—acknowledging the multitude of meanings—and strengthening the design–science connection.  As we reach the `end' of this endeavor, we find reassurance in Berliner's thoughtful reflection:
\begin{quote}
\textit{``Hard-to-do science is what the social scientists do and, in particular, it is what we educational researchers do''}~\cite[p.18]{berliner2002comment}.
\end{quote}
 
\section{Limitations of the study} 
While this study provides valuable insights, we acknowledge the following limitations which may need to be considered when evaluating our findings.

(i) The quality of group discussions among students in a physics lab depends on various factors: individual prior knowledge, motivation, communication skills, group composition, role assignment, group dynamics, physical environment, time constraints, cultural context, peer pressure, and confidence levels~\cite{schmidt2000factors}. We did not consider these factors, nor would it have been practically possible. (ii) A major portion of the data was gathered through group discussions which, while rich in real-time thinking, often lack the depth of written work, potentially skewing our results~\cite{rivard2000effect}. But then, discussions are more dynamic and the evolution of human thought is more evident than in written reports. (iii) Large enrollment courses with multiple TAs introduce variables, particularly in labs. The TA's influence on student output through the multi-week activity is unknown. Instructional factors like clarity, support, and feedback also play a role~\cite{huffmyer2019graduate}. (iv) The findings are specific to the 14 groups in this study. Different or larger samples might yield different results. (v) Findings are specific to the given prompts and student-groups' interpretation of the same. The wording may have unintentionally biased responses towards or away from design or science. (vi) The nature of the ED task likely influenced student motivation and engagement, impacting the data~\cite{schmidt2000factors, white2019validating}. (vii) Given that the definitions of `design', `science', `technology', and `engineering'~\cite{radder2009science} will forever remain open to debate (for good reasons), our findings depend solely on our coding schema, and are specific to our context. (viii) While not a direct limitation, students did not develop a physical setup due to time and resource constraints - typical aspects to consider in large enrollment courses. This study focused on the ED process and model solution. Results may be expected to differ with physical materials and technology use.

Despite these limitations, the study offers valuable insights into student groups' thinking patterns, potentially reflecting broader trends in the entire cohort. These findings will guide future instructional interventions to address gaps and enhance student learning.

\section{Implications}

The study reveals that while integrating engineering design and science based on the ED process model holds substantial educational value, there are gaps in the pedagogical strategies that need to be addressed. Our findings suggest that further research should focus on identifying appropriate design-based problems, implementing effective instructional interventions, developing pedagogical strategies and thoughtful scaffolds, creating relevant assessments, and training graduate teaching assistants to support design-based learning. 

Additionally, we detailed our qualitative data analysis process invoking the Gioia and MIRACLE frameworks, thereby contributing to the literature on the use of qualitative methods in physics education research.

Transitioning from an initial focus on `gap' to `connection' later, also reflects a move away from what researchers term `deficit thinking'~\cite{davis2019deficit} on our part. A deeper understanding of design and science can help educators plan their classroom interventions more effectively. Our view is that it may be overly simplistic to assume human thinking occurs in isolated silos of `design' and `science'. Instead, strengthening the connection between these domains and deepening both science and design thinking would benefit students. We fondly hope this exploration has further revealed that the topics of `design' and `science' are as ripe as ever for philosophical reflection, at least within STEM environments~\cite{shiner2012blurred}.  

Instead of holding a dichotomous view towards design and science, a more holistic integrated perspective would be pragmatic. If design~\textit{is} art, then the words of Eisner and Powell resonate profoundly: 
\begin{quote}
\textit{``As we see it, the art in science inspires, motivates, and enriches the pursuit of inquiry; indeed, for good work to be done, artistry appears inevitable''}~\cite[p.157]{eisner2002art}.
\end{quote}

\section{Future Work}
This study is based on data collected during Fall 2021. In subsequent semesters, we refined the prompts and scaffolds to further strengthen the design-science connection in students' thinking. These refinements included the integration of Jupyter Notebooks~\cite{jupyter2024}, enabling students to combine text, images, and Python code within a single platform. This approach provides a rich context for students to develop computational skills. 

Building on our findings and drawing inspiration from recent `Ways of Thinking' (WoT) frameworks~\cite{english2023ways, slavit2019stem, slavit2021student, dalal2021developing}, we plan a two-part series. The first part will present a novel STEM Ways of Thinking framework~\cite{subramaniam2024presentingstemwaysthinking}, while the second part will demonstrate its application to fresh data from engineering design-based physics projects. Among the aims of this framework is to enhance the design-science connection by providing a structured approach that supports educators, researchers, and students alike. For educators, it offers a systematic lens for planning, implementing, and assessing design-based science activities, facilitating a deeper understanding of students' design and science thinking. Researchers can leverage the framework to analyze student artifacts more effectively, gaining insights into the integration of design and science across diverse contexts. For students, the framework serves as a guide, fostering stronger, more coherent connections between design and science, and encouraging a more integrated approach to problem-solving. By expanding on current work, the STEM Ways of Thinking framework offers a more integrated perspective that highlights how disciplinary knowledge, problem-solving strategies, and iterative design processes interact, enabling a deeper understanding of how these elements can be effectively connected in educational and practical contexts.

\section{Acknowledgments}
We sincerely thank the anonymous referees for their insightful and encouraging comments on both our earlier publications and this article. Their feedback has significantly contributed to making this paper more coherent, detailed, and well-structured. We owe our gratitude to all the authors referenced, for they provided us with innumerable moments of much needed inspiration. ChatGPT-4o, MS Copilot, and Perplexity AI were employed by the lead author only to `wordsmith' passages, and as a `learning partner', but not for any data analysis. A special thanks to Dr. Amogh Sirnoorkar and Winter Allen for their insights in the preparation of this manuscript. The lead author assumes full responsibility for any errors and omissions.

This work is supported in part by U.S. National Science Foundation grant DUE-2021389. The opinions expressed are those of the authors and not of the Foundation.

\bibliography{references}

\begin{thebibliography}{141}%
\makeatletter
\providecommand \@ifxundefined [1]{%
 \@ifx{#1\undefined}
}%
\providecommand \@ifnum [1]{%
 \ifnum #1\expandafter \@firstoftwo
 \else \expandafter \@secondoftwo
 \fi
}%
\providecommand \@ifx [1]{%
 \ifx #1\expandafter \@firstoftwo
 \else \expandafter \@secondoftwo
 \fi
}%
\providecommand \natexlab [1]{#1}%
\providecommand \enquote  [1]{``#1''}%
\providecommand \bibnamefont  [1]{#1}%
\providecommand \bibfnamefont [1]{#1}%
\providecommand \citenamefont [1]{#1}%
\providecommand \href@noop [0]{\@secondoftwo}%
\providecommand \href [0]{\begingroup \@sanitize@url \@href}%
\providecommand \@href[1]{\@@startlink{#1}\@@href}%
\providecommand \@@href[1]{\endgroup#1\@@endlink}%
\providecommand \@sanitize@url [0]{\catcode `\\12\catcode `\$12\catcode `\&12\catcode `\#12\catcode `\^12\catcode `\_12\catcode `\%12\relax}%
\providecommand \@@startlink[1]{}%
\providecommand \@@endlink[0]{}%
\providecommand \url  [0]{\begingroup\@sanitize@url \@url }%
\providecommand \@url [1]{\endgroup\@href {#1}{\urlprefix }}%
\providecommand \urlprefix  [0]{URL }%
\providecommand \Eprint [0]{\href }%
\providecommand \doibase [0]{https://doi.org/}%
\providecommand \selectlanguage [0]{\@gobble}%
\providecommand \bibinfo  [0]{\@secondoftwo}%
\providecommand \bibfield  [0]{\@secondoftwo}%
\providecommand \translation [1]{[#1]}%
\providecommand \BibitemOpen [0]{}%
\providecommand \bibitemStop [0]{}%
\providecommand \bibitemNoStop [0]{.\EOS\space}%
\providecommand \EOS [0]{\spacefactor3000\relax}%
\providecommand \BibitemShut  [1]{\csname bibitem#1\endcsname}%
\let\auto@bib@innerbib\@empty
%</preamble>
\bibitem [{\citenamefont {Olson}\ and\ \citenamefont {Riordan}(2012)}]{pcast_2012}%
  \BibitemOpen
  \bibfield  {author} {\bibinfo {author} {\bibfnamefont {S.}~\bibnamefont {Olson}}\ and\ \bibinfo {author} {\bibfnamefont {D.~G.}\ \bibnamefont {Riordan}},\ }\bibfield  {title} {\bibinfo {title} {Engage to excel: producing one million additional college graduates with degrees in science, technology, engineering, and mathematics. report to the president.},\ }\href@noop {} {\bibfield  {journal} {\bibinfo  {journal} {Executive office of the president}\ } (\bibinfo {year} {2012})}\BibitemShut {NoStop}%
\bibitem [{\citenamefont {Kelley}\ and\ \citenamefont {Knowles}(2016)}]{kelley2016conceptual}%
  \BibitemOpen
  \bibfield  {author} {\bibinfo {author} {\bibfnamefont {T.~R.}\ \bibnamefont {Kelley}}\ and\ \bibinfo {author} {\bibfnamefont {J.~G.}\ \bibnamefont {Knowles}},\ }\bibfield  {title} {\bibinfo {title} {A conceptual framework for integrated stem education},\ }\href@noop {} {\bibfield  {journal} {\bibinfo  {journal} {International Journal of STEM education}\ }\textbf {\bibinfo {volume} {3}},\ \bibinfo {pages} {1} (\bibinfo {year} {2016})}\BibitemShut {NoStop}%
\bibitem [{\citenamefont {Roehrig}\ \emph {et~al.}(2021)\citenamefont {Roehrig}, \citenamefont {Dare}, \citenamefont {Ellis},\ and\ \citenamefont {Ring-Whalen}}]{roehrig2021beyond}%
  \BibitemOpen
  \bibfield  {author} {\bibinfo {author} {\bibfnamefont {G.~H.}\ \bibnamefont {Roehrig}}, \bibinfo {author} {\bibfnamefont {E.~A.}\ \bibnamefont {Dare}}, \bibinfo {author} {\bibfnamefont {J.~A.}\ \bibnamefont {Ellis}},\ and\ \bibinfo {author} {\bibfnamefont {E.}~\bibnamefont {Ring-Whalen}},\ }\bibfield  {title} {\bibinfo {title} {Beyond the basics: A detailed conceptual framework of integrated stem},\ }\href@noop {} {\bibfield  {journal} {\bibinfo  {journal} {Disciplinary and Interdisciplinary Science Education Research}\ }\textbf {\bibinfo {volume} {3}},\ \bibinfo {pages} {1} (\bibinfo {year} {2021})}\BibitemShut {NoStop}%
\bibitem [{\citenamefont {Schweingruber}\ \emph {et~al.}(2012)\citenamefont {Schweingruber}, \citenamefont {Nielsen},\ and\ \citenamefont {Singer}}]{nrc_dber_2012}%
  \BibitemOpen
  \bibfield  {author} {\bibinfo {author} {\bibfnamefont {H.~A.}\ \bibnamefont {Schweingruber}}, \bibinfo {author} {\bibfnamefont {N.~R.}\ \bibnamefont {Nielsen}},\ and\ \bibinfo {author} {\bibfnamefont {S.~R.}\ \bibnamefont {Singer}},\ }\href@noop {} {\emph {\bibinfo {title} {Discipline-based education research: Understanding and improving learning in undergraduate science and engineering}}}\ (\bibinfo  {publisher} {National Academies Press},\ \bibinfo {year} {2012})\BibitemShut {NoStop}%
\bibitem [{\citenamefont {Cooper}\ \emph {et~al.}(2015)\citenamefont {Cooper}, \citenamefont {Caballero}, \citenamefont {Ebert-May}, \citenamefont {Fata-Hartley}, \citenamefont {Jardeleza}, \citenamefont {Krajcik}, \citenamefont {Laverty}, \citenamefont {Matz}, \citenamefont {Posey},\ and\ \citenamefont {Underwood}}]{cooper2015challenge}%
  \BibitemOpen
  \bibfield  {author} {\bibinfo {author} {\bibfnamefont {M.~M.}\ \bibnamefont {Cooper}}, \bibinfo {author} {\bibfnamefont {M.~D.}\ \bibnamefont {Caballero}}, \bibinfo {author} {\bibfnamefont {D.}~\bibnamefont {Ebert-May}}, \bibinfo {author} {\bibfnamefont {C.~L.}\ \bibnamefont {Fata-Hartley}}, \bibinfo {author} {\bibfnamefont {S.~E.}\ \bibnamefont {Jardeleza}}, \bibinfo {author} {\bibfnamefont {J.~S.}\ \bibnamefont {Krajcik}}, \bibinfo {author} {\bibfnamefont {J.~T.}\ \bibnamefont {Laverty}}, \bibinfo {author} {\bibfnamefont {R.~L.}\ \bibnamefont {Matz}}, \bibinfo {author} {\bibfnamefont {L.~A.}\ \bibnamefont {Posey}},\ and\ \bibinfo {author} {\bibfnamefont {S.~M.}\ \bibnamefont {Underwood}},\ }\bibfield  {title} {\bibinfo {title} {Challenge faculty to transform stem learning},\ }\href@noop {} {\bibfield  {journal} {\bibinfo  {journal} {Science}\ }\textbf {\bibinfo {volume} {350}},\ \bibinfo {pages} {281} (\bibinfo {year} {2015})}\BibitemShut {NoStop}%
\bibitem [{\citenamefont {Chao}\ \emph {et~al.}(2017)\citenamefont {Chao}, \citenamefont {Xie}, \citenamefont {Nourian}, \citenamefont {Chen}, \citenamefont {Bailey}, \citenamefont {Goldstein}, \citenamefont {Purzer}, \citenamefont {Adams},\ and\ \citenamefont {Tutwiler}}]{chao2017bridging}%
  \BibitemOpen
  \bibfield  {author} {\bibinfo {author} {\bibfnamefont {J.}~\bibnamefont {Chao}}, \bibinfo {author} {\bibfnamefont {C.}~\bibnamefont {Xie}}, \bibinfo {author} {\bibfnamefont {S.}~\bibnamefont {Nourian}}, \bibinfo {author} {\bibfnamefont {G.}~\bibnamefont {Chen}}, \bibinfo {author} {\bibfnamefont {S.}~\bibnamefont {Bailey}}, \bibinfo {author} {\bibfnamefont {M.~H.}\ \bibnamefont {Goldstein}}, \bibinfo {author} {\bibfnamefont {S.}~\bibnamefont {Purzer}}, \bibinfo {author} {\bibfnamefont {R.~S.}\ \bibnamefont {Adams}},\ and\ \bibinfo {author} {\bibfnamefont {M.~S.}\ \bibnamefont {Tutwiler}},\ }\bibfield  {title} {\bibinfo {title} {Bridging the design-science gap with tools: Science learning and design behaviors in a simulated environment for engineering design},\ }\href@noop {} {\bibfield  {journal} {\bibinfo  {journal} {Journal of Research in Science Teaching}\ }\textbf {\bibinfo {volume} {54}},\ \bibinfo {pages} {1049} (\bibinfo {year} {2017})}\BibitemShut {NoStop}%
\bibitem [{\citenamefont {{National Research Council}}(2013)}]{NGSS2013}%
  \BibitemOpen
  \bibfield  {author} {\bibinfo {author} {\bibnamefont {{National Research Council}}},\ }\href {https://doi.org/10.17226/18290} {\emph {\bibinfo {title} {Next Generation Science Standards: For States, By States}}}\ (\bibinfo  {publisher} {The National Academies Press},\ \bibinfo {address} {Washington, DC},\ \bibinfo {year} {2013})\BibitemShut {NoStop}%
\bibitem [{\citenamefont {Vattam}\ and\ \citenamefont {Kolodner}(2008)}]{vattam2008foundations}%
  \BibitemOpen
  \bibfield  {author} {\bibinfo {author} {\bibfnamefont {S.~S.}\ \bibnamefont {Vattam}}\ and\ \bibinfo {author} {\bibfnamefont {J.~L.}\ \bibnamefont {Kolodner}},\ }\bibfield  {title} {\bibinfo {title} {On foundations of technological support for addressing challenges facing design-based science learning},\ }\href@noop {} {\bibfield  {journal} {\bibinfo  {journal} {Pragmatics \& Cognition}\ }\textbf {\bibinfo {volume} {16}},\ \bibinfo {pages} {406} (\bibinfo {year} {2008})}\BibitemShut {NoStop}%
\bibitem [{\citenamefont {Kolodner}\ \emph {et~al.}(2003)\citenamefont {Kolodner}, \citenamefont {Camp}, \citenamefont {Crismond}, \citenamefont {Fasse}, \citenamefont {Gray}, \citenamefont {Holbrook}, \citenamefont {Puntambekar},\ and\ \citenamefont {Ryan}}]{kolod_punt_2003}%
  \BibitemOpen
  \bibfield  {author} {\bibinfo {author} {\bibfnamefont {J.~L.}\ \bibnamefont {Kolodner}}, \bibinfo {author} {\bibfnamefont {P.~J.}\ \bibnamefont {Camp}}, \bibinfo {author} {\bibfnamefont {D.}~\bibnamefont {Crismond}}, \bibinfo {author} {\bibfnamefont {B.}~\bibnamefont {Fasse}}, \bibinfo {author} {\bibfnamefont {J.}~\bibnamefont {Gray}}, \bibinfo {author} {\bibfnamefont {J.}~\bibnamefont {Holbrook}}, \bibinfo {author} {\bibfnamefont {S.}~\bibnamefont {Puntambekar}},\ and\ \bibinfo {author} {\bibfnamefont {M.}~\bibnamefont {Ryan}},\ }\bibfield  {title} {\bibinfo {title} {Problem-based learning meets case-based reasoning in the middle-school science classroom: Putting learning by design (tm) into practice},\ }\href@noop {} {\bibfield  {journal} {\bibinfo  {journal} {The journal of the learning sciences}\ }\textbf {\bibinfo {volume} {12}},\ \bibinfo {pages} {495} (\bibinfo {year} {2003})}\BibitemShut {NoStop}%
\bibitem [{\citenamefont {Berland}\ \emph {et~al.}(2013)\citenamefont {Berland}, \citenamefont {Martin}, \citenamefont {Ko}, \citenamefont {Peacock}, \citenamefont {Rudolph},\ and\ \citenamefont {Golubski}}]{berland2013student}%
  \BibitemOpen
  \bibfield  {author} {\bibinfo {author} {\bibfnamefont {L.~K.}\ \bibnamefont {Berland}}, \bibinfo {author} {\bibfnamefont {T.~H.}\ \bibnamefont {Martin}}, \bibinfo {author} {\bibfnamefont {P.}~\bibnamefont {Ko}}, \bibinfo {author} {\bibfnamefont {S.~B.}\ \bibnamefont {Peacock}}, \bibinfo {author} {\bibfnamefont {J.~J.}\ \bibnamefont {Rudolph}},\ and\ \bibinfo {author} {\bibfnamefont {C.}~\bibnamefont {Golubski}},\ }\bibfield  {title} {\bibinfo {title} {Student learning in challenge-based engineering curricula},\ }\href@noop {} {\bibfield  {journal} {\bibinfo  {journal} {Journal of Pre-College Engineering Education Research (J-PEER)}\ }\textbf {\bibinfo {volume} {3}},\ \bibinfo {pages} {5} (\bibinfo {year} {2013})}\BibitemShut {NoStop}%
\bibitem [{\citenamefont {Capobianco}\ \emph {et~al.}(2018)\citenamefont {Capobianco}, \citenamefont {DeLisi},\ and\ \citenamefont {Radloff}}]{capobianco2018characterizing}%
  \BibitemOpen
  \bibfield  {author} {\bibinfo {author} {\bibfnamefont {B.~M.}\ \bibnamefont {Capobianco}}, \bibinfo {author} {\bibfnamefont {J.}~\bibnamefont {DeLisi}},\ and\ \bibinfo {author} {\bibfnamefont {J.}~\bibnamefont {Radloff}},\ }\bibfield  {title} {\bibinfo {title} {Characterizing elementary teachers’ enactment of high-leverage practices through engineering design-based science instruction},\ }\href@noop {} {\bibfield  {journal} {\bibinfo  {journal} {Science Education}\ }\textbf {\bibinfo {volume} {102}},\ \bibinfo {pages} {342} (\bibinfo {year} {2018})}\BibitemShut {NoStop}%
\bibitem [{\citenamefont {Shekoyan}\ and\ \citenamefont {Etkina}(2007)}]{shekoyan2007introducing}%
  \BibitemOpen
  \bibfield  {author} {\bibinfo {author} {\bibfnamefont {V.}~\bibnamefont {Shekoyan}}\ and\ \bibinfo {author} {\bibfnamefont {E.}~\bibnamefont {Etkina}},\ }\bibfield  {title} {\bibinfo {title} {Introducing ill-structured problems in introductory physics recitations},\ }in\ \href@noop {} {\emph {\bibinfo {booktitle} {AIP Conference Proceedings}}},\ Vol.\ \bibinfo {volume} {951}\ (\bibinfo {organization} {American Institute of Physics},\ \bibinfo {year} {2007})\ pp.\ \bibinfo {pages} {192--195}\BibitemShut {NoStop}%
\bibitem [{\citenamefont {Subramaniam}\ \emph {et~al.}(2023{\natexlab{a}})\citenamefont {Subramaniam}, \citenamefont {Bralin}, \citenamefont {Morphew}, \citenamefont {Rebello},\ and\ \citenamefont {Rebello}}]{subramaniam2023narst}%
  \BibitemOpen
  \bibfield  {author} {\bibinfo {author} {\bibfnamefont {R.}~\bibnamefont {Subramaniam}}, \bibinfo {author} {\bibfnamefont {A.}~\bibnamefont {Bralin}}, \bibinfo {author} {\bibfnamefont {J.}~\bibnamefont {Morphew}}, \bibinfo {author} {\bibfnamefont {C.}~\bibnamefont {Rebello}},\ and\ \bibinfo {author} {\bibfnamefont {N.}~\bibnamefont {Rebello}},\ }\bibfield  {title} {\bibinfo {title} {Characterizing student thinking and evidence-based reasoning during an engineering design activity in introductory physics.},\ }in\ \href@noop {} {\emph {\bibinfo {booktitle} {Proceedings of the National Association for Research in Science Teaching}}},\ \bibinfo {editor} {edited by\ \bibinfo {editor} {\bibnamefont {NARST}}}\ (\bibinfo {address} {Chicago},\ \bibinfo {year} {2023})\BibitemShut {NoStop}%
\bibitem [{\citenamefont {Subramaniam}\ \emph {et~al.}(2023{\natexlab{b}})\citenamefont {Subramaniam}, \citenamefont {Bralin}, \citenamefont {Morphew}, \citenamefont {Rebello},\ and\ \citenamefont {Rebello}}]{ravi_perc_2023}%
  \BibitemOpen
  \bibfield  {author} {\bibinfo {author} {\bibfnamefont {R.}~\bibnamefont {Subramaniam}}, \bibinfo {author} {\bibfnamefont {A.}~\bibnamefont {Bralin}}, \bibinfo {author} {\bibfnamefont {J.}~\bibnamefont {Morphew}}, \bibinfo {author} {\bibfnamefont {C.}~\bibnamefont {Rebello}},\ and\ \bibinfo {author} {\bibfnamefont {N.~S.}\ \bibnamefont {Rebello}},\ }\bibfield  {title} {\bibinfo {title} {Characterizing the 'design-science gap' in an engineering design-based laboratory unit in an introductory physics course for future engineers},\ }in\ \href {https://doi.org/10.1119/perc.2023.pr.Subramaniam} {\emph {\bibinfo {booktitle} {2023 Physics Education Research Conference Proceedings}}}\ (\bibinfo {year} {2023})\ pp.\ \bibinfo {pages} {344--349}\BibitemShut {NoStop}%
\bibitem [{\citenamefont {Otero}\ and\ \citenamefont {Harlow}(2009)}]{otero2009getting}%
  \BibitemOpen
  \bibfield  {author} {\bibinfo {author} {\bibfnamefont {V.~K.}\ \bibnamefont {Otero}}\ and\ \bibinfo {author} {\bibfnamefont {D.~B.}\ \bibnamefont {Harlow}},\ }\bibfield  {title} {\bibinfo {title} {Getting started in qualitative physics education research},\ }\href@noop {} {\bibfield  {journal} {\bibinfo  {journal} {Reviews in PER}\ }\textbf {\bibinfo {volume} {2}} (\bibinfo {year} {2009})}\BibitemShut {NoStop}%
\bibitem [{\citenamefont {Braun}\ and\ \citenamefont {Clarke}(2006)}]{braun2006using}%
  \BibitemOpen
  \bibfield  {author} {\bibinfo {author} {\bibfnamefont {V.}~\bibnamefont {Braun}}\ and\ \bibinfo {author} {\bibfnamefont {V.}~\bibnamefont {Clarke}},\ }\bibfield  {title} {\bibinfo {title} {Using thematic analysis in psychology},\ }\href@noop {} {\bibfield  {journal} {\bibinfo  {journal} {Qualitative research in psychology}\ }\textbf {\bibinfo {volume} {3}},\ \bibinfo {pages} {77} (\bibinfo {year} {2006})}\BibitemShut {NoStop}%
\bibitem [{\citenamefont {Brink}(1987)}]{brink1987reliability}%
  \BibitemOpen
  \bibfield  {author} {\bibinfo {author} {\bibfnamefont {P.~J.}\ \bibnamefont {Brink}},\ }\href@noop {} {\bibinfo {title} {On reliability and validity in qualitative research}} (\bibinfo {year} {1987})\BibitemShut {NoStop}%
\bibitem [{\citenamefont {Stahl}\ and\ \citenamefont {King}(2020)}]{stahl2020expanding}%
  \BibitemOpen
  \bibfield  {author} {\bibinfo {author} {\bibfnamefont {N.~A.}\ \bibnamefont {Stahl}}\ and\ \bibinfo {author} {\bibfnamefont {J.~R.}\ \bibnamefont {King}},\ }\bibfield  {title} {\bibinfo {title} {Expanding approaches for research: Understanding and using trustworthiness in qualitative research},\ }\href@noop {} {\bibfield  {journal} {\bibinfo  {journal} {Journal of developmental education}\ }\textbf {\bibinfo {volume} {44}},\ \bibinfo {pages} {26} (\bibinfo {year} {2020})}\BibitemShut {NoStop}%
\bibitem [{\citenamefont {Ponterotto}(2006)}]{ponterotto2006brief}%
  \BibitemOpen
  \bibfield  {author} {\bibinfo {author} {\bibfnamefont {J.~G.}\ \bibnamefont {Ponterotto}},\ }\bibfield  {title} {\bibinfo {title} {Brief note on the origins, evolution, and meaning of the qualitative research concept thick description},\ }\href@noop {} {\bibfield  {journal} {\bibinfo  {journal} {The qualitative report}\ }\textbf {\bibinfo {volume} {11}},\ \bibinfo {pages} {538} (\bibinfo {year} {2006})}\BibitemShut {NoStop}%
\bibitem [{\citenamefont {Gioia}\ \emph {et~al.}(2013)\citenamefont {Gioia}, \citenamefont {Corley},\ and\ \citenamefont {Hamilton}}]{gioia2013seeking}%
  \BibitemOpen
  \bibfield  {author} {\bibinfo {author} {\bibfnamefont {D.~A.}\ \bibnamefont {Gioia}}, \bibinfo {author} {\bibfnamefont {K.~G.}\ \bibnamefont {Corley}},\ and\ \bibinfo {author} {\bibfnamefont {A.~L.}\ \bibnamefont {Hamilton}},\ }\bibfield  {title} {\bibinfo {title} {Seeking qualitative rigor in inductive research: Notes on the gioia methodology},\ }\href@noop {} {\bibfield  {journal} {\bibinfo  {journal} {Organizational research methods}\ }\textbf {\bibinfo {volume} {16}},\ \bibinfo {pages} {15} (\bibinfo {year} {2013})}\BibitemShut {NoStop}%
\bibitem [{\citenamefont {{National Research Council}}\ \emph {et~al.}(2013)\citenamefont {{National Research Council}}, \citenamefont {{Division on Engineering and Physical Sciences}}, \citenamefont {{Board on Physics and Astronomy}},\ and\ \citenamefont {{Committee on Undergraduate Physics Education Research and Implementation}}}]{national2013adapting}%
  \BibitemOpen
  \bibfield  {author} {\bibinfo {author} {\bibnamefont {{National Research Council}}}, \bibinfo {author} {\bibnamefont {{Division on Engineering and Physical Sciences}}}, \bibinfo {author} {\bibnamefont {{Board on Physics and Astronomy}}},\ and\ \bibinfo {author} {\bibnamefont {{Committee on Undergraduate Physics Education Research and Implementation}}},\ }\href@noop {} {\emph {\bibinfo {title} {Adapting to a Changing World: Challenges and Opportunities in Undergraduate Physics Education}}}\ (\bibinfo  {publisher} {National Academies Press},\ \bibinfo {year} {2013})\BibitemShut {NoStop}%
\bibitem [{\citenamefont {Council}\ \emph {et~al.}(2012)\citenamefont {Council} \emph {et~al.}}]{NRC2012}%
  \BibitemOpen
  \bibfield  {author} {\bibinfo {author} {\bibfnamefont {N.~R.}\ \bibnamefont {Council}} \emph {et~al.},\ }\href@noop {} {\emph {\bibinfo {title} {A framework for K-12 science education: Practices, crosscutting concepts, and core ideas}}}\ (\bibinfo  {publisher} {National Academies Press},\ \bibinfo {year} {2012})\BibitemShut {NoStop}%
\bibitem [{\citenamefont {Malmqvist}\ \emph {et~al.}(2004)\citenamefont {Malmqvist}, \citenamefont {Young}, \citenamefont {Hallstr{\"o}m}, \citenamefont {Kuttenkeuler}, \citenamefont {Svensson} \emph {et~al.}}]{malmqvist2004lessons}%
  \BibitemOpen
  \bibfield  {author} {\bibinfo {author} {\bibfnamefont {J.}~\bibnamefont {Malmqvist}}, \bibinfo {author} {\bibfnamefont {P.~Y.}\ \bibnamefont {Young}}, \bibinfo {author} {\bibfnamefont {S.}~\bibnamefont {Hallstr{\"o}m}}, \bibinfo {author} {\bibfnamefont {J.}~\bibnamefont {Kuttenkeuler}}, \bibinfo {author} {\bibfnamefont {T.}~\bibnamefont {Svensson}}, \emph {et~al.},\ }\bibfield  {title} {\bibinfo {title} {Lessons learned from design-build-test-based project courses},\ }in\ \href@noop {} {\emph {\bibinfo {booktitle} {DS 32: Proceedings of DESIGN 2004, the 8th International Design Conference, Dubrovnik, Croatia}}}\ (\bibinfo {year} {2004})\ pp.\ \bibinfo {pages} {665--672}\BibitemShut {NoStop}%
\bibitem [{\citenamefont {Redish}\ and\ \citenamefont {Smith}(2008)}]{redish2008looking}%
  \BibitemOpen
  \bibfield  {author} {\bibinfo {author} {\bibfnamefont {E.~F.}\ \bibnamefont {Redish}}\ and\ \bibinfo {author} {\bibfnamefont {K.~A.}\ \bibnamefont {Smith}},\ }\bibfield  {title} {\bibinfo {title} {Looking beyond content: Skill development for engineers},\ }\href@noop {} {\bibfield  {journal} {\bibinfo  {journal} {Journal of Engineering Education}\ }\textbf {\bibinfo {volume} {97}},\ \bibinfo {pages} {295} (\bibinfo {year} {2008})}\BibitemShut {NoStop}%
\bibitem [{\citenamefont {Pintrich}(2004)}]{pintrich2004conceptual}%
  \BibitemOpen
  \bibfield  {author} {\bibinfo {author} {\bibfnamefont {P.~R.}\ \bibnamefont {Pintrich}},\ }\bibfield  {title} {\bibinfo {title} {A conceptual framework for assessing motivation and self-regulated learning in college students},\ }\href@noop {} {\bibfield  {journal} {\bibinfo  {journal} {Educational psychology review}\ }\textbf {\bibinfo {volume} {16}},\ \bibinfo {pages} {385} (\bibinfo {year} {2004})}\BibitemShut {NoStop}%
\bibitem [{\citenamefont {Campione}\ \emph {et~al.}(2013)\citenamefont {Campione}, \citenamefont {Shapiro},\ and\ \citenamefont {Brown}}]{campione2013forms}%
  \BibitemOpen
  \bibfield  {author} {\bibinfo {author} {\bibfnamefont {J.~C.}\ \bibnamefont {Campione}}, \bibinfo {author} {\bibfnamefont {A.~M.}\ \bibnamefont {Shapiro}},\ and\ \bibinfo {author} {\bibfnamefont {A.~L.}\ \bibnamefont {Brown}},\ }\bibfield  {title} {\bibinfo {title} {Forms of transfer in a community of learners: Flexible learning and understanding},\ }in\ \href@noop {} {\emph {\bibinfo {booktitle} {Teaching for transfer}}}\ (\bibinfo  {publisher} {Routledge},\ \bibinfo {year} {2013})\ pp.\ \bibinfo {pages} {35--68}\BibitemShut {NoStop}%
\bibitem [{\citenamefont {Davidowitz}\ and\ \citenamefont {Rollnick}(2003)}]{davidowitz2003enabling}%
  \BibitemOpen
  \bibfield  {author} {\bibinfo {author} {\bibfnamefont {B.}~\bibnamefont {Davidowitz}}\ and\ \bibinfo {author} {\bibfnamefont {M.}~\bibnamefont {Rollnick}},\ }\bibfield  {title} {\bibinfo {title} {Enabling metacognition in the laboratory: A case study of four second year university chemistry students},\ }\href@noop {} {\bibfield  {journal} {\bibinfo  {journal} {Research in Science Education}\ }\textbf {\bibinfo {volume} {33}},\ \bibinfo {pages} {43} (\bibinfo {year} {2003})}\BibitemShut {NoStop}%
\bibitem [{\citenamefont {Hofstein}\ \emph {et~al.}(2004)\citenamefont {Hofstein}, \citenamefont {Shore},\ and\ \citenamefont {Kipnis}}]{hofsteinchem2004}%
  \BibitemOpen
  \bibfield  {author} {\bibinfo {author} {\bibfnamefont {A.}~\bibnamefont {Hofstein}}, \bibinfo {author} {\bibfnamefont {R.}~\bibnamefont {Shore}},\ and\ \bibinfo {author} {\bibfnamefont {M.}~\bibnamefont {Kipnis}},\ }\bibfield  {title} {\bibinfo {title} {Providing high school chemistry students with opportunities to develop learning skills in an inquiry-type laboratory: A case study},\ }\href@noop {} {\bibfield  {journal} {\bibinfo  {journal} {International Journal of Science Education}\ }\textbf {\bibinfo {volume} {26}},\ \bibinfo {pages} {47} (\bibinfo {year} {2004})}\BibitemShut {NoStop}%
\bibitem [{\citenamefont {Gee}(2014)}]{gee2014discourse}%
  \BibitemOpen
  \bibfield  {author} {\bibinfo {author} {\bibfnamefont {J.~P.}\ \bibnamefont {Gee}},\ }\href@noop {} {\emph {\bibinfo {title} {An introduction to discourse analysis: Theory and method}}}\ (\bibinfo  {publisher} {routledge},\ \bibinfo {year} {2014})\BibitemShut {NoStop}%
\bibitem [{\citenamefont {Schnotz}\ \emph {et~al.}(2009)\citenamefont {Schnotz}, \citenamefont {Fries},\ and\ \citenamefont {Horz}}]{schnotz2009some}%
  \BibitemOpen
  \bibfield  {author} {\bibinfo {author} {\bibfnamefont {W.}~\bibnamefont {Schnotz}}, \bibinfo {author} {\bibfnamefont {S.}~\bibnamefont {Fries}},\ and\ \bibinfo {author} {\bibfnamefont {H.}~\bibnamefont {Horz}},\ }\bibfield  {title} {\bibinfo {title} {Some motivational aspects of cognitive load theory},\ }\href@noop {} {\bibfield  {journal} {\bibinfo  {journal} {Contemporary motivation research: From global to local perspectives}\ ,\ \bibinfo {pages} {86}} (\bibinfo {year} {2009})}\BibitemShut {NoStop}%
\bibitem [{\citenamefont {Karelina}\ and\ \citenamefont {Etkina}(2007)}]{karelina2007acting}%
  \BibitemOpen
  \bibfield  {author} {\bibinfo {author} {\bibfnamefont {A.}~\bibnamefont {Karelina}}\ and\ \bibinfo {author} {\bibfnamefont {E.}~\bibnamefont {Etkina}},\ }\bibfield  {title} {\bibinfo {title} {Acting like a physicist: Student approach study to experimental design},\ }\href@noop {} {\bibfield  {journal} {\bibinfo  {journal} {Physical Review Special Topics-Physics Education Research}\ }\textbf {\bibinfo {volume} {3}},\ \bibinfo {pages} {020106} (\bibinfo {year} {2007})}\BibitemShut {NoStop}%
\bibitem [{\citenamefont {May}(2023)}]{may2023historical}%
  \BibitemOpen
  \bibfield  {author} {\bibinfo {author} {\bibfnamefont {J.~M.}\ \bibnamefont {May}},\ }\bibfield  {title} {\bibinfo {title} {Historical analysis of innovation and research in physics instructional laboratories: Recurring themes and future directions},\ }\href@noop {} {\bibfield  {journal} {\bibinfo  {journal} {Physical Review Physics Education Research}\ }\textbf {\bibinfo {volume} {19}},\ \bibinfo {pages} {020168} (\bibinfo {year} {2023})}\BibitemShut {NoStop}%
\bibitem [{\citenamefont {Capobianco}\ \emph {et~al.}(2013)\citenamefont {Capobianco}, \citenamefont {Nyquist},\ and\ \citenamefont {Tyrie}}]{capobianco2013shedding}%
  \BibitemOpen
  \bibfield  {author} {\bibinfo {author} {\bibfnamefont {B.~M.}\ \bibnamefont {Capobianco}}, \bibinfo {author} {\bibfnamefont {C.}~\bibnamefont {Nyquist}},\ and\ \bibinfo {author} {\bibfnamefont {N.}~\bibnamefont {Tyrie}},\ }\bibfield  {title} {\bibinfo {title} {Shedding light on engineering design},\ }\href@noop {} {\bibfield  {journal} {\bibinfo  {journal} {Science and Children}\ }\textbf {\bibinfo {volume} {50}},\ \bibinfo {pages} {58} (\bibinfo {year} {2013})}\BibitemShut {NoStop}%
\bibitem [{\citenamefont {Dringenberg}\ and\ \citenamefont {Purzer}(2018)}]{dringenberg2018experiences}%
  \BibitemOpen
  \bibfield  {author} {\bibinfo {author} {\bibfnamefont {E.}~\bibnamefont {Dringenberg}}\ and\ \bibinfo {author} {\bibfnamefont {{\c{S}}.}~\bibnamefont {Purzer}},\ }\bibfield  {title} {\bibinfo {title} {Experiences of first-year engineering students working on ill-structured problems in teams},\ }\href@noop {} {\bibfield  {journal} {\bibinfo  {journal} {Journal of Engineering Education}\ }\textbf {\bibinfo {volume} {107}},\ \bibinfo {pages} {442} (\bibinfo {year} {2018})}\BibitemShut {NoStop}%
\bibitem [{\citenamefont {Cross}(1982)}]{cross1982designerly}%
  \BibitemOpen
  \bibfield  {author} {\bibinfo {author} {\bibfnamefont {N.}~\bibnamefont {Cross}},\ }\bibfield  {title} {\bibinfo {title} {Designerly ways of knowing},\ }\href@noop {} {\bibfield  {journal} {\bibinfo  {journal} {Design studies}\ }\textbf {\bibinfo {volume} {3}},\ \bibinfo {pages} {221} (\bibinfo {year} {1982})}\BibitemShut {NoStop}%
\bibitem [{\citenamefont {Puntambekar}\ and\ \citenamefont {Kolodner}(2005)}]{puntambekar2005toward}%
  \BibitemOpen
  \bibfield  {author} {\bibinfo {author} {\bibfnamefont {S.}~\bibnamefont {Puntambekar}}\ and\ \bibinfo {author} {\bibfnamefont {J.~L.}\ \bibnamefont {Kolodner}},\ }\bibfield  {title} {\bibinfo {title} {Toward implementing distributed scaffolding: Helping students learn science from design},\ }\href@noop {} {\bibfield  {journal} {\bibinfo  {journal} {Journal of Research in Science Teaching: The Official Journal of the National Association for Research in Science Teaching}\ }\textbf {\bibinfo {volume} {42}},\ \bibinfo {pages} {185} (\bibinfo {year} {2005})}\BibitemShut {NoStop}%
\bibitem [{\citenamefont {Engle}\ \emph {et~al.}(2012)\citenamefont {Engle}, \citenamefont {Lam}, \citenamefont {Meyer},\ and\ \citenamefont {Nix}}]{engle2012does}%
  \BibitemOpen
  \bibfield  {author} {\bibinfo {author} {\bibfnamefont {R.~A.}\ \bibnamefont {Engle}}, \bibinfo {author} {\bibfnamefont {D.~P.}\ \bibnamefont {Lam}}, \bibinfo {author} {\bibfnamefont {X.~S.}\ \bibnamefont {Meyer}},\ and\ \bibinfo {author} {\bibfnamefont {S.~E.}\ \bibnamefont {Nix}},\ }\bibfield  {title} {\bibinfo {title} {How does expansive framing promote transfer? several proposed explanations and a research agenda for investigating them},\ }\href@noop {} {\bibfield  {journal} {\bibinfo  {journal} {Educational Psychologist}\ }\textbf {\bibinfo {volume} {47}},\ \bibinfo {pages} {215} (\bibinfo {year} {2012})}\BibitemShut {NoStop}%
\bibitem [{\citenamefont {Sirnoorkar}\ \emph {et~al.}(2016)\citenamefont {Sirnoorkar}, \citenamefont {Mazumdar},\ and\ \citenamefont {Kumar}}]{sirnoorkar2016students}%
  \BibitemOpen
  \bibfield  {author} {\bibinfo {author} {\bibfnamefont {A.}~\bibnamefont {Sirnoorkar}}, \bibinfo {author} {\bibfnamefont {A.}~\bibnamefont {Mazumdar}},\ and\ \bibinfo {author} {\bibfnamefont {A.}~\bibnamefont {Kumar}},\ }\bibfield  {title} {\bibinfo {title} {Students’ epistemic understanding of mathematical derivations in physics},\ }\href@noop {} {\bibfield  {journal} {\bibinfo  {journal} {European Journal of Physics}\ }\textbf {\bibinfo {volume} {38}},\ \bibinfo {pages} {015703} (\bibinfo {year} {2016})}\BibitemShut {NoStop}%
\bibitem [{\citenamefont {Sirnoorkar}\ \emph {et~al.}(2020)\citenamefont {Sirnoorkar}, \citenamefont {Mazumdar},\ and\ \citenamefont {Kumar}}]{sirnoorkar2020towards}%
  \BibitemOpen
  \bibfield  {author} {\bibinfo {author} {\bibfnamefont {A.}~\bibnamefont {Sirnoorkar}}, \bibinfo {author} {\bibfnamefont {A.}~\bibnamefont {Mazumdar}},\ and\ \bibinfo {author} {\bibfnamefont {A.}~\bibnamefont {Kumar}},\ }\bibfield  {title} {\bibinfo {title} {Towards a content-based epistemic measure in physics},\ }\href@noop {} {\bibfield  {journal} {\bibinfo  {journal} {Physical Review Physics Education Research}\ }\textbf {\bibinfo {volume} {16}},\ \bibinfo {pages} {010103} (\bibinfo {year} {2020})}\BibitemShut {NoStop}%
\bibitem [{\citenamefont {Radloff}\ and\ \citenamefont {Chase}(2018)}]{radloff2018using}%
  \BibitemOpen
  \bibfield  {author} {\bibinfo {author} {\bibfnamefont {J.~D.}\ \bibnamefont {Radloff}}\ and\ \bibinfo {author} {\bibfnamefont {A.}~\bibnamefont {Chase}},\ }\bibfield  {title} {\bibinfo {title} {Using expansive framing to enhance personal relevancy and engagement in science},\ }\href@noop {} {\bibfield  {journal} {\bibinfo  {journal} {The Hoosier Science Teacher}\ }\textbf {\bibinfo {volume} {41}},\ \bibinfo {pages} {29} (\bibinfo {year} {2018})}\BibitemShut {NoStop}%
\bibitem [{\citenamefont {Mestre}(2006)}]{mestre2006transfer}%
  \BibitemOpen
  \bibfield  {author} {\bibinfo {author} {\bibfnamefont {J.~P.}\ \bibnamefont {Mestre}},\ }\href@noop {} {\emph {\bibinfo {title} {Transfer of learning from a modern multidisciplinary perspective}}}\ (\bibinfo  {publisher} {IAP},\ \bibinfo {year} {2006})\BibitemShut {NoStop}%
\bibitem [{\citenamefont {Engle}(2006)}]{engle2006framing}%
  \BibitemOpen
  \bibfield  {author} {\bibinfo {author} {\bibfnamefont {R.~A.}\ \bibnamefont {Engle}},\ }\bibfield  {title} {\bibinfo {title} {Framing interactions to foster generative learning: A situative explanation of transfer in a community of learners classroom},\ }\href@noop {} {\bibfield  {journal} {\bibinfo  {journal} {The journal of the learning sciences}\ }\textbf {\bibinfo {volume} {15}},\ \bibinfo {pages} {451} (\bibinfo {year} {2006})}\BibitemShut {NoStop}%
\bibitem [{\citenamefont {Chase}\ \emph {et~al.}(2019)\citenamefont {Chase}, \citenamefont {Malkiewich},\ and\ \citenamefont {S~Kumar}}]{chase2019learning}%
  \BibitemOpen
  \bibfield  {author} {\bibinfo {author} {\bibfnamefont {C.~C.}\ \bibnamefont {Chase}}, \bibinfo {author} {\bibfnamefont {L.}~\bibnamefont {Malkiewich}},\ and\ \bibinfo {author} {\bibfnamefont {A.}~\bibnamefont {S~Kumar}},\ }\bibfield  {title} {\bibinfo {title} {Learning to notice science concepts in engineering activities and transfer situations},\ }\href@noop {} {\bibfield  {journal} {\bibinfo  {journal} {Science Education}\ }\textbf {\bibinfo {volume} {103}},\ \bibinfo {pages} {440} (\bibinfo {year} {2019})}\BibitemShut {NoStop}%
\bibitem [{\citenamefont {Razzouk}\ and\ \citenamefont {Shute}(2012)}]{razzouk2012design}%
  \BibitemOpen
  \bibfield  {author} {\bibinfo {author} {\bibfnamefont {R.}~\bibnamefont {Razzouk}}\ and\ \bibinfo {author} {\bibfnamefont {V.}~\bibnamefont {Shute}},\ }\bibfield  {title} {\bibinfo {title} {What is design thinking and why is it important?},\ }\href@noop {} {\bibfield  {journal} {\bibinfo  {journal} {Review of educational research}\ }\textbf {\bibinfo {volume} {82}},\ \bibinfo {pages} {330} (\bibinfo {year} {2012})}\BibitemShut {NoStop}%
\bibitem [{\citenamefont {Li}\ \emph {et~al.}(2019)\citenamefont {Li}, \citenamefont {Schoenfeld}, \citenamefont {Disessa}, \citenamefont {Graesser}, \citenamefont {Benson}, \citenamefont {English},\ and\ \citenamefont {Duschl}}]{li2019design}%
  \BibitemOpen
  \bibfield  {author} {\bibinfo {author} {\bibfnamefont {Y.}~\bibnamefont {Li}}, \bibinfo {author} {\bibfnamefont {A.~H.}\ \bibnamefont {Schoenfeld}}, \bibinfo {author} {\bibfnamefont {A.~A.}\ \bibnamefont {Disessa}}, \bibinfo {author} {\bibfnamefont {A.~C.}\ \bibnamefont {Graesser}}, \bibinfo {author} {\bibfnamefont {L.~C.}\ \bibnamefont {Benson}}, \bibinfo {author} {\bibfnamefont {L.~D.}\ \bibnamefont {English}},\ and\ \bibinfo {author} {\bibfnamefont {R.~A.}\ \bibnamefont {Duschl}},\ }\href@noop {} {\bibinfo {title} {Design and design thinking in stem education}} (\bibinfo {year} {2019})\BibitemShut {NoStop}%
\bibitem [{\citenamefont {Dalsgaard}(2014)}]{dalsgaard2014pragmatism}%
  \BibitemOpen
  \bibfield  {author} {\bibinfo {author} {\bibfnamefont {P.}~\bibnamefont {Dalsgaard}},\ }\bibfield  {title} {\bibinfo {title} {Pragmatism and design thinking.},\ }\href@noop {} {\bibfield  {journal} {\bibinfo  {journal} {International Journal of design}\ }\textbf {\bibinfo {volume} {8}} (\bibinfo {year} {2014})}\BibitemShut {NoStop}%
\bibitem [{\citenamefont {English}(2023)}]{english2023ways}%
  \BibitemOpen
  \bibfield  {author} {\bibinfo {author} {\bibfnamefont {L.~D.}\ \bibnamefont {English}},\ }\bibfield  {title} {\bibinfo {title} {Ways of thinking in stem-based problem solving},\ }\href@noop {} {\bibfield  {journal} {\bibinfo  {journal} {ZDM--Mathematics Education}\ }\textbf {\bibinfo {volume} {55}},\ \bibinfo {pages} {1219} (\bibinfo {year} {2023})}\BibitemShut {NoStop}%
\bibitem [{\citenamefont {Owen}(2007)}]{owen2007design}%
  \BibitemOpen
  \bibfield  {author} {\bibinfo {author} {\bibfnamefont {C.}~\bibnamefont {Owen}},\ }\bibfield  {title} {\bibinfo {title} {Design thinking: Notes on its nature and use},\ }\href@noop {} {\bibfield  {journal} {\bibinfo  {journal} {Design research quarterly}\ }\textbf {\bibinfo {volume} {2}},\ \bibinfo {pages} {16} (\bibinfo {year} {2007})}\BibitemShut {NoStop}%
\bibitem [{\citenamefont {Kimbell}(2011)}]{kimbell2011rethinking}%
  \BibitemOpen
  \bibfield  {author} {\bibinfo {author} {\bibfnamefont {L.}~\bibnamefont {Kimbell}},\ }\bibfield  {title} {\bibinfo {title} {Rethinking design thinking: Part i},\ }\href@noop {} {\bibfield  {journal} {\bibinfo  {journal} {Design and culture}\ }\textbf {\bibinfo {volume} {3}},\ \bibinfo {pages} {285} (\bibinfo {year} {2011})}\BibitemShut {NoStop}%
\bibitem [{\citenamefont {Lehrer}\ and\ \citenamefont {Schauble}(2006)}]{lehrer2006scientific}%
  \BibitemOpen
  \bibfield  {author} {\bibinfo {author} {\bibfnamefont {R.}~\bibnamefont {Lehrer}}\ and\ \bibinfo {author} {\bibfnamefont {L.}~\bibnamefont {Schauble}},\ }\bibfield  {title} {\bibinfo {title} {Scientific thinking and science literacy},\ }\href@noop {} {\bibfield  {journal} {\bibinfo  {journal} {Handbook of child psychology}\ }\textbf {\bibinfo {volume} {4}},\ \bibinfo {pages} {153} (\bibinfo {year} {2006})}\BibitemShut {NoStop}%
\bibitem [{\citenamefont {Slavit}\ \emph {et~al.}(2021)\citenamefont {Slavit}, \citenamefont {Grace},\ and\ \citenamefont {Lesseig}}]{slavit2021student}%
  \BibitemOpen
  \bibfield  {author} {\bibinfo {author} {\bibfnamefont {D.}~\bibnamefont {Slavit}}, \bibinfo {author} {\bibfnamefont {E.}~\bibnamefont {Grace}},\ and\ \bibinfo {author} {\bibfnamefont {K.}~\bibnamefont {Lesseig}},\ }\bibfield  {title} {\bibinfo {title} {Student ways of thinking in stem contexts: A focus on claim making and reasoning},\ }\href@noop {} {\bibfield  {journal} {\bibinfo  {journal} {School Science and Mathematics}\ }\textbf {\bibinfo {volume} {121}},\ \bibinfo {pages} {466} (\bibinfo {year} {2021})}\BibitemShut {NoStop}%
\bibitem [{\citenamefont {Denick}\ \emph {et~al.}(2013)\citenamefont {Denick}, \citenamefont {Dringenberg}, \citenamefont {Fayyaz}, \citenamefont {Nelson}, \citenamefont {Pitterson}, \citenamefont {Tolbert}, \citenamefont {Yatchmeneff},\ and\ \citenamefont {Cardella}}]{denick2013stem}%
  \BibitemOpen
  \bibfield  {author} {\bibinfo {author} {\bibfnamefont {D.}~\bibnamefont {Denick}}, \bibinfo {author} {\bibfnamefont {E.}~\bibnamefont {Dringenberg}}, \bibinfo {author} {\bibfnamefont {F.}~\bibnamefont {Fayyaz}}, \bibinfo {author} {\bibfnamefont {L.}~\bibnamefont {Nelson}}, \bibinfo {author} {\bibfnamefont {N.}~\bibnamefont {Pitterson}}, \bibinfo {author} {\bibfnamefont {D.}~\bibnamefont {Tolbert}}, \bibinfo {author} {\bibfnamefont {M.}~\bibnamefont {Yatchmeneff}},\ and\ \bibinfo {author} {\bibfnamefont {M.}~\bibnamefont {Cardella}},\ }\bibfield  {title} {\bibinfo {title} {Stem thinking in informal environments: Integration and recommendations for formal settings},\ }in\ \href@noop {} {\emph {\bibinfo {booktitle} {2013 ASEE IL-IN Section Conference}}}\ (\bibinfo {year} {2013})\ pp.\ \bibinfo {pages} {1--17}\BibitemShut {NoStop}%
\bibitem [{\citenamefont {Lehrer}(2007)}]{lehrer2007scientific}%
  \BibitemOpen
  \bibfield  {author} {\bibinfo {author} {\bibfnamefont {R.}~\bibnamefont {Lehrer}},\ }\href@noop {} {\bibinfo {title} {Scientific thinking and science literacy. in, l. schauble, handbook of child psychology}} (\bibinfo {year} {2007})\BibitemShut {NoStop}%
\bibitem [{\citenamefont {Janou{\v{s}}kov{\'a}}\ \emph {et~al.}(2023)\citenamefont {Janou{\v{s}}kov{\'a}}, \citenamefont {Pyskat{\'a}~Rathousk{\'a}}, \citenamefont {{\v{Z}}{\'a}k},\ and\ \citenamefont {Urv{\'a}lkov{\'a}}}]{janouvskova2023scientific}%
  \BibitemOpen
  \bibfield  {author} {\bibinfo {author} {\bibfnamefont {S.}~\bibnamefont {Janou{\v{s}}kov{\'a}}}, \bibinfo {author} {\bibfnamefont {L.}~\bibnamefont {Pyskat{\'a}~Rathousk{\'a}}}, \bibinfo {author} {\bibfnamefont {V.}~\bibnamefont {{\v{Z}}{\'a}k}},\ and\ \bibinfo {author} {\bibfnamefont {E.~S.}\ \bibnamefont {Urv{\'a}lkov{\'a}}},\ }\bibfield  {title} {\bibinfo {title} {The scientific thinking and reasoning framework and its applicability to manufacturing and services firms in natural sciences},\ }\href@noop {} {\bibfield  {journal} {\bibinfo  {journal} {Research in Science \& Technological Education}\ }\textbf {\bibinfo {volume} {41}},\ \bibinfo {pages} {653} (\bibinfo {year} {2023})}\BibitemShut {NoStop}%
\bibitem [{\citenamefont {Lee}\ \emph {et~al.}(2013)\citenamefont {Lee}, \citenamefont {Quinn},\ and\ \citenamefont {Vald{\'e}s}}]{lee2013science}%
  \BibitemOpen
  \bibfield  {author} {\bibinfo {author} {\bibfnamefont {O.}~\bibnamefont {Lee}}, \bibinfo {author} {\bibfnamefont {H.}~\bibnamefont {Quinn}},\ and\ \bibinfo {author} {\bibfnamefont {G.}~\bibnamefont {Vald{\'e}s}},\ }\bibfield  {title} {\bibinfo {title} {Science and language for english language learners in relation to next generation science standards and with implications for common core state standards for english language arts and mathematics},\ }\href@noop {} {\bibfield  {journal} {\bibinfo  {journal} {Educational Researcher}\ }\textbf {\bibinfo {volume} {42}},\ \bibinfo {pages} {223} (\bibinfo {year} {2013})}\BibitemShut {NoStop}%
\bibitem [{\citenamefont {Luna}\ \emph {et~al.}(2018)\citenamefont {Luna}, \citenamefont {Selmer},\ and\ \citenamefont {Rye}}]{luna2018teachers}%
  \BibitemOpen
  \bibfield  {author} {\bibinfo {author} {\bibfnamefont {M.~J.}\ \bibnamefont {Luna}}, \bibinfo {author} {\bibfnamefont {S.~J.}\ \bibnamefont {Selmer}},\ and\ \bibinfo {author} {\bibfnamefont {J.~A.}\ \bibnamefont {Rye}},\ }\bibfield  {title} {\bibinfo {title} {Teachers’ noticing of students’ thinking in science through classroom artifacts: In what ways are science and engineering practices evident?},\ }\href@noop {} {\bibfield  {journal} {\bibinfo  {journal} {Journal of Science Teacher Education}\ }\textbf {\bibinfo {volume} {29}},\ \bibinfo {pages} {148} (\bibinfo {year} {2018})}\BibitemShut {NoStop}%
\bibitem [{\citenamefont {Etkina}\ and\ \citenamefont {Planin{\v{s}}i{\v{c}}}(2014)}]{etkina2014thinking}%
  \BibitemOpen
  \bibfield  {author} {\bibinfo {author} {\bibfnamefont {E.}~\bibnamefont {Etkina}}\ and\ \bibinfo {author} {\bibfnamefont {G.}~\bibnamefont {Planin{\v{s}}i{\v{c}}}},\ }\bibfield  {title} {\bibinfo {title} {Thinking like a scientist},\ }\href@noop {} {\bibfield  {journal} {\bibinfo  {journal} {Physics world}\ }\textbf {\bibinfo {volume} {27}},\ \bibinfo {pages} {48} (\bibinfo {year} {2014})}\BibitemShut {NoStop}%
\bibitem [{\citenamefont {Firetto}\ \emph {et~al.}(2023)\citenamefont {Firetto}, \citenamefont {Starrett},\ and\ \citenamefont {Jordan}}]{firetto2023embracing}%
  \BibitemOpen
  \bibfield  {author} {\bibinfo {author} {\bibfnamefont {C.~M.}\ \bibnamefont {Firetto}}, \bibinfo {author} {\bibfnamefont {E.}~\bibnamefont {Starrett}},\ and\ \bibinfo {author} {\bibfnamefont {M.~E.}\ \bibnamefont {Jordan}},\ }\bibfield  {title} {\bibinfo {title} {Embracing a culture of talk: Stem teachers’ engagement in small-group discussions about photovoltaics},\ }\href@noop {} {\bibfield  {journal} {\bibinfo  {journal} {International Journal of STEM Education}\ }\textbf {\bibinfo {volume} {10}},\ \bibinfo {pages} {50} (\bibinfo {year} {2023})}\BibitemShut {NoStop}%
\bibitem [{\citenamefont {Bennett}\ \emph {et~al.}(2010)\citenamefont {Bennett}, \citenamefont {Hogarth}, \citenamefont {Lubben}, \citenamefont {Campbell},\ and\ \citenamefont {Robinson}}]{bennett2010talking}%
  \BibitemOpen
  \bibfield  {author} {\bibinfo {author} {\bibfnamefont {J.}~\bibnamefont {Bennett}}, \bibinfo {author} {\bibfnamefont {S.}~\bibnamefont {Hogarth}}, \bibinfo {author} {\bibfnamefont {F.}~\bibnamefont {Lubben}}, \bibinfo {author} {\bibfnamefont {B.}~\bibnamefont {Campbell}},\ and\ \bibinfo {author} {\bibfnamefont {A.}~\bibnamefont {Robinson}},\ }\bibfield  {title} {\bibinfo {title} {Talking science: The research evidence on the use of small group discussions in science teaching},\ }\href@noop {} {\bibfield  {journal} {\bibinfo  {journal} {International Journal of Science Education}\ }\textbf {\bibinfo {volume} {32}},\ \bibinfo {pages} {69} (\bibinfo {year} {2010})}\BibitemShut {NoStop}%
\bibitem [{\citenamefont {Wieman}(2014)}]{wieman2014similarities}%
  \BibitemOpen
  \bibfield  {author} {\bibinfo {author} {\bibfnamefont {C.~E.}\ \bibnamefont {Wieman}},\ }\bibfield  {title} {\bibinfo {title} {The similarities between research in education and research in the hard sciences},\ }\href@noop {} {\bibfield  {journal} {\bibinfo  {journal} {Educational Researcher}\ }\textbf {\bibinfo {volume} {43}},\ \bibinfo {pages} {12} (\bibinfo {year} {2014})}\BibitemShut {NoStop}%
\bibitem [{\citenamefont {Wilkinson}\ \emph {et~al.}(2010)\citenamefont {Wilkinson}, \citenamefont {Soter},\ and\ \citenamefont {Murphy}}]{wilkinson2010developing}%
  \BibitemOpen
  \bibfield  {author} {\bibinfo {author} {\bibfnamefont {I.}~\bibnamefont {Wilkinson}}, \bibinfo {author} {\bibfnamefont {A.}~\bibnamefont {Soter}},\ and\ \bibinfo {author} {\bibfnamefont {P.}~\bibnamefont {Murphy}},\ }\bibfield  {title} {\bibinfo {title} {Developing a model of quality talk about literary text},\ }\href@noop {} {\bibfield  {journal} {\bibinfo  {journal} {Bringing reading research to life}\ ,\ \bibinfo {pages} {142}} (\bibinfo {year} {2010})}\BibitemShut {NoStop}%
\bibitem [{\citenamefont {Alev}(2010)}]{alev2010perceived}%
  \BibitemOpen
  \bibfield  {author} {\bibinfo {author} {\bibfnamefont {N.}~\bibnamefont {Alev}},\ }\bibfield  {title} {\bibinfo {title} {Perceived values of reading and writing in learning physics in secondary classrooms},\ }\href@noop {} {\bibfield  {journal} {\bibinfo  {journal} {Scientific Research and Essays}\ }\textbf {\bibinfo {volume} {5}},\ \bibinfo {pages} {1333} (\bibinfo {year} {2010})}\BibitemShut {NoStop}%
\bibitem [{\citenamefont {Atman}\ \emph {et~al.}(2007)\citenamefont {Atman}, \citenamefont {Adams}, \citenamefont {Cardella}, \citenamefont {Turns}, \citenamefont {Mosborg},\ and\ \citenamefont {Saleem}}]{atman2007engineering}%
  \BibitemOpen
  \bibfield  {author} {\bibinfo {author} {\bibfnamefont {C.~J.}\ \bibnamefont {Atman}}, \bibinfo {author} {\bibfnamefont {R.~S.}\ \bibnamefont {Adams}}, \bibinfo {author} {\bibfnamefont {M.~E.}\ \bibnamefont {Cardella}}, \bibinfo {author} {\bibfnamefont {J.}~\bibnamefont {Turns}}, \bibinfo {author} {\bibfnamefont {S.}~\bibnamefont {Mosborg}},\ and\ \bibinfo {author} {\bibfnamefont {J.}~\bibnamefont {Saleem}},\ }\bibfield  {title} {\bibinfo {title} {Engineering design processes: A comparison of students and expert practitioners},\ }\href@noop {} {\bibfield  {journal} {\bibinfo  {journal} {Journal of engineering education}\ }\textbf {\bibinfo {volume} {96}},\ \bibinfo {pages} {359} (\bibinfo {year} {2007})}\BibitemShut {NoStop}%
\bibitem [{\citenamefont {Mosborg}\ \emph {et~al.}(2005)\citenamefont {Mosborg}, \citenamefont {Adams}, \citenamefont {Kim}, \citenamefont {Atman}, \citenamefont {Turns},\ and\ \citenamefont {Cardella}}]{mosborg2005conceptions}%
  \BibitemOpen
  \bibfield  {author} {\bibinfo {author} {\bibfnamefont {S.}~\bibnamefont {Mosborg}}, \bibinfo {author} {\bibfnamefont {R.}~\bibnamefont {Adams}}, \bibinfo {author} {\bibfnamefont {R.}~\bibnamefont {Kim}}, \bibinfo {author} {\bibfnamefont {C.}~\bibnamefont {Atman}}, \bibinfo {author} {\bibfnamefont {J.}~\bibnamefont {Turns}},\ and\ \bibinfo {author} {\bibfnamefont {M.}~\bibnamefont {Cardella}},\ }\bibfield  {title} {\bibinfo {title} {Conceptions of the engineering design process: An expert study of advanced practicing professionals},\ }in\ \href@noop {} {\emph {\bibinfo {booktitle} {2005 Annual Conference}}}\ (\bibinfo {year} {2005})\ pp.\ \bibinfo {pages} {10--337}\BibitemShut {NoStop}%
\bibitem [{\citenamefont {Unknown}(2007)}]{wikimedia_iamge_gorilla}%
  \BibitemOpen
  \bibfield  {author} {\bibinfo {author} {\bibnamefont {Unknown}},\ }\href {https://commons.wikimedia.org/w/index.php?curid=7384043} {\bibinfo {title} {Cross-river-gorilla}} (\bibinfo {year} {2007}),\ \bibinfo {note} {accessed: 2024-08-09}\BibitemShut {NoStop}%
\bibitem [{\citenamefont {Contributors}(2024)}]{wikipedia_image_congo}%
  \BibitemOpen
  \bibfield  {author} {\bibinfo {author} {\bibfnamefont {W.}~\bibnamefont {Contributors}},\ }\href {https://en.wikipedia.org/wiki/Congo_River} {\bibinfo {title} {Congo river}} (\bibinfo {year} {2024}),\ \bibinfo {note} {accessed: 2024-08-09}\BibitemShut {NoStop}%
\bibitem [{\citenamefont {Chabay}\ and\ \citenamefont {Sherwood}(2015)}]{chabay2015matter}%
  \BibitemOpen
  \bibfield  {author} {\bibinfo {author} {\bibfnamefont {R.}~\bibnamefont {Chabay}}\ and\ \bibinfo {author} {\bibfnamefont {B.}~\bibnamefont {Sherwood}},\ }\href {https://books.google.com/books?id=Gz4HBgAAQBAJ} {\emph {\bibinfo {title} {Matter and Interactions}}}\ (\bibinfo  {publisher} {Wiley},\ \bibinfo {year} {2015})\BibitemShut {NoStop}%
\bibitem [{\citenamefont {Lopez}\ and\ \citenamefont {Whitehead}(2013)}]{lopez2013sampling}%
  \BibitemOpen
  \bibfield  {author} {\bibinfo {author} {\bibfnamefont {V.}~\bibnamefont {Lopez}}\ and\ \bibinfo {author} {\bibfnamefont {D.}~\bibnamefont {Whitehead}},\ }\bibfield  {title} {\bibinfo {title} {Sampling data and data collection in qualitative research},\ }\href@noop {} {\bibfield  {journal} {\bibinfo  {journal} {Nursing \& midwifery research: Methods and appraisal for evidence-based practice}\ }\textbf {\bibinfo {volume} {123}},\ \bibinfo {pages} {140} (\bibinfo {year} {2013})}\BibitemShut {NoStop}%
\bibitem [{\citenamefont {{PASCO Scientific}}(2024{\natexlab{a}})}]{pasco_me5300}%
  \BibitemOpen
  \bibfield  {author} {\bibinfo {author} {\bibnamefont {{PASCO Scientific}}},\ }\href@noop {} {\bibinfo {title} {{Engineering Statics Course (ME-5300)}}},\ \bibinfo {howpublished} {\url{https://www.pasco.com/products/bundles/physics/me-5300}} (\bibinfo {year} {2024}{\natexlab{a}}),\ \bibinfo {note} {accessed: 2024-09-03}\BibitemShut {NoStop}%
\bibitem [{\citenamefont {{PASCO Scientific}}(2024{\natexlab{b}})}]{pasco_capstone}%
  \BibitemOpen
  \bibfield  {author} {\bibinfo {author} {\bibnamefont {{PASCO Scientific}}},\ }\href@noop {} {\bibinfo {title} {{PASCO Capstone Software}}},\ \bibinfo {howpublished} {\url{https://www.pasco.com/products/software/capstone}} (\bibinfo {year} {2024}{\natexlab{b}}),\ \bibinfo {note} {accessed: 2024-09-03}\BibitemShut {NoStop}%
\bibitem [{PhE(2021)}]{PhET}%
  \BibitemOpen
  \href@noop {} {\bibinfo {title} {{PhET Interactive Simulations - }projectile motion}},\ \bibinfo {howpublished} {\url{http:https://phet.colorado.edu/en/simulations/projectile-motion}} (\bibinfo {year} {2021}),\ \bibinfo {note} {accessed: 2023-10-30}\BibitemShut {NoStop}%
\bibitem [{\citenamefont {{VPython Community}}(2024)}]{vpython}%
  \BibitemOpen
  \bibfield  {author} {\bibinfo {author} {\bibnamefont {{VPython Community}}},\ }\href@noop {} {\bibinfo {title} {{VPython: 3D Programming for Ordinary Mortals}}},\ \bibinfo {howpublished} {\url{https://vpython.org/}} (\bibinfo {year} {2024}),\ \bibinfo {note} {accessed: 2024-09-03}\BibitemShut {NoStop}%
\bibitem [{\citenamefont {Saldana}(2009)}]{saldana2009introduction}%
  \BibitemOpen
  \bibfield  {author} {\bibinfo {author} {\bibfnamefont {J.}~\bibnamefont {Saldana}},\ }\bibfield  {title} {\bibinfo {title} {An introduction to codes and coding},\ }\href@noop {} {\bibfield  {journal} {\bibinfo  {journal} {The coding manual for qualitative researchers}\ }\textbf {\bibinfo {volume} {3}} (\bibinfo {year} {2009})}\BibitemShut {NoStop}%
\bibitem [{\citenamefont {O'Dwyer}(2004)}]{odwyer_qlr_messy_intimate}%
  \BibitemOpen
  \bibfield  {author} {\bibinfo {author} {\bibfnamefont {B.}~\bibnamefont {O'Dwyer}},\ }\bibfield  {title} {\bibinfo {title} {Qualitative data analysis: illuminating a process for transforming a ‘messy’but ‘attractive’‘nuisance’},\ }in\ \href@noop {} {\emph {\bibinfo {booktitle} {The real life guide to accounting research}}}\ (\bibinfo  {publisher} {Elsevier},\ \bibinfo {year} {2004})\ pp.\ \bibinfo {pages} {391--407}\BibitemShut {NoStop}%
\bibitem [{\citenamefont {Morse}\ \emph {et~al.}(2002)\citenamefont {Morse}, \citenamefont {Barrett}, \citenamefont {Mayan}, \citenamefont {Olson},\ and\ \citenamefont {Spiers}}]{morse2002verification}%
  \BibitemOpen
  \bibfield  {author} {\bibinfo {author} {\bibfnamefont {J.~M.}\ \bibnamefont {Morse}}, \bibinfo {author} {\bibfnamefont {M.}~\bibnamefont {Barrett}}, \bibinfo {author} {\bibfnamefont {M.}~\bibnamefont {Mayan}}, \bibinfo {author} {\bibfnamefont {K.}~\bibnamefont {Olson}},\ and\ \bibinfo {author} {\bibfnamefont {J.}~\bibnamefont {Spiers}},\ }\bibfield  {title} {\bibinfo {title} {Verification strategies for establishing reliability and validity in qualitative research},\ }\href@noop {} {\bibfield  {journal} {\bibinfo  {journal} {International journal of qualitative methods}\ }\textbf {\bibinfo {volume} {1}},\ \bibinfo {pages} {13} (\bibinfo {year} {2002})}\BibitemShut {NoStop}%
\bibitem [{\citenamefont {Subramaniam}\ \emph {et~al.}(2024{\natexlab{a}})\citenamefont {Subramaniam}, \citenamefont {L'Fontaine}, \citenamefont {Bralin}, \citenamefont {Morphew}, \citenamefont {Rebello},\ and\ \citenamefont {Rebello}}]{ravi_perc_2024}%
  \BibitemOpen
  \bibfield  {author} {\bibinfo {author} {\bibfnamefont {R.}~\bibnamefont {Subramaniam}}, \bibinfo {author} {\bibfnamefont {C.}~\bibnamefont {L'Fontaine}}, \bibinfo {author} {\bibfnamefont {A.}~\bibnamefont {Bralin}}, \bibinfo {author} {\bibfnamefont {J.}~\bibnamefont {Morphew}}, \bibinfo {author} {\bibfnamefont {C.}~\bibnamefont {Rebello}},\ and\ \bibinfo {author} {\bibfnamefont {N.~S.}\ \bibnamefont {Rebello}},\ }\bibfield  {title} {\bibinfo {title} {Characterising {STEM} {W}ays of {T}hinking in engineering design ({ED})-based tasks},\ }in\ \href@noop {} {\emph {\bibinfo {booktitle} {Physics Education Research Conference Proceedings}}}\ (\bibinfo {year} {2024})\ \bibinfo {note} {({I}n press)}\BibitemShut {NoStop}%
\bibitem [{\citenamefont {Nihas}(2020)}]{nihas2020}%
  \BibitemOpen
  \bibfield  {author} {\bibinfo {author} {\bibfnamefont {P.}~\bibnamefont {Nihas}},\ }\bibfield  {title} {\bibinfo {title} {Chapter 7: Qualitative analysis},\ }in\ \href@noop {} {\emph {\bibinfo {booktitle} {Research design and methods: an applied guide for the scholar-practitioner}}}\ (\bibinfo  {publisher} {Sage Publications},\ \bibinfo {year} {2020})\ pp.\ \bibinfo {pages} {99--112}\BibitemShut {NoStop}%
\bibitem [{\citenamefont {Elliott}(2018)}]{elliott2018thinking}%
  \BibitemOpen
  \bibfield  {author} {\bibinfo {author} {\bibfnamefont {V.}~\bibnamefont {Elliott}},\ }\bibfield  {title} {\bibinfo {title} {Thinking about the coding process in qualitative data analysis},\ }\href@noop {} {\bibfield  {journal} {\bibinfo  {journal} {Qualitative report}\ }\textbf {\bibinfo {volume} {23}} (\bibinfo {year} {2018})}\BibitemShut {NoStop}%
\bibitem [{\citenamefont {Guest}\ \emph {et~al.}(2012)\citenamefont {Guest}, \citenamefont {MacQueen},\ and\ \citenamefont {Namey}}]{guest2012introduction}%
  \BibitemOpen
  \bibfield  {author} {\bibinfo {author} {\bibfnamefont {G.}~\bibnamefont {Guest}}, \bibinfo {author} {\bibfnamefont {K.~M.}\ \bibnamefont {MacQueen}},\ and\ \bibinfo {author} {\bibfnamefont {E.~E.}\ \bibnamefont {Namey}},\ }\bibfield  {title} {\bibinfo {title} {Introduction to applied thematic analysis},\ }\href@noop {} {\bibfield  {journal} {\bibinfo  {journal} {Applied thematic analysis}\ }\textbf {\bibinfo {volume} {3}},\ \bibinfo {pages} {1} (\bibinfo {year} {2012})}\BibitemShut {NoStop}%
\bibitem [{\citenamefont {Wa-Mbaleka}(2020)}]{wa2020researcher}%
  \BibitemOpen
  \bibfield  {author} {\bibinfo {author} {\bibfnamefont {S.}~\bibnamefont {Wa-Mbaleka}},\ }\bibfield  {title} {\bibinfo {title} {The researcher as an instrument},\ }in\ \href@noop {} {\emph {\bibinfo {booktitle} {Computer Supported Qualitative Research: New Trends on Qualitative Research (WCQR2019) 4}}}\ (\bibinfo {organization} {Springer},\ \bibinfo {year} {2020})\ pp.\ \bibinfo {pages} {33--41}\BibitemShut {NoStop}%
\bibitem [{\citenamefont {Creswell}\ and\ \citenamefont {Creswell}(2017)}]{creswell2017research}%
  \BibitemOpen
  \bibfield  {author} {\bibinfo {author} {\bibfnamefont {J.~W.}\ \bibnamefont {Creswell}}\ and\ \bibinfo {author} {\bibfnamefont {J.~D.}\ \bibnamefont {Creswell}},\ }\href@noop {} {\emph {\bibinfo {title} {Research Design: Qualitative, Quantitative, and Mixed Methods Approaches}}},\ \bibinfo {edition} {5th}\ ed.\ (\bibinfo  {publisher} {Sage Publications},\ \bibinfo {address} {Thousand Oaks, CA},\ \bibinfo {year} {2017})\ \bibinfo {note} {john W. Creswell: Department of Family Medicine, University of Michigan, J. David Creswell: Department of Psychology, Carnegie Mellon University}\BibitemShut {NoStop}%
\bibitem [{\citenamefont {Guba}(1981)}]{guba1981criteria}%
  \BibitemOpen
  \bibfield  {author} {\bibinfo {author} {\bibfnamefont {E.~G.}\ \bibnamefont {Guba}},\ }\bibfield  {title} {\bibinfo {title} {Criteria for assessing the trustworthiness of naturalistic inquiries},\ }\href@noop {} {\bibfield  {journal} {\bibinfo  {journal} {Ectj}\ }\textbf {\bibinfo {volume} {29}},\ \bibinfo {pages} {75} (\bibinfo {year} {1981})}\BibitemShut {NoStop}%
\bibitem [{\citenamefont {Patton}(2002)}]{patton2002qualitative}%
  \BibitemOpen
  \bibfield  {author} {\bibinfo {author} {\bibfnamefont {M.~Q.}\ \bibnamefont {Patton}},\ }\href@noop {} {\emph {\bibinfo {title} {Qualitative Research \& Evaluation Methods}}},\ \bibinfo {edition} {3rd}\ ed.\ (\bibinfo  {publisher} {Sage Publications},\ \bibinfo {year} {2002})\BibitemShut {NoStop}%
\bibitem [{\citenamefont {Perplexity~AI}(2024)}]{perplexity_ai}%
  \BibitemOpen
  \bibfield  {author} {\bibinfo {author} {\bibfnamefont {I.}~\bibnamefont {Perplexity~AI}},\ }\href {https://www.perplexity.ai/} {\bibinfo {title} {Perplexity ai}} (\bibinfo {year} {2024}),\ \bibinfo {note} {accessed: 2024-12-02}\BibitemShut {NoStop}%
\bibitem [{\citenamefont {OpenAI}(2024)}]{openai_chatgpt}%
  \BibitemOpen
  \bibfield  {author} {\bibinfo {author} {\bibnamefont {OpenAI}},\ }\href {https://chat.openai.com/} {\bibinfo {title} {Chatgpt}} (\bibinfo {year} {2024}),\ \bibinfo {note} {accessed: 2024-12-02}\BibitemShut {NoStop}%
\bibitem [{\citenamefont {Microsoft}(2024)}]{ms_copilot}%
  \BibitemOpen
  \bibfield  {author} {\bibinfo {author} {\bibnamefont {Microsoft}},\ }\href {https://copilot.microsoft.com/} {\bibinfo {title} {Microsoft copilot}} (\bibinfo {year} {2024}),\ \bibinfo {note} {accessed: 2024-12-02}\BibitemShut {NoStop}%
\bibitem [{\citenamefont {McDonald}\ \emph {et~al.}(2019)\citenamefont {McDonald}, \citenamefont {Schoenebeck},\ and\ \citenamefont {Forte}}]{mcdonald2019reliability}%
  \BibitemOpen
  \bibfield  {author} {\bibinfo {author} {\bibfnamefont {N.}~\bibnamefont {McDonald}}, \bibinfo {author} {\bibfnamefont {S.}~\bibnamefont {Schoenebeck}},\ and\ \bibinfo {author} {\bibfnamefont {A.}~\bibnamefont {Forte}},\ }\bibfield  {title} {\bibinfo {title} {Reliability and inter-rater reliability in qualitative research: Norms and guidelines for cscw and hci practice},\ }\href@noop {} {\bibfield  {journal} {\bibinfo  {journal} {Proceedings of the ACM on human-computer interaction}\ }\textbf {\bibinfo {volume} {3}},\ \bibinfo {pages} {1} (\bibinfo {year} {2019})}\BibitemShut {NoStop}%
\bibitem [{\citenamefont {O’Connor}\ and\ \citenamefont {Joffe}(2020)}]{oconnor_icr}%
  \BibitemOpen
  \bibfield  {author} {\bibinfo {author} {\bibfnamefont {C.}~\bibnamefont {O’Connor}}\ and\ \bibinfo {author} {\bibfnamefont {H.}~\bibnamefont {Joffe}},\ }\bibfield  {title} {\bibinfo {title} {Intercoder reliability in qualitative research: debates and practical guidelines},\ }\href@noop {} {\bibfield  {journal} {\bibinfo  {journal} {International journal of qualitative methods}\ }\textbf {\bibinfo {volume} {19}},\ \bibinfo {pages} {1609406919899220} (\bibinfo {year} {2020})}\BibitemShut {NoStop}%
\bibitem [{\citenamefont {Younas}\ \emph {et~al.}(2023)\citenamefont {Younas}, \citenamefont {F{\`a}bregues}, \citenamefont {Durante}, \citenamefont {Escalante}, \citenamefont {Inayat},\ and\ \citenamefont {Ali}}]{younas2023proposing}%
  \BibitemOpen
  \bibfield  {author} {\bibinfo {author} {\bibfnamefont {A.}~\bibnamefont {Younas}}, \bibinfo {author} {\bibfnamefont {S.}~\bibnamefont {F{\`a}bregues}}, \bibinfo {author} {\bibfnamefont {A.}~\bibnamefont {Durante}}, \bibinfo {author} {\bibfnamefont {E.~L.}\ \bibnamefont {Escalante}}, \bibinfo {author} {\bibfnamefont {S.}~\bibnamefont {Inayat}},\ and\ \bibinfo {author} {\bibfnamefont {P.}~\bibnamefont {Ali}},\ }\bibfield  {title} {\bibinfo {title} {Proposing the “miracle” narrative framework for providing thick description in qualitative research},\ }\href@noop {} {\bibfield  {journal} {\bibinfo  {journal} {International Journal of Qualitative Methods}\ }\textbf {\bibinfo {volume} {22}},\ \bibinfo {pages} {16094069221147162} (\bibinfo {year} {2023})}\BibitemShut {NoStop}%
\bibitem [{\citenamefont {Leeds-Hurwitz}(2019)}]{leeds-hurwitz2019}%
  \BibitemOpen
  \bibfield  {author} {\bibinfo {author} {\bibfnamefont {W.}~\bibnamefont {Leeds-Hurwitz}},\ }\href {https://doi.org/10.4135/9781526421036765746} {\emph {\bibinfo {title} {Sage Research Methods Foundations}}}\ (\bibinfo  {publisher} {SAGE Publications Ltd},\ \bibinfo {address} {London},\ \bibinfo {year} {2019})\ \bibinfo {note} {accessed: 2024-08-23}\BibitemShut {NoStop}%
\bibitem [{\citenamefont {Urquhart}(2022)}]{urquhart2022grounded}%
  \BibitemOpen
  \bibfield  {author} {\bibinfo {author} {\bibfnamefont {C.}~\bibnamefont {Urquhart}},\ }\href@noop {} {\emph {\bibinfo {title} {Grounded theory for qualitative research: A practical guide}}}\ (\bibinfo  {publisher} {Sage publications},\ \bibinfo {year} {2022})\BibitemShut {NoStop}%
\bibitem [{\citenamefont {Peel}(2020)}]{peel2020beginner}%
  \BibitemOpen
  \bibfield  {author} {\bibinfo {author} {\bibfnamefont {K.~L.}\ \bibnamefont {Peel}},\ }\bibfield  {title} {\bibinfo {title} {A beginner’s guide to applied educational research using thematic analysis},\ }\href@noop {} {\bibfield  {journal} {\bibinfo  {journal} {Practical Assessment Research and Evaluation}\ }\textbf {\bibinfo {volume} {25}} (\bibinfo {year} {2020})}\BibitemShut {NoStop}%
\bibitem [{\citenamefont {Maxwell}\ and\ \citenamefont {Chmiel}(2014)}]{maxwell2014generalization}%
  \BibitemOpen
  \bibfield  {author} {\bibinfo {author} {\bibfnamefont {J.~A.}\ \bibnamefont {Maxwell}}\ and\ \bibinfo {author} {\bibfnamefont {M.}~\bibnamefont {Chmiel}},\ }\bibfield  {title} {\bibinfo {title} {Generalization in and from qualitative analysis},\ }\href@noop {} {\bibfield  {journal} {\bibinfo  {journal} {The SAGE handbook of qualitative data analysis}\ }\textbf {\bibinfo {volume} {7}},\ \bibinfo {pages} {540} (\bibinfo {year} {2014})}\BibitemShut {NoStop}%
\bibitem [{\citenamefont {Kaufman}\ and\ \citenamefont {McCuish}(2002)}]{kaufman2002getting}%
  \BibitemOpen
  \bibfield  {author} {\bibinfo {author} {\bibfnamefont {J.~J.}\ \bibnamefont {Kaufman}}\ and\ \bibinfo {author} {\bibfnamefont {J.~D.}\ \bibnamefont {McCuish}},\ }\bibfield  {title} {\bibinfo {title} {Getting better solutions with brainstorming},\ }in\ \href@noop {} {\emph {\bibinfo {booktitle} {SAVE International Annual Conference Proceedings}}},\ Vol.~\bibinfo {volume} {37}\ (\bibinfo {year} {2002})\BibitemShut {NoStop}%
\bibitem [{\citenamefont {Dalal}\ \emph {et~al.}(2021)\citenamefont {Dalal}, \citenamefont {Carberry}, \citenamefont {Archambault} \emph {et~al.}}]{dalal2021developing}%
  \BibitemOpen
  \bibfield  {author} {\bibinfo {author} {\bibfnamefont {M.}~\bibnamefont {Dalal}}, \bibinfo {author} {\bibfnamefont {A.}~\bibnamefont {Carberry}}, \bibinfo {author} {\bibfnamefont {L.}~\bibnamefont {Archambault}}, \emph {et~al.},\ }\bibfield  {title} {\bibinfo {title} {Developing a ways of thinking framework for engineering education research},\ }\href@noop {} {\bibfield  {journal} {\bibinfo  {journal} {Studies in Engineering Education}\ }\textbf {\bibinfo {volume} {1}},\ \bibinfo {pages} {108} (\bibinfo {year} {2021})}\BibitemShut {NoStop}%
\bibitem [{\citenamefont {Slavit}\ \emph {et~al.}(2019)\citenamefont {Slavit}, \citenamefont {Grace},\ and\ \citenamefont {Lesseig}}]{slavit2019stem}%
  \BibitemOpen
  \bibfield  {author} {\bibinfo {author} {\bibfnamefont {D.}~\bibnamefont {Slavit}}, \bibinfo {author} {\bibfnamefont {E.}~\bibnamefont {Grace}},\ and\ \bibinfo {author} {\bibfnamefont {K.}~\bibnamefont {Lesseig}},\ }\bibfield  {title} {\bibinfo {title} {Stem ways of thinking.},\ }\href@noop {} {\bibfield  {journal} {\bibinfo  {journal} {North American Chapter of the International Group for the Psychology of Mathematics Education}\ } (\bibinfo {year} {2019})}\BibitemShut {NoStop}%
\bibitem [{\citenamefont {Hammer}(1995)}]{hammer1995student}%
  \BibitemOpen
  \bibfield  {author} {\bibinfo {author} {\bibfnamefont {D.}~\bibnamefont {Hammer}},\ }\bibfield  {title} {\bibinfo {title} {Student inquiry in a physics class discussion},\ }\href@noop {} {\bibfield  {journal} {\bibinfo  {journal} {Cognition and Instruction}\ }\textbf {\bibinfo {volume} {13}},\ \bibinfo {pages} {401} (\bibinfo {year} {1995})}\BibitemShut {NoStop}%
\bibitem [{\citenamefont {Greenhalgh}(2005)}]{greenhalgh2005can}%
  \BibitemOpen
  \bibfield  {author} {\bibinfo {author} {\bibfnamefont {T.}~\bibnamefont {Greenhalgh}},\ }\href@noop {} {\bibinfo {title} {Can ‘anecdote’ever be research?}} (\bibinfo {year} {2005})\BibitemShut {NoStop}%
\bibitem [{\citenamefont {{NSF Consulting}}(2024)}]{nsfconsulting2024}%
  \BibitemOpen
  \bibfield  {author} {\bibinfo {author} {\bibnamefont {{NSF Consulting}}},\ }\href {https://nsfconsulting.com.au/the-plural-of-anecdote-data/} {\bibinfo {title} {The plural of anecdote is not data}} (\bibinfo {year} {2024}),\ \bibinfo {note} {accessed: 2024-08-10}\BibitemShut {NoStop}%
\bibitem [{\citenamefont {{Kauffman Foundation}}(2024)}]{kauffman2024}%
  \BibitemOpen
  \bibfield  {author} {\bibinfo {author} {\bibnamefont {{Kauffman Foundation}}},\ }\href {https://www.kauffman.org/currents/the-plural-of-anecdote-is-data/} {\bibinfo {title} {The plural of anecdote is data}} (\bibinfo {year} {2024}),\ \bibinfo {note} {accessed: 2024-08-10}\BibitemShut {NoStop}%
\bibitem [{\citenamefont {Moye}\ \emph {et~al.}(2014)\citenamefont {Moye}, \citenamefont {DUGGER}, \citenamefont {WILLIAM},\ and\ \citenamefont {STARK-WEATHER}}]{moye2014learning}%
  \BibitemOpen
  \bibfield  {author} {\bibinfo {author} {\bibfnamefont {J.~J.}\ \bibnamefont {Moye}}, \bibinfo {author} {\bibfnamefont {J.}~\bibnamefont {DUGGER}}, \bibinfo {author} {\bibfnamefont {E.}~\bibnamefont {WILLIAM}},\ and\ \bibinfo {author} {\bibfnamefont {K.~N.}\ \bibnamefont {STARK-WEATHER}},\ }\bibfield  {title} {\bibinfo {title} {Is" learning by doing" important? a study of doing-based learning.},\ }\href@noop {} {\bibfield  {journal} {\bibinfo  {journal} {Technology \& Engineering Teacher}\ }\textbf {\bibinfo {volume} {74}} (\bibinfo {year} {2014})}\BibitemShut {NoStop}%
\bibitem [{\citenamefont {Flick}(1993)}]{flick1993meanings}%
  \BibitemOpen
  \bibfield  {author} {\bibinfo {author} {\bibfnamefont {L.~B.}\ \bibnamefont {Flick}},\ }\bibfield  {title} {\bibinfo {title} {The meanings of hands-on science},\ }\href@noop {} {\bibfield  {journal} {\bibinfo  {journal} {Journal of Science Teacher Education}\ }\textbf {\bibinfo {volume} {4}},\ \bibinfo {pages} {1} (\bibinfo {year} {1993})}\BibitemShut {NoStop}%
\bibitem [{\citenamefont {Darrah}\ \emph {et~al.}(2014)\citenamefont {Darrah}, \citenamefont {Humbert}, \citenamefont {Finstein}, \citenamefont {Simon},\ and\ \citenamefont {Hopkins}}]{darrah2014virtual}%
  \BibitemOpen
  \bibfield  {author} {\bibinfo {author} {\bibfnamefont {M.}~\bibnamefont {Darrah}}, \bibinfo {author} {\bibfnamefont {R.}~\bibnamefont {Humbert}}, \bibinfo {author} {\bibfnamefont {J.}~\bibnamefont {Finstein}}, \bibinfo {author} {\bibfnamefont {M.}~\bibnamefont {Simon}},\ and\ \bibinfo {author} {\bibfnamefont {J.}~\bibnamefont {Hopkins}},\ }\bibfield  {title} {\bibinfo {title} {Are virtual labs as effective as hands-on labs for undergraduate physics? a comparative study at two major universities},\ }\href@noop {} {\bibfield  {journal} {\bibinfo  {journal} {Journal of science education and technology}\ }\textbf {\bibinfo {volume} {23}},\ \bibinfo {pages} {803} (\bibinfo {year} {2014})}\BibitemShut {NoStop}%
\bibitem [{\citenamefont {Ekmekci}\ and\ \citenamefont {Gulacar}(2015)}]{ekmekci2015case}%
  \BibitemOpen
  \bibfield  {author} {\bibinfo {author} {\bibfnamefont {A.}~\bibnamefont {Ekmekci}}\ and\ \bibinfo {author} {\bibfnamefont {O.}~\bibnamefont {Gulacar}},\ }\bibfield  {title} {\bibinfo {title} {A case study for comparing the effectiveness of a computer simulation and a hands-on activity on learning electric circuits},\ }\href@noop {} {\bibfield  {journal} {\bibinfo  {journal} {Eurasia Journal of Mathematics, Science and Technology Education}\ }\textbf {\bibinfo {volume} {11}},\ \bibinfo {pages} {765} (\bibinfo {year} {2015})}\BibitemShut {NoStop}%
\bibitem [{\citenamefont {Fudenberg}\ \emph {et~al.}(2014)\citenamefont {Fudenberg}, \citenamefont {Levine} \emph {et~al.}}]{fudenberg2014learning}%
  \BibitemOpen
  \bibfield  {author} {\bibinfo {author} {\bibfnamefont {D.}~\bibnamefont {Fudenberg}}, \bibinfo {author} {\bibfnamefont {D.~K.}\ \bibnamefont {Levine}}, \emph {et~al.},\ }\bibfield  {title} {\bibinfo {title} {Learning with recency bias},\ }\href@noop {} {\bibfield  {journal} {\bibinfo  {journal} {Proceedings of the National Academy of Sciences}\ }\textbf {\bibinfo {volume} {111}},\ \bibinfo {pages} {10826} (\bibinfo {year} {2014})}\BibitemShut {NoStop}%
\bibitem [{\citenamefont {Siverling}\ \emph {et~al.}(2021)\citenamefont {Siverling}, \citenamefont {Moore}, \citenamefont {Suazo-Flores}, \citenamefont {Mathis},\ and\ \citenamefont {Guzey}}]{siverling2021initiates}%
  \BibitemOpen
  \bibfield  {author} {\bibinfo {author} {\bibfnamefont {E.~A.}\ \bibnamefont {Siverling}}, \bibinfo {author} {\bibfnamefont {T.~J.}\ \bibnamefont {Moore}}, \bibinfo {author} {\bibfnamefont {E.}~\bibnamefont {Suazo-Flores}}, \bibinfo {author} {\bibfnamefont {C.~A.}\ \bibnamefont {Mathis}},\ and\ \bibinfo {author} {\bibfnamefont {S.~S.}\ \bibnamefont {Guzey}},\ }\bibfield  {title} {\bibinfo {title} {What initiates evidence-based reasoning?: Situations that prompt students to support their design ideas and decisions},\ }\href@noop {} {\bibfield  {journal} {\bibinfo  {journal} {Journal of Engineering Education}\ }\textbf {\bibinfo {volume} {110}},\ \bibinfo {pages} {294} (\bibinfo {year} {2021})}\BibitemShut {NoStop}%
\bibitem [{\citenamefont {Gretton}\ \emph {et~al.}(2017)\citenamefont {Gretton}, \citenamefont {Bridges},\ and\ \citenamefont {Fraser}}]{gretton2017transforming}%
  \BibitemOpen
  \bibfield  {author} {\bibinfo {author} {\bibfnamefont {A.~L.}\ \bibnamefont {Gretton}}, \bibinfo {author} {\bibfnamefont {T.}~\bibnamefont {Bridges}},\ and\ \bibinfo {author} {\bibfnamefont {J.~M.}\ \bibnamefont {Fraser}},\ }\bibfield  {title} {\bibinfo {title} {Transforming physics educator identities: Tas help tas become teaching professionals},\ }\href@noop {} {\bibfield  {journal} {\bibinfo  {journal} {American Journal of Physics}\ }\textbf {\bibinfo {volume} {85}},\ \bibinfo {pages} {381} (\bibinfo {year} {2017})}\BibitemShut {NoStop}%
\bibitem [{\citenamefont {Stang}\ and\ \citenamefont {Roll}(2014)}]{stang2014interactions}%
  \BibitemOpen
  \bibfield  {author} {\bibinfo {author} {\bibfnamefont {J.~B.}\ \bibnamefont {Stang}}\ and\ \bibinfo {author} {\bibfnamefont {I.}~\bibnamefont {Roll}},\ }\bibfield  {title} {\bibinfo {title} {Interactions between teaching assistants and students boost engagement in physics labs},\ }\href@noop {} {\bibfield  {journal} {\bibinfo  {journal} {Physical Review Special Topics-Physics Education Research}\ }\textbf {\bibinfo {volume} {10}},\ \bibinfo {pages} {020117} (\bibinfo {year} {2014})}\BibitemShut {NoStop}%
\bibitem [{\citenamefont {Mashood}\ \emph {et~al.}(2020)\citenamefont {Mashood}, \citenamefont {Kumar},\ and\ \citenamefont {Mazumdar}}]{mashood2020approximations}%
  \BibitemOpen
  \bibfield  {author} {\bibinfo {author} {\bibfnamefont {K.}~\bibnamefont {Mashood}}, \bibinfo {author} {\bibfnamefont {A.}~\bibnamefont {Kumar}},\ and\ \bibinfo {author} {\bibfnamefont {A.}~\bibnamefont {Mazumdar}},\ }\bibfield  {title} {\bibinfo {title} {Approximations in physics: A pedagogic perspective.},\ }\href@noop {} {\bibfield  {journal} {\bibinfo  {journal} {Resonance: Journal of Science Education}\ }\textbf {\bibinfo {volume} {25}} (\bibinfo {year} {2020})}\BibitemShut {NoStop}%
\bibitem [{\citenamefont {Nuere}\ \emph {et~al.}(2022)\citenamefont {Nuere}, \citenamefont {Ruiz~G{\'o}mez},\ and\ \citenamefont {de~Miguel~{\'A}lvarez}}]{nuere2022sketch}%
  \BibitemOpen
  \bibfield  {author} {\bibinfo {author} {\bibfnamefont {S.}~\bibnamefont {Nuere}}, \bibinfo {author} {\bibfnamefont {M.~E.}\ \bibnamefont {Ruiz~G{\'o}mez}},\ and\ \bibinfo {author} {\bibfnamefont {L.}~\bibnamefont {de~Miguel~{\'A}lvarez}},\ }\bibfield  {title} {\bibinfo {title} {Sketch as a tool of thought in art and science},\ }\href@noop {} {\bibfield  {journal} {\bibinfo  {journal} {[Journal Name]}\ } (\bibinfo {year} {2022})}\BibitemShut {NoStop}%
\bibitem [{\citenamefont {Slattery}\ and\ \citenamefont {Langerock}(2002)}]{slattery2002blurring}%
  \BibitemOpen
  \bibfield  {author} {\bibinfo {author} {\bibfnamefont {P.}~\bibnamefont {Slattery}}\ and\ \bibinfo {author} {\bibfnamefont {N.}~\bibnamefont {Langerock}},\ }\bibfield  {title} {\bibinfo {title} {Blurring art and science: Synthetical moments on the borders},\ }\href@noop {} {\bibfield  {journal} {\bibinfo  {journal} {Curriculum Inquiry}\ }\textbf {\bibinfo {volume} {32}},\ \bibinfo {pages} {349} (\bibinfo {year} {2002})}\BibitemShut {NoStop}%
\bibitem [{\citenamefont {Sherin}(2001)}]{sherin2001students}%
  \BibitemOpen
  \bibfield  {author} {\bibinfo {author} {\bibfnamefont {B.~L.}\ \bibnamefont {Sherin}},\ }\bibfield  {title} {\bibinfo {title} {How students understand physics equations},\ }\href@noop {} {\bibfield  {journal} {\bibinfo  {journal} {Cognition and instruction}\ }\textbf {\bibinfo {volume} {19}},\ \bibinfo {pages} {479} (\bibinfo {year} {2001})}\BibitemShut {NoStop}%
\bibitem [{\citenamefont {Liu}\ and\ \citenamefont {Fang}(2016)}]{liu2016student}%
  \BibitemOpen
  \bibfield  {author} {\bibinfo {author} {\bibfnamefont {G.}~\bibnamefont {Liu}}\ and\ \bibinfo {author} {\bibfnamefont {N.}~\bibnamefont {Fang}},\ }\bibfield  {title} {\bibinfo {title} {Student misconceptions about force and acceleration in physics and engineering mechanics education},\ }\href@noop {} {\bibfield  {journal} {\bibinfo  {journal} {International Journal of Engineering Education}\ }\textbf {\bibinfo {volume} {32}},\ \bibinfo {pages} {19} (\bibinfo {year} {2016})}\BibitemShut {NoStop}%
\bibitem [{\citenamefont {Harte}(1988)}]{harte1988consider}%
  \BibitemOpen
  \bibfield  {author} {\bibinfo {author} {\bibfnamefont {J.}~\bibnamefont {Harte}},\ }\href@noop {} {\emph {\bibinfo {title} {Consider a spherical cow: A course in environmental problem solving}}}\ (\bibinfo  {publisher} {University Science Books},\ \bibinfo {year} {1988})\BibitemShut {NoStop}%
\bibitem [{\citenamefont {Verostek}\ \emph {et~al.}(2022)\citenamefont {Verostek}, \citenamefont {Griston}, \citenamefont {Botello},\ and\ \citenamefont {Zwickl}}]{verostek2022making}%
  \BibitemOpen
  \bibfield  {author} {\bibinfo {author} {\bibfnamefont {M.}~\bibnamefont {Verostek}}, \bibinfo {author} {\bibfnamefont {M.}~\bibnamefont {Griston}}, \bibinfo {author} {\bibfnamefont {J.}~\bibnamefont {Botello}},\ and\ \bibinfo {author} {\bibfnamefont {B.}~\bibnamefont {Zwickl}},\ }\bibfield  {title} {\bibinfo {title} {Making expert processes visible: How and why theorists use assumptions and analogies in their research},\ }\href@noop {} {\bibfield  {journal} {\bibinfo  {journal} {Physical Review Physics Education Research}\ }\textbf {\bibinfo {volume} {18}},\ \bibinfo {pages} {020143} (\bibinfo {year} {2022})}\BibitemShut {NoStop}%
\bibitem [{\citenamefont {Sirnoorkar}\ and\ \citenamefont {Laverty}(2023)}]{sirnoorkar2023analyzing}%
  \BibitemOpen
  \bibfield  {author} {\bibinfo {author} {\bibfnamefont {A.}~\bibnamefont {Sirnoorkar}}\ and\ \bibinfo {author} {\bibfnamefont {J.~T.}\ \bibnamefont {Laverty}},\ }\bibfield  {title} {\bibinfo {title} {Analyzing students’ assumptions to varying degree of prompting during problem solving},\ }\href@noop {} {\bibfield  {journal} {\bibinfo  {journal} {Physics Education Research Proceedings}\ ,\ \bibinfo {pages} {302}} (\bibinfo {year} {2023})}\BibitemShut {NoStop}%
\bibitem [{\citenamefont {Kuo}\ \emph {et~al.}(2013)\citenamefont {Kuo}, \citenamefont {Hull}, \citenamefont {Gupta},\ and\ \citenamefont {Elby}}]{kuo2013students}%
  \BibitemOpen
  \bibfield  {author} {\bibinfo {author} {\bibfnamefont {E.}~\bibnamefont {Kuo}}, \bibinfo {author} {\bibfnamefont {M.~M.}\ \bibnamefont {Hull}}, \bibinfo {author} {\bibfnamefont {A.}~\bibnamefont {Gupta}},\ and\ \bibinfo {author} {\bibfnamefont {A.}~\bibnamefont {Elby}},\ }\bibfield  {title} {\bibinfo {title} {How students blend conceptual and formal mathematical reasoning in solving physics problems},\ }\href@noop {} {\bibfield  {journal} {\bibinfo  {journal} {Science Education}\ }\textbf {\bibinfo {volume} {97}},\ \bibinfo {pages} {32} (\bibinfo {year} {2013})}\BibitemShut {NoStop}%
\bibitem [{\citenamefont {Pines}\ \emph {et~al.}(2002)\citenamefont {Pines}, \citenamefont {Nowak}, \citenamefont {Alnajjar}, \citenamefont {Gould},\ and\ \citenamefont {Bernardete}}]{pines2002integrating}%
  \BibitemOpen
  \bibfield  {author} {\bibinfo {author} {\bibfnamefont {D.}~\bibnamefont {Pines}}, \bibinfo {author} {\bibfnamefont {M.}~\bibnamefont {Nowak}}, \bibinfo {author} {\bibfnamefont {H.}~\bibnamefont {Alnajjar}}, \bibinfo {author} {\bibfnamefont {L.}~\bibnamefont {Gould}},\ and\ \bibinfo {author} {\bibfnamefont {D.}~\bibnamefont {Bernardete}},\ }\bibfield  {title} {\bibinfo {title} {Integrating science and math into the freshman engineering design course},\ }in\ \href@noop {} {\emph {\bibinfo {booktitle} {2002 Annual Conference}}}\ (\bibinfo {year} {2002})\ pp.\ \bibinfo {pages} {7--701}\BibitemShut {NoStop}%
\bibitem [{\citenamefont {Narode}(2011)}]{narode2011math}%
  \BibitemOpen
  \bibfield  {author} {\bibinfo {author} {\bibfnamefont {R.~B.}\ \bibnamefont {Narode}},\ }\bibfield  {title} {\bibinfo {title} {‘‘math in a can’’: Teaching mathematics and engineering design},\ }\href@noop {} {\bibfield  {journal} {\bibinfo  {journal} {Journal of Pre-College Engineering Education Research (J-PEER)}\ }\textbf {\bibinfo {volume} {1}},\ \bibinfo {pages} {3} (\bibinfo {year} {2011})}\BibitemShut {NoStop}%
\bibitem [{\citenamefont {Bush}\ \emph {et~al.}(2018)\citenamefont {Bush}, \citenamefont {Karp}, \citenamefont {Cox}, \citenamefont {Cook}, \citenamefont {Albanese},\ and\ \citenamefont {Karp}}]{bush2018design}%
  \BibitemOpen
  \bibfield  {author} {\bibinfo {author} {\bibfnamefont {S.~B.}\ \bibnamefont {Bush}}, \bibinfo {author} {\bibfnamefont {K.~S.}\ \bibnamefont {Karp}}, \bibinfo {author} {\bibfnamefont {R.}~\bibnamefont {Cox}}, \bibinfo {author} {\bibfnamefont {K.~L.}\ \bibnamefont {Cook}}, \bibinfo {author} {\bibfnamefont {J.}~\bibnamefont {Albanese}},\ and\ \bibinfo {author} {\bibfnamefont {M.}~\bibnamefont {Karp}},\ }\bibfield  {title} {\bibinfo {title} {Design thinking framework: Shaping powerful mathematics},\ }\href@noop {} {\bibfield  {journal} {\bibinfo  {journal} {Mathematics Teaching in the Middle School}\ }\textbf {\bibinfo {volume} {23}},\ \bibinfo {pages} {e1} (\bibinfo {year} {2018})}\BibitemShut {NoStop}%
\bibitem [{\citenamefont {Lambert}\ \emph {et~al.}(2021)\citenamefont {Lambert}, \citenamefont {Imm}, \citenamefont {Schuck}, \citenamefont {Choi},\ and\ \citenamefont {McNiff}}]{lambert2021udl}%
  \BibitemOpen
  \bibfield  {author} {\bibinfo {author} {\bibfnamefont {R.}~\bibnamefont {Lambert}}, \bibinfo {author} {\bibfnamefont {K.}~\bibnamefont {Imm}}, \bibinfo {author} {\bibfnamefont {R.}~\bibnamefont {Schuck}}, \bibinfo {author} {\bibfnamefont {S.}~\bibnamefont {Choi}},\ and\ \bibinfo {author} {\bibfnamefont {A.}~\bibnamefont {McNiff}},\ }\bibfield  {title} {\bibinfo {title} {" udl is the what, design thinking is the how:" designing for differentiation in mathematics.},\ }\href@noop {} {\bibfield  {journal} {\bibinfo  {journal} {Mathematics Teacher Education and Development}\ }\textbf {\bibinfo {volume} {23}},\ \bibinfo {pages} {54} (\bibinfo {year} {2021})}\BibitemShut {NoStop}%
\bibitem [{\citenamefont {Hacker}\ and\ \citenamefont {Dunlosky}(2003)}]{hacker2003not}%
  \BibitemOpen
  \bibfield  {author} {\bibinfo {author} {\bibfnamefont {D.~J.}\ \bibnamefont {Hacker}}\ and\ \bibinfo {author} {\bibfnamefont {J.}~\bibnamefont {Dunlosky}},\ }\bibfield  {title} {\bibinfo {title} {Not all metacognition is created equal.},\ }\href@noop {} {\bibfield  {journal} {\bibinfo  {journal} {New Directions for Teaching \& Learning}\ }\textbf {\bibinfo {volume} {2003}} (\bibinfo {year} {2003})}\BibitemShut {NoStop}%
\bibitem [{\citenamefont {Kanarev}(2011)}]{kanarevnobel}%
  \BibitemOpen
  \bibfield  {author} {\bibinfo {author} {\bibfnamefont {P.~M.}\ \bibnamefont {Kanarev}},\ }\bibfield  {title} {\bibinfo {title} {Nobel result in physics has been obtained by the trial and error method},\ }\href@noop {} {\bibfield  {journal} {\bibinfo  {journal} {Proceedings of the NPA}\ }\textbf {\bibinfo {volume} {1}} (\bibinfo {year} {2011})},\ \bibinfo {note} {professor Emeritus of Physics, Pushkin Str. 11, Apt. 19, 350063 Krasnodar, RUSSIA, e-mail: kanphil@mail.ru}\BibitemShut {NoStop}%
\bibitem [{\citenamefont {Wills}(1996)}]{wills1996trial}%
  \BibitemOpen
  \bibfield  {author} {\bibinfo {author} {\bibfnamefont {I.}~\bibnamefont {Wills}},\ }\bibfield  {title} {\bibinfo {title} {Trial and error},\ }\href@noop {} {\bibfield  {journal} {\bibinfo  {journal} {Journal Name}\ } (\bibinfo {year} {1996})}\BibitemShut {NoStop}%
\bibitem [{\citenamefont {Bell}(2007)}]{bell2007teachers}%
  \BibitemOpen
  \bibfield  {author} {\bibinfo {author} {\bibfnamefont {D.~M.}\ \bibnamefont {Bell}},\ }\bibfield  {title} {\bibinfo {title} {Do teachers think that methods are dead?},\ }\href@noop {} {\bibfield  {journal} {\bibinfo  {journal} {ELT journal}\ }\textbf {\bibinfo {volume} {61}},\ \bibinfo {pages} {135} (\bibinfo {year} {2007})}\BibitemShut {NoStop}%
\bibitem [{\citenamefont {Wilson}(1994)}]{wilson1994there}%
  \BibitemOpen
  \bibfield  {author} {\bibinfo {author} {\bibfnamefont {S.~M.}\ \bibnamefont {Wilson}},\ }\href@noop {} {\emph {\bibinfo {title} {Is there a method in this madness?}}}\ (\bibinfo  {publisher} {Citeseer},\ \bibinfo {year} {1994})\BibitemShut {NoStop}%
\bibitem [{\citenamefont {Rajagopalan}(2008)}]{rajagopalan2008madness}%
  \BibitemOpen
  \bibfield  {author} {\bibinfo {author} {\bibfnamefont {K.}~\bibnamefont {Rajagopalan}},\ }\bibfield  {title} {\bibinfo {title} {From madness in method to method in madness},\ }\href@noop {} {\bibfield  {journal} {\bibinfo  {journal} {ELT Journal}\ }\textbf {\bibinfo {volume} {62}},\ \bibinfo {pages} {84} (\bibinfo {year} {2008})}\BibitemShut {NoStop}%
\bibitem [{\citenamefont {Kember}(2003)}]{kember2003control}%
  \BibitemOpen
  \bibfield  {author} {\bibinfo {author} {\bibfnamefont {D.}~\bibnamefont {Kember}},\ }\bibfield  {title} {\bibinfo {title} {To control or not to control: The question of whether experimental designs are appropriate for evaluating teaching innovations in higher education},\ }\href@noop {} {\bibfield  {journal} {\bibinfo  {journal} {Assessment \& Evaluation in Higher Education}\ }\textbf {\bibinfo {volume} {28}},\ \bibinfo {pages} {89} (\bibinfo {year} {2003})}\BibitemShut {NoStop}%
\bibitem [{\citenamefont {Kpamma}\ \emph {et~al.}(2017)\citenamefont {Kpamma}, \citenamefont {Adjei-Kumi}, \citenamefont {Ayarkwa},\ and\ \citenamefont {Adinyira}}]{kpamma2017participatory}%
  \BibitemOpen
  \bibfield  {author} {\bibinfo {author} {\bibfnamefont {Z.~E.}\ \bibnamefont {Kpamma}}, \bibinfo {author} {\bibfnamefont {T.}~\bibnamefont {Adjei-Kumi}}, \bibinfo {author} {\bibfnamefont {J.}~\bibnamefont {Ayarkwa}},\ and\ \bibinfo {author} {\bibfnamefont {E.}~\bibnamefont {Adinyira}},\ }\bibfield  {title} {\bibinfo {title} {Participatory design, wicked problems, choosing by advantages},\ }\href@noop {} {\bibfield  {journal} {\bibinfo  {journal} {Engineering, Construction and Architectural Management}\ }\textbf {\bibinfo {volume} {24}},\ \bibinfo {pages} {289} (\bibinfo {year} {2017})}\BibitemShut {NoStop}%
\bibitem [{\citenamefont {Hathcock}\ \emph {et~al.}(2015)\citenamefont {Hathcock}, \citenamefont {Dickerson}, \citenamefont {Eckhoff},\ and\ \citenamefont {Katsioloudis}}]{hathcock2015scaffolding}%
  \BibitemOpen
  \bibfield  {author} {\bibinfo {author} {\bibfnamefont {S.~J.}\ \bibnamefont {Hathcock}}, \bibinfo {author} {\bibfnamefont {D.~L.}\ \bibnamefont {Dickerson}}, \bibinfo {author} {\bibfnamefont {A.}~\bibnamefont {Eckhoff}},\ and\ \bibinfo {author} {\bibfnamefont {P.}~\bibnamefont {Katsioloudis}},\ }\bibfield  {title} {\bibinfo {title} {Scaffolding for creative product possibilities in a design-based stem activity},\ }\href@noop {} {\bibfield  {journal} {\bibinfo  {journal} {Research in science education}\ }\textbf {\bibinfo {volume} {45}},\ \bibinfo {pages} {727} (\bibinfo {year} {2015})}\BibitemShut {NoStop}%
\bibitem [{\citenamefont {Berliner}(2002)}]{berliner2002comment}%
  \BibitemOpen
  \bibfield  {author} {\bibinfo {author} {\bibfnamefont {D.~C.}\ \bibnamefont {Berliner}},\ }\bibfield  {title} {\bibinfo {title} {Comment: Educational research: The hardest science of all},\ }\href@noop {} {\bibfield  {journal} {\bibinfo  {journal} {Educational researcher}\ }\textbf {\bibinfo {volume} {31}},\ \bibinfo {pages} {18} (\bibinfo {year} {2002})}\BibitemShut {NoStop}%
\bibitem [{\citenamefont {Schmidt}\ and\ \citenamefont {Moust}(2000)}]{schmidt2000factors}%
  \BibitemOpen
  \bibfield  {author} {\bibinfo {author} {\bibfnamefont {H.~G.}\ \bibnamefont {Schmidt}}\ and\ \bibinfo {author} {\bibfnamefont {J.~H.}\ \bibnamefont {Moust}},\ }\bibfield  {title} {\bibinfo {title} {Factors affecting small-group tutorial learning: A review of research},\ }\href@noop {} {\bibfield  {journal} {\bibinfo  {journal} {Problem-based learning}\ ,\ \bibinfo {pages} {19}} (\bibinfo {year} {2000})}\BibitemShut {NoStop}%
\bibitem [{\citenamefont {Rivard}\ and\ \citenamefont {Straw}(2000)}]{rivard2000effect}%
  \BibitemOpen
  \bibfield  {author} {\bibinfo {author} {\bibfnamefont {L.~P.}\ \bibnamefont {Rivard}}\ and\ \bibinfo {author} {\bibfnamefont {S.~B.}\ \bibnamefont {Straw}},\ }\bibfield  {title} {\bibinfo {title} {The effect of talk and writing on learning science: An exploratory study},\ }\href@noop {} {\bibfield  {journal} {\bibinfo  {journal} {Science education}\ }\textbf {\bibinfo {volume} {84}},\ \bibinfo {pages} {566} (\bibinfo {year} {2000})}\BibitemShut {NoStop}%
\bibitem [{\citenamefont {Huffmyer}\ and\ \citenamefont {Lemus}(2019)}]{huffmyer2019graduate}%
  \BibitemOpen
  \bibfield  {author} {\bibinfo {author} {\bibfnamefont {A.~S.}\ \bibnamefont {Huffmyer}}\ and\ \bibinfo {author} {\bibfnamefont {J.~D.}\ \bibnamefont {Lemus}},\ }\bibfield  {title} {\bibinfo {title} {Graduate ta teaching behaviors impact student achievement in a research-based undergraduate science course},\ }\href@noop {} {\bibfield  {journal} {\bibinfo  {journal} {Journal of College Science Teaching}\ }\textbf {\bibinfo {volume} {48}},\ \bibinfo {pages} {56} (\bibinfo {year} {2019})}\BibitemShut {NoStop}%
\bibitem [{\citenamefont {White}\ \emph {et~al.}(2019)\citenamefont {White}, \citenamefont {Svihla}, \citenamefont {Chen}, \citenamefont {Hynson}, \citenamefont {Drackert}, \citenamefont {James}, \citenamefont {Saul},\ and\ \citenamefont {Megli}}]{white2019validating}%
  \BibitemOpen
  \bibfield  {author} {\bibinfo {author} {\bibfnamefont {L.}~\bibnamefont {White}}, \bibinfo {author} {\bibfnamefont {V.}~\bibnamefont {Svihla}}, \bibinfo {author} {\bibfnamefont {Y.}~\bibnamefont {Chen}}, \bibinfo {author} {\bibfnamefont {T.}~\bibnamefont {Hynson}}, \bibinfo {author} {\bibfnamefont {I.}~\bibnamefont {Drackert}}, \bibinfo {author} {\bibfnamefont {J.}~\bibnamefont {James}}, \bibinfo {author} {\bibfnamefont {C.}~\bibnamefont {Saul}},\ and\ \bibinfo {author} {\bibfnamefont {A.}~\bibnamefont {Megli}},\ }\bibfield  {title} {\bibinfo {title} {Validating a measure of problem-framing ability to support evidence-based teaching practice},\ }in\ \href@noop {} {\emph {\bibinfo {booktitle} {Proceedings of the ASEE 126th Annual Conference and Exhibition}}}\ (\bibinfo {year} {2019})\BibitemShut {NoStop}%
\bibitem [{\citenamefont {Radder}(2009)}]{radder2009science}%
  \BibitemOpen
  \bibfield  {author} {\bibinfo {author} {\bibfnamefont {H.}~\bibnamefont {Radder}},\ }\bibfield  {title} {\bibinfo {title} {Science, technology and the science-technology relationship},\ }in\ \href@noop {} {\emph {\bibinfo {booktitle} {Philosophy of technology and engineering sciences}}}\ (\bibinfo  {publisher} {Elsevier},\ \bibinfo {year} {2009})\ pp.\ \bibinfo {pages} {65--91}\BibitemShut {NoStop}%
\bibitem [{\citenamefont {Davis}\ and\ \citenamefont {Museus}(2019)}]{davis2019deficit}%
  \BibitemOpen
  \bibfield  {author} {\bibinfo {author} {\bibfnamefont {L.~P.}\ \bibnamefont {Davis}}\ and\ \bibinfo {author} {\bibfnamefont {S.~D.}\ \bibnamefont {Museus}},\ }\bibfield  {title} {\bibinfo {title} {What is deficit thinking? an analysis of conceptualizations of deficit thinking and implications for scholarly research},\ }\href@noop {} {\bibfield  {journal} {\bibinfo  {journal} {NCID Currents}\ }\textbf {\bibinfo {volume} {1}} (\bibinfo {year} {2019})}\BibitemShut {NoStop}%
\bibitem [{\citenamefont {Shiner}(2012)}]{shiner2012blurred}%
  \BibitemOpen
  \bibfield  {author} {\bibinfo {author} {\bibfnamefont {L.}~\bibnamefont {Shiner}},\ }\bibfield  {title} {\bibinfo {title} {“blurred boundaries”? rethinking the concept of craft and its relation to art and design},\ }\href@noop {} {\bibfield  {journal} {\bibinfo  {journal} {Philosophy Compass}\ }\textbf {\bibinfo {volume} {7}},\ \bibinfo {pages} {230} (\bibinfo {year} {2012})}\BibitemShut {NoStop}%
\bibitem [{\citenamefont {Eisner}\ and\ \citenamefont {Powell}(2002)}]{eisner2002art}%
  \BibitemOpen
  \bibfield  {author} {\bibinfo {author} {\bibfnamefont {E.}~\bibnamefont {Eisner}}\ and\ \bibinfo {author} {\bibfnamefont {K.}~\bibnamefont {Powell}},\ }\bibfield  {title} {\bibinfo {title} {Art in science?},\ }\href@noop {} {\bibfield  {journal} {\bibinfo  {journal} {Curriculum Inquiry}\ }\textbf {\bibinfo {volume} {32}},\ \bibinfo {pages} {131} (\bibinfo {year} {2002})}\BibitemShut {NoStop}%
\bibitem [{\citenamefont {Team}(2024)}]{jupyter2024}%
  \BibitemOpen
  \bibfield  {author} {\bibinfo {author} {\bibfnamefont {J.~D.}\ \bibnamefont {Team}},\ }\href {https://jupyter.org/} {\bibinfo {title} {Jupyter}} (\bibinfo {year} {2024}),\ \bibinfo {note} {accessed: 2024-12-02}\BibitemShut {NoStop}%
\bibitem [{\citenamefont {Subramaniam}\ \emph {et~al.}(2024{\natexlab{b}})\citenamefont {Subramaniam}, \citenamefont {Morphew}, \citenamefont {Rebello},\ and\ \citenamefont {Rebello}}]{subramaniam2024presentingstemwaysthinking}%
  \BibitemOpen
  \bibfield  {author} {\bibinfo {author} {\bibfnamefont {R.~C.}\ \bibnamefont {Subramaniam}}, \bibinfo {author} {\bibfnamefont {J.~W.}\ \bibnamefont {Morphew}}, \bibinfo {author} {\bibfnamefont {C.~M.}\ \bibnamefont {Rebello}},\ and\ \bibinfo {author} {\bibfnamefont {N.~S.}\ \bibnamefont {Rebello}},\ }\href {https://arxiv.org/abs/2411.11654} {\bibinfo {title} {Presenting a stem ways of thinking framework for engineering design-based physics problems}} (\bibinfo {year} {2024}{\natexlab{b}}),\ \Eprint {https://arxiv.org/abs/2411.11654} {arXiv:2411.11654 [physics.ed-ph]} \BibitemShut {NoStop}%
\end{thebibliography}%
\clearpage
\appendix
\onecolumngrid

\appendix
\onecolumngrid
\section*{Appendix A} 
\renewcommand{\arraystretch}{1}
\begin{table*}[htbp]
\begin{center}
\captionsetup{justification=raggedright,belowskip=0pt} % Adjusted caption setup
\caption{Stage 2 Coding. Multi-layered coding schema for coding transcripts (lab 06 - 09). Data Reduction is inspired by the Gioia Chart \cite[p.21]{gioia2013seeking}. The I-order codes are closest to the raw data (transcripts). Example statements for the codes (not given here for want of space) may be evident either as actual quotes or through descriptions in the `Findings \&\ Discussion' section.
\newline{DST - Design Thinking; SCT - Science Thinking; DA - Design Aspects; CC - Criteria and Constraints; ME - Metrics; SH - Stakeholders; PCP - Physics Concepts \&\ Principles; MAT - Mathematical Thinking; ES - Environmental Science; MS - Material Science. 
}}
\begin{ruledtabular}
\label{tab:stage_2_code_transcripts} 
\begin{tabular}{p{0.10\linewidth} p{0.15\linewidth} p{0.6\linewidth}}
\textbf{III-order code} & \textbf{II-order code} & \textbf{I-order code} \\
\hline
DST & DA & Economics - Cost \\
    &    & Mechanisms / Design Detail - no / low mention \\
    &    & Mechanisms / Design Detail - moderate / high detail \\
    &    & Ensure safe landing \\
    &    & Environmentally safe \\
    &    & Design limitations \\
    &    & Incorrect / inaccurate statements \\
    & CC  & Payload weight \\
    &    & 150 m - distance \\
    &    & Minimum C-footprint \\
    &    & Habitat Integrity \\
    & ME  & Delivery accuracy  \\
    &    & Payload dimensions \\
    &    & Carbon footprint \\
    &    & Habitat Integrity \\
    & SH & Engineers / Human workers \\
    &    & Flora and Fauna \\
    &    & Others  \\
\hline
SCT  & PCP  & Thinking about the path; Trajectory calculations.\\
    &    & Physics principles - low / no detail  \\
    &    & Physics principles - elaborated  \\
    &    & Physics terms / facts / ideas/concepts - low / no detail \\
    &    & Physics terms / facts/ideas / concepts - elaborated \\
    &    & Conceptually incorrect / factually incorrect statements.  \\
    & MAT  & Stated, in low / no detail. \\
    &    & Presented in some detail. \\
    & ES & Mere mention - direct / indirect. \\
    & MS & Mere mention - direct / indirect.  \\   
\end{tabular}
\end{ruledtabular}
\end{center}
\vspace{-0.5cm}
\end{table*}

\section*{Appendix B}
\renewcommand{\arraystretch}{1}
\begin{table*}[htbp]
\begin{center}
\captionsetup{justification=raggedright,belowskip=0pt} % Adjusted caption setup
\caption{Stage 2 Coding. Multi-layered coding schema for coding written reports (Lab 10). Data Reduction is inspired by the Gioia Chart \cite[p.21]{gioia2013seeking}. The I-order codes are closest to the raw data (written reports). Example statements for the codes (not given here for want of space) may be evident either as actual quotes or through descriptions in the `Findings \&\ Discussion' section.
\newline{DST - Design Thinking; SCT - Science Thinking; DA - Design Aspects; CC - Criteria and Constraints; ME - Metrics; SH - Stakeholders; DL-Design Limitations; AA - Assumptions \&\ Approximations; ; FC - Physics Facts \&\ Concepts; ; PP - Physics Principles (Momentum, Energy, Angular Momentum Principles); MAT - Mathematical Thinking; ES - Environmental Science; MS - Material Science; IT - Iterations; SK - Sketch/Diagram. 
}}
\begin{ruledtabular}
\label{tab:stage_2_code_reports} 
\begin{tabular}{p{0.10\linewidth} p{0.15\linewidth} p{0.6\linewidth}}
\textbf{III-order code} & \textbf{II-order code} & \textbf{I-order code} \\
\hline
DST & SH, CC, ME & Does the written work have anything significantly new/different as compared to the group discussions? \\
    &  IT-Level 1  & Team states skeletal details.   There is very little meaningful iteration. There's nothing new compared to what's seen in the group's prior discussions. \\
    &  IT-Level 2  & Team states skeletal details.   There is some meaningful iteration. There are a few changes the group has effected in the report building on the prior discussions.  \\
    &  IT-Level 3  & Team provides significantly more details, not seen in the prior discussions. \\
    &  SK-Level 1  & The sketch is skeletal, with barely anything new to the text / earlier discussions. \\
    &  SK-Level 2  & The sketch has new information not seen in the text / earlier discussions. \\
    &  SK-Level 3  & The sketch has significant information beyond text / earlier discussions. \\
    &  AA-Level 1  & Vague statements.(see Section~\ref{sec:V.F(iii)} \\
    &  AA-Level 2  & Cursory / generic statements. (see Section~\ref{sec:V.F(iii)}  \\
    &  AA-Level 3  & Thoughtful statements. (see Section~\ref{sec:V.F(iii)}\\
    &  DL-Level 1  & Vague statements. \\
    &  DL-Level 2  & Cursory / generic statements.  \\
    &  DL-Level 3  & Thoughtful statements. \\
\hline
SCT &  FC-Level 1   & Mostly a  repetition of what was present in the prior discussions. \\
    &  FC-Level 2   & There were a few new facts, but nothing significant. \\
    &  FC-Level 3   & Several new ideas were presented, but not elaborated. \\
    &  FC-Level 4   & Several new ideas were presented and elaborated. \\
    &  PP-Level 1  & Mostly repetition of what was present in the prior discussions. \\
    &  PP-Level 2  & Team elaborated on the principles used in prior discussions. \\
    &  SK-Level 1  & The sketch is skeletal, with barely anything new to the text / earlier discussions. \\
    &  SK-Level 2  & The sketch has new information not seen in the text / earlier discussions. \\
    &  SK-Level 3  & The sketch has significant information beyond text / earlier discussions. \\
    &  AA-Level 1  & Vague statements. (see Section~\ref{sec:V.F(iii)}\\
    &  AA-Level 2  & Cursory / generic statements. (see Section~\ref{sec:V.F(iii)} \\
    &  AA-Level 3  & Thoughtful statements. (see Section~\ref{sec:V.F(iii)}\\
    &  ES          & Nothing new when compared to the discussions.   \\
    &  MAT-Level 1  & No math. \\
    &  MAT-Level 2  & Some equations / formulas in text.   \\
    &  MAT-Level 3  & Some equations / formulas in sketch.  \\
  
\end{tabular}
\end{ruledtabular}
\end{center}
\vspace{-0.5cm}
\end{table*}

\end{document}